\documentclass[11pt]{article}
\usepackage[latin1 ]{inputenc}
\usepackage[dvips ]{graphicx}
\usepackage{amsmath}
\usepackage{amssymb}
\usepackage{placeins}	
\usepackage{axodraw}
\usepackage{paragraph}
\begin{document}
\begin{center}
\Large\bf
\vspace*{1.5cm}
Neutralino annihilation processes in the minimal supergravity model\\
\end{center}
\vspace{0.6cm}
\begin{center}
\large\bf An undergraduate project report
\end{center}
\vspace{0.6cm}
\begin{center}
Tim Stefaniak\footnote{Exchange student from University of Göttingen, Germany. Electronic address: \tt tim.stefaniak@fysast.uu.se}\\[0.4cm]
{\sl Supervisor:} Dr. Nazila Mahmoudi\\[0.4cm]
{\sl High Energy Physics, Dept. of Physics and Astronomy, Uppsala University, Box 535, SE-75121 Uppsala, Sweden}\\
\vspace{0.6cm}
{June 2008}
\end{center}
\vspace{0.6cm}
\begin{abstract}
We study all tree-level neutralino annihilation processes concerning energy-dependence at the mSUGRA benchmark point SPS1a using tree-level calculators {\scshape{MadGraph}} and {\scshape{CompHEP}} as well as {\scshape{FormCalc}}. We further investigate the total tree-level neutralino annihilation cross section by scanning over the mSUGRA parameters and identifying the dominant channels in every region. The results for the cross sections are presented in the $m_{1/2}$-$m_{0}$ planes for different values of $\tan \beta$ and $A_0$. We identify as well the resonances due to the exchange of the heavy Higgs bosons $A^0$ and $H^0$ in the s-channel of many neutralino annihilation processes, where the total cross section increases significantly if these Higgs bosons can be produced close to on-shell. Finally, we provide a few comments on the neutralino annihilation processes at loop-level.
\end{abstract}
\newpage
\section{Introduction}
In recent years great efforts have been performed in the development of theories beyond the standard model of particle physics. One of the most promising theories is supersymmetry (SUSY). In supersymmetric models many new particles are introduced and some of them are expected to be found at the upcoming collider LHC at CERN.\\
Supersymmetry is able to solve the most severe problems of the standard model in a quite natural way. In addition it may supply a promising candidate for dark matter under the assumption of an additional discrete symmetry, i.e. R-parity conservation. Under this symmetry SUSY-particles can only be created and annihilated in pairs, so that supersymmetric decay chains lead to a stable lightest supersymmetric particle (LSP), which is the candidate for dark matter.
In most models this LSP is the lightest neutralino. Assuming this particle to be the weakly interacting particle the dark matter of the universe is made of, the results from measurements of the dark matter relic density from the cosmological observations by WMAP \cite{WMAP} can be used to put constraints on supersymmetric models \cite{WMAPconstraints,constraints2}. For these purposes it is important to know how the neutralinos may annihilate themselves.\\
\\
In this report we study the cross section of neutralino annihilation processes at tree-level in the model of minimal supergravity (mSUGRA), which we review briefly in section \ref{sec_mSUGRA}. After some remarks on the neutralino annihilation cross sections and the connection to cosmology in section \ref{sec_cosmo}, we focus on annihilation processes at tree-level in section \ref{sec_treelevel}. Here we discuss first the possible final states and Feynman diagrams (\ref{sec_tree1}). Then a brief review on the tree-level calculators {\scshape{CompHEP}} \cite{CompHEP} and {\scshape{MadGraph}} \cite{MadGraph} is given (\ref{sec_tree2}). We study the different tree-level neutralino annihilation processes at the mSUGRA benchmark point SPS1a in section \ref{sec_tree3} concerning energy dependence and channel dominance. 
The mSUGRA parameter dependence of the total annihilation cross section at tree-level is analyzed in section \ref{sec_tree4}. Here we point out the main contributing channels for specific regions in the parameter space. In Appendix A we study in detail the contributions and interferences of the different diagrams in the process $\chi\chi \rightarrow Z^0Z^0$, trying to get a better understanding of the energy dependence plots presented in section \ref{sec_treeenergy}. We summarize our results in section \ref{summary} and give an outlook on future investigations including loop-level corrections.
\section{Definition of the supersymmetric model}\label{sec_mSUGRA}
This study on neutralino annihilation processes is evaluated and interpreted in the minimal supergravity (mSUGRA) model. This model is defined by only 4 independent parameters plus a sign (in addition to the 18 standard model parameters):
\begin{eqnarray}
m_0, \,m_{1/2},\, A_0, \,\tan \beta, \, \mathrm{sign} (\mu)
\end{eqnarray}
where $m_0$ is the soft symmetry breaking scalar mass (universal for all flavors, at the scale $M_{GUT}$ of Grand Unification), $m_{1/2}$ the universal supersymmetry breaking gaugino mass, $A_0$ the universal supersymmetry breaking trilinear scalar interactions, $\tan\beta$ the ratio of the two Higgs doublet vacuum expectation values and $\mathrm{sign}(\mu)$ the sign of the supersymmetric higgsino mass parameter.\\
\\
In this model, R-Parity is assumed to be conserved, i.e. there is a discrete symmetry $R=(-1)^{3(B-L) + 2S}$ of the theory, where $B$ and $L$ are baryon and lepton number operators and $S$ the spin. This imposes, that $R=1$ for standard model particles and $R=-1$ for their superpartners, which leads consequently to a stable lightest supersymmetric particle (LSP), since the decay of SUSY-particles to standard model particles is forbidden. In mSUGRA, the LSP is most likely the lightest neutralino $\chi_1^0$, in the following sometimes only denoted by $\chi$. 
\section{Neutralino annihilation cross sections}\label{sec_cosmo}
The annihilation of neutralino pairs plays a central role in neutralino cosmology. Its cross section is needed for calculations of the cosmological neutralino relic abundance, the flux of energetic neutrinos from neutralino annihilation in the Sun and Earth and fluxes of anomalous cosmic rays produced by neutralino annihilation in the galactic halo. For these purposes, it is generally sufficient to expand\footnote{This approximation is valid unless the neutralino annihilation takes place near a pole in the cross section, as shown in \cite{griestseckel}.} the annihilation cross section $\sigma_A$ in the nonrelativistic limit ($v \rightarrow 0$, where $v$ is the neutralino-neutralino relative velocity),
\begin{eqnarray}
\sigma_A v = a + bv^2 + \mathcal{O}(v^4), \label{sigmacosmo}
\end{eqnarray}
where $a$ is the s-wave contribution at zero relative velocity and $b$ contains contributions from both the s- and p-waves. The relative velocity of neutralinos in the galactic halo, sun and earth is of order $\mathcal{O}(10^{-3})$, so only the $a$ term in Eq. (\ref{sigmacosmo}) is needed for calculations involving relic neutralinos. When neutralino interactions freeze out in the early Universe, their relative velocities are approximately $v \simeq \frac{1}{2}$, so both the $a$ and $b$ terms are generally needed for relic-abundance calculations. For more accurate calculations the term of order $\mathcal{O}(v^4)$ is also included, but generally, one calculates the thermal average
\begin{eqnarray}
\langle \sigma_{ij}v_{ij}\rangle = \frac{\int d^3\mathbf{p}_i d^3\mathbf{p}_j \, f_if_j \sigma_{ij} v_{ij}}{\int d^3 \mathbf{p}_i d^3\mathbf{p}_j \, f_if_j},
\end{eqnarray}
where $\mathbf{p}_i$ is the three-momentum of the incoming particle $i$, $f_i$ is its equilibrium distribution function and $v_{ij}$ is the relative velocity of the interacting particles \cite{relicdensity}.

\section{Neutralino annihilation processes at tree-level}\label{sec_treelevel}
\subsection{Neutralino annihilation processes}\label{sec_tree1}
There are numerous final states into which the neutralinos can annihilate. The most important of these are those which appear at lowest order in pertubation theory, i.e. the final states which occur at tree level. Specifically, these are fermion-antifermion pairs $f\bar f$ (where $f$ are the standard model leptons and quarks), $W^+W^-$, $Z^0Z^0$, $W^+H^-$, $W^-H^+$, $Z^0A^0$, $Z^0H^0$, $Z^0h^0$, $H^+H^-$ and all six combinations of $A^0$, $h^0$ and $H^0$. There are various Feynman diagrams contributing to every channel and the cross-sections for all these tree-level processes have been calculated and are summarized in \cite{SDM}. Moreover, automatic calculators such as CompHEP \cite{CompHEP}, MadGraph \cite{MadGraph} and FormCalc \cite{FormCalc} are able to calculate these cross-sections since the MSSM is implemented in these programs. In the following sections we will take a closer look at the different final states which are classified as fermionic final states, weak gauge-boson final states and final states containing Higgs bosons.
\subsubsection{Fermionic final states}
The annihilation of neutralinos to a fermion-antifermion pair reveals several important features. First, with the expectation that the neutralino mass $m_\chi$ is of order or greater than 10 GeV, light fermion final states will always be an open annihilation channel. For many interesting neutralino masses, other channels will be closed or suppressed, so that fermionic final states are often the only open channels. Second, there is a suppression of the cross sections to fermionic final states in the limit of zero relative velocity due to helicity constraints: since neutralinos are Majorana particles, i.e. they are their own anti-particles, their spins have to be opposite in a relative s-wave due to Fermi statistics. Therefore, the fermions in the final state must have opposite spins as well which implies that the amplitude for this process carries a factor of the fermion mass $m_f$ due to the helicity flip. Thus, the cross-section is of order $m_f^2/m_\chi^2$. So the suppression of light fermionic final states is due purely to the fact, that these fermions are relatively light compared to the energy scale $2m_\chi$ of the process. Of course there is no suppression of the top-quark final states if it is open. In fact, in models where the neutralino is heavier than the top-quark, this annihilation channel is expected to be the dominant one since the cross section depends on the square of the fermion mass.
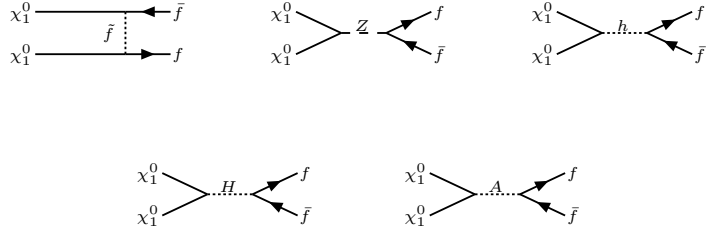
\begin{figure}[h!]
{\begin{center}
\unitlength=1.0 pt
\SetScale{1.0}
\SetWidth{0.7}      % line    size control
\tiny    %  letter  size control
{} \qquad\allowbreak
%  diagram # 5
\begin{picture}(79,60)(0,10)
\Text(13.0,57.0)[r]{$\chi^0_1$}
\Line(14.0,57.0)(48.0,57.0)
\Text(66.0,57.0)[l]{$\bar{f}$}
\ArrowLine(65.0,57.0)(48.0,57.0)
\Text(44.0,49.0)[r]{$\tilde{f}$}
\DashLine(48.0,57.0)(48.0,41.0){1.0}
\Text(13.0,41.0)[r]{$\chi^0_1$}
\Line(14.0,41.0)(48.0,41.0)
\Text(66.0,41.0)[l]{$f$}
\ArrowLine(48.0,41.0)(65.0,41.0)
\end{picture} 
{} \qquad\allowbreak
%  diagram # 1
\begin{picture}(79,60)(0,10)
\Text(13.0,57.0)[r]{$\chi^0_1$}
\Line(14.0,57.0)(31.0,49.0)
\Text(13.0,41.0)[r]{$\chi^0_1$}
\Line(14.0,41.0)(31.0,49.0)
\Text(39.0,50.0)[b]{$Z$}
\DashLine(31.0,49.0)(48.0,49.0){3.0}
\Text(66.0,57.0)[l]{$f$}
\ArrowLine(48.0,49.0)(65.0,57.0)
\Text(66.0,41.0)[l]{$\bar{f}$}
\ArrowLine(65.0,41.0)(48.0,49.0)
\end{picture} 
{} \qquad\allowbreak
%  diagram # 2
\begin{picture}(79,60)(0,10)
\Text(13.0,57.0)[r]{$\chi^0_1$}
\Line(14.0,57.0)(31.0,49.0)
\Text(13.0,41.0)[r]{$\chi^0_1$}
\Line(14.0,41.0)(31.0,49.0)
\Text(39.0,50.0)[b]{$h$}
\DashLine(31.0,49.0)(48.0,49.0){1.0}
\Text(66.0,57.0)[l]{$f$}
\ArrowLine(48.0,49.0)(65.0,57.0)
\Text(66.0,41.0)[l]{$\bar{f}$}
\ArrowLine(65.0,41.0)(48.0,49.0)
\end{picture} \\
{} \qquad\allowbreak
\begin{picture}(79,40)(0,30)
\Text(13.0,57.0)[r]{$\chi^0_1$}
\Line(14.0,57.0)(31.0,49.0)
\Text(13.0,41.0)[r]{$\chi^0_1$}
\Line(14.0,41.0)(31.0,49.0)
\Text(39.0,50.0)[b]{$H$}
\DashLine(31.0,49.0)(48.0,49.0){1.0}
\Text(66.0,57.0)[l]{$f$}
\ArrowLine(48.0,49.0)(65.0,57.0)
\Text(66.0,41.0)[l]{$\bar{f}$}
\ArrowLine(65.0,41.0)(48.0,49.0)
\end{picture} \
{} \qquad\allowbreak
%  diagram # 4
\begin{picture}(79,40)(0,30)
\Text(13.0,57.0)[r]{$\chi^0_1$}
\Line(14.0,57.0)(31.0,49.0)
\Text(13.0,41.0)[r]{$\chi^0_1$}
\Line(14.0,41.0)(31.0,49.0)
\Text(39.0,50.0)[b]{$A$}
\DashLine(31.0,49.0)(48.0,49.0){1.0}
\Text(66.0,57.0)[l]{$f$}
\ArrowLine(48.0,49.0)(65.0,57.0)
\Text(66.0,41.0)[l]{$\bar{f}$}
\ArrowLine(65.0,41.0)(48.0,49.0)
\end{picture} 
\end{center}
}
\caption{Diagrams contributing to neutralino annihilation into fermions.}\label{ffdiagrams}
\end{figure}\\
The diagrams for the neutralino-neutralino annihilation to the $f\bar f$ final state are shown in Fig. \ref{ffdiagrams}. The cross section contains contributions from t- and u-channel exchange of all sfermions $\tilde f$ which can couple to $f$ and from s-channel exchange of $Z^0$ and all three neutral Higgs bosons. In general, there are six sfermions which share the same charge and mix among themselves. All these sfermions can therefore couple to $f$ in the t- and u-channels, but in our case it is sufficient only to consider the exchange of the superpartner of the right- and left-handed final state fermions. Since there are s-channel diagrams exchanging massive particles we expect resonance effects if the energy is sufficient to produce these particles on-shell. However, we will approximate all fermions except those of the third generation as massless. Since the Higgs particles couple to the mass, s-channel exchange of these will only happen for the third generation.
\subsubsection{Weak gauge-boson final states}
Annihilation to weak gauge-bosons will occur if the neutralino mass is high enough for such channels to be open ($m_\chi > m_W$) or if the colliding energy is high enough. There is no s-wave suppression mechanism for these processes and thus they can be very important in the limit $v \rightarrow 0$ if the neutralino is heavy enough to produce weak gauge-bosons in the final state.\\
\begin{figure}[h!]
{\begin{center}
\unitlength=1.0 pt
\SetScale{1.0}
\SetWidth{0.7}      % line    size control
\tiny    %  letter  size control
{} \qquad\allowbreak
%  diagram # 1
\begin{picture}(79,40)(0,30)
\Text(13.0,57.0)[r]{$\chi^0_1$}
\Line(14.0,57.0)(48.0,57.0)
\Text(66.0,57.0)[l]{$W^-$}
\DashArrowLine(65.0,57.0)(48.0,57.0){3.0}
\Text(44.0,49.0)[r]{$\chi^+_{n}$}
\ArrowLine(48.0,57.0)(48.0,41.0)
\Text(13.0,41.0)[r]{$\chi^0_1$}
\Line(14.0,41.0)(48.0,41.0)
\Text(66.0,41.0)[l]{$W^+$}
\DashArrowLine(48.0,41.0)(65.0,41.0){3.0}
\end{picture} \
{} \qquad\allowbreak
%  diagram # 5
\begin{picture}(79,40)(0,30)
\Text(13.0,57.0)[r]{$\chi^0_1$}
\Line(14.0,57.0)(31.0,49.0)
\Text(13.0,41.0)[r]{$\chi^0_1$}
\Line(14.0,41.0)(31.0,49.0)
\Text(39.0,50.0)[b]{$Z$}
\DashLine(31.0,49.0)(48.0,49.0){3.0}
\Text(66.0,57.0)[l]{$W^+$}
\DashArrowLine(48.0,49.0)(65.0,57.0){3.0}
\Text(66.0,41.0)[l]{$W^-$}
\DashArrowLine(65.0,41.0)(48.0,49.0){3.0}
\end{picture} \
{} \qquad\allowbreak
%  diagram # 6
\begin{picture}(79,40)(0,30)
\Text(13.0,57.0)[r]{$\chi^0_1$}
\Line(14.0,57.0)(31.0,49.0)
\Text(13.0,41.0)[r]{$\chi^0_1$}
\Line(14.0,41.0)(31.0,49.0)
\Text(39.0,50.0)[b]{$h$}
\DashLine(31.0,49.0)(48.0,49.0){1.0}
\Text(66.0,57.0)[l]{$W^+$}
\DashArrowLine(48.0,49.0)(65.0,57.0){3.0}
\Text(66.0,41.0)[l]{$W^-$}
\DashArrowLine(65.0,41.0)(48.0,49.0){3.0}
\end{picture} \
{} \qquad\allowbreak
%  diagram # 7
\begin{picture}(79,40)(0,30)
\Text(13.0,57.0)[r]{$\chi^0_1$}
\Line(14.0,57.0)(31.0,49.0)
\Text(13.0,41.0)[r]{$\chi^0_1$}
\Line(14.0,41.0)(31.0,49.0)
\Text(39.0,50.0)[b]{$H$}
\DashLine(31.0,49.0)(48.0,49.0){1.0}
\Text(66.0,57.0)[l]{$W^+$}
\DashArrowLine(48.0,49.0)(65.0,57.0){3.0}
\Text(66.0,41.0)[l]{$W^-$}
\DashArrowLine(65.0,41.0)(48.0,49.0){3.0}
\end{picture}
\end{center}
}
\caption{Diagrams of the process $\chi \chi \rightarrow W^+W^-$.}\label{wwdiagrams}
{\begin{center}
\unitlength=1.0 pt
\SetScale{1.0}
\SetWidth{0.7}      % line    size control
\tiny    %  letter  size control
{} \qquad\allowbreak
%  diagram # 1
\begin{picture}(79,40)(0,30)
\Text(13.0,57.0)[r]{$\chi^0_1$}
\Line(14.0,57.0)(48.0,57.0)
\Text(66.0,57.0)[l]{$Z$}
\DashLine(48.0,57.0)(65.0,57.0){3.0}
\Text(47.0,49.0)[r]{$\chi^0_{n}$}
\Line(48.0,57.0)(48.0,41.0)
\Text(13.0,41.0)[r]{$\chi^0_1$}
\Line(14.0,41.0)(48.0,41.0)
\Text(66.0,41.0)[l]{$Z$}
\DashLine(48.0,41.0)(65.0,41.0){3.0}
\end{picture} \
{} \qquad\allowbreak
%  diagram # 2
\begin{picture}(79,40)(0,30)
\Text(13.0,57.0)[r]{$\chi^0_1$}
\Line(14.0,57.0)(31.0,49.0)
\Text(13.0,41.0)[r]{$\chi^0_1$}
\Line(14.0,41.0)(31.0,49.0)
\Text(39.0,50.0)[b]{$h$}
\DashLine(31.0,49.0)(48.0,49.0){1.0}
\Text(66.0,57.0)[l]{$Z$}
\DashLine(48.0,49.0)(65.0,57.0){3.0}
\Text(66.0,41.0)[l]{$Z$}
\DashLine(48.0,49.0)(65.0,41.0){3.0}
\end{picture} \
{} \qquad\allowbreak
%  diagram # 3
\begin{picture}(79,40)(0,30)
\Text(13.0,57.0)[r]{$\chi^0_1$}
\Line(14.0,57.0)(31.0,49.0)
\Text(13.0,41.0)[r]{$\chi^0_1$}
\Line(14.0,41.0)(31.0,49.0)
\Text(39.0,50.0)[b]{$H$}
\DashLine(31.0,49.0)(48.0,49.0){1.0}
\Text(66.0,57.0)[l]{$Z$}
\DashLine(48.0,49.0)(65.0,57.0){3.0}
\Text(66.0,41.0)[l]{$Z$}
\DashLine(48.0,49.0)(65.0,41.0){3.0}
\end{picture}
\end{center}}
\caption{Diagrams of the process $\chi\chi \rightarrow Z^0Z^0$.}\label{zzdiagrams}
\end{figure}\\
The diagrams for the neutralino-neutralino annihilation to weak gauge-boson final states $W^+W^-$ and $Z^0Z^0$ are given in Fig. \ref{wwdiagrams} and Fig. \ref{zzdiagrams} respectively. Annihilation to $W^+W^-$ pairs occurs by s-channel exchange of $Z^0$, $h^0$ and $H^0$ bosons and by t- and u-channel exchange of charginos. Neutralinos annihilate to $Z^0Z^0$ pairs via t- and u-channel exchange of neutralinos and by s-channel exchange of $H^0$ and $h^0$ bosons. In the $v\rightarrow 0$ limit the cross section is determined solely by the t- and u- channel, i.e. the exchange of charginos and neutralinos for $W^+W^-$ and $Z^0Z^0$ respectively.
\subsubsection{Final states containing Higgs bosons}
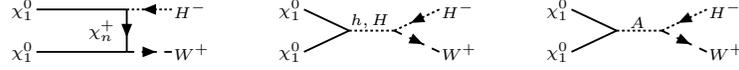
\begin{figure}[h!]
{\begin{center}
\unitlength=1.0 pt
\SetScale{1.0}
\SetWidth{0.7}      % line    size control
\tiny    %  letter  size control
{} \qquad\allowbreak
%  diagram # 1
\begin{picture}(79,40)(0,30)
\Text(13.0,57.0)[r]{$\chi^0_1$}
\Line(14.0,57.0)(48.0,57.0)
\Text(66.0,57.0)[l]{$H^-$}
\DashArrowLine(65.0,57.0)(48.0,57.0){1.0}
\Text(44.0,49.0)[r]{$\chi^+_n$}
\ArrowLine(48.0,57.0)(48.0,41.0)
\Text(13.0,41.0)[r]{$\chi^0_1$}
\Line(14.0,41.0)(48.0,41.0)
\Text(66.0,41.0)[l]{$W^+$}
\DashArrowLine(48.0,41.0)(65.0,41.0){3.0}
\end{picture} \
{} \qquad\allowbreak
%  diagram # 5
\begin{picture}(79,40)(0,30)
\Text(13.0,57.0)[r]{$\chi^0_1$}
\Line(14.0,57.0)(31.0,49.0)
\Text(13.0,41.0)[r]{$\chi^0_1$}
\Line(14.0,41.0)(31.0,49.0)
\Text(39.0,50.0)[b]{$h,H$}
\DashLine(31.0,49.0)(48.0,49.0){1.0}
\Text(66.0,57.0)[l]{$H^-$}
\DashArrowLine(65.0,57.0)(48.0,49.0){1.0}
\Text(66.0,41.0)[l]{$W^+$}
\DashArrowLine(48.0,49.0)(65.0,41.0){3.0}
\end{picture} \
{} \qquad\allowbreak
\begin{picture}(79,40)(0,30)
\Text(13.0,57.0)[r]{$\chi^0_1$}
\Line(14.0,57.0)(31.0,49.0)
\Text(13.0,41.0)[r]{$\chi^0_1$}
\Line(14.0,41.0)(31.0,49.0)
\Text(39.0,50.0)[b]{$A$}
\DashLine(31.0,49.0)(48.0,49.0){1.0}
\Text(66.0,57.0)[l]{$H^-$}
\DashArrowLine(65.0,57.0)(48.0,49.0){1.0}
\Text(66.0,41.0)[l]{$W^+$}
\DashArrowLine(48.0,49.0)(65.0,41.0){3.0}
\end{picture} \
\end{center}
}\caption{Diagrams for the process $\chi\chi\rightarrow W^+H^-$ (the charge-conjugate process is similar).}\label{HWdiagrams}
\end{figure}

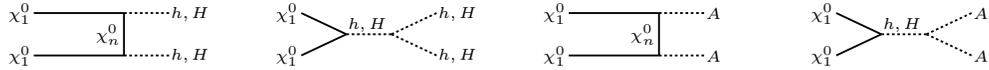
\begin{figure}[h!]
{\begin{center}
\unitlength=1.0 pt
\SetScale{1.0}
\SetWidth{0.7}      % line    size control
\tiny    %  letter  size control
{} \qquad\allowbreak
%  diagram # 1
\begin{picture}(79,40)(0,30)
\Text(13.0,57.0)[r]{$\chi^0_1$}
\Line(14.0,57.0)(48.0,57.0)
\Text(66.0,57.0)[l]{$h,H$}
\DashLine(48.0,57.0)(65.0,57.0){1.0}
\Text(47.0,49.0)[r]{$\chi^0_n$}
\Line(48.0,57.0)(48.0,41.0)
\Text(13.0,41.0)[r]{$\chi^0_1$}
\Line(14.0,41.0)(48.0,41.0)
\Text(66.0,41.0)[l]{$h,H$}
\DashLine(48.0,41.0)(65.0,41.0){1.0}
\end{picture} \
{} \qquad\allowbreak
%  diagram # 2
\begin{picture}(79,40)(0,30)
\Text(13.0,57.0)[r]{$\chi^0_1$}
\Line(14.0,57.0)(31.0,49.0)
\Text(13.0,41.0)[r]{$\chi^0_1$}
\Line(14.0,41.0)(31.0,49.0)
\Text(39.0,50.0)[b]{$h,H$}
\DashLine(31.0,49.0)(48.0,49.0){1.0}
\Text(66.0,57.0)[l]{$h,H$}
\DashLine(48.0,49.0)(65.0,57.0){1.0}
\Text(66.0,41.0)[l]{$h,H$}
\DashLine(48.0,49.0)(65.0,41.0){1.0}
\end{picture} \
{} \qquad\allowbreak
%  diagram # 1
\begin{picture}(79,40)(0,30)
\Text(13.0,57.0)[r]{$\chi^0_1$}
\Line(14.0,57.0)(48.0,57.0)
\Text(66.0,57.0)[l]{$A$}
\DashLine(48.0,57.0)(65.0,57.0){1.0}
\Text(47.0,49.0)[r]{$\chi^0_n$}
\Line(48.0,57.0)(48.0,41.0)
\Text(13.0,41.0)[r]{$\chi^0_1$}
\Line(14.0,41.0)(48.0,41.0)
\Text(66.0,41.0)[l]{$A$}
\DashLine(48.0,41.0)(65.0,41.0){1.0}
\end{picture} \
{} \qquad\allowbreak
%  diagram # 2
\begin{picture}(79,40)(0,30)
\Text(13.0,57.0)[r]{$\chi^0_1$}
\Line(14.0,57.0)(31.0,49.0)
\Text(13.0,41.0)[r]{$\chi^0_1$}
\Line(14.0,41.0)(31.0,49.0)
\Text(39.0,50.0)[b]{$h,H$}
\DashLine(31.0,49.0)(48.0,49.0){1.0}
\Text(66.0,57.0)[l]{$A$}
\DashLine(48.0,49.0)(65.0,57.0){1.0}
\Text(66.0,41.0)[l]{$A$}
\DashLine(48.0,49.0)(65.0,41.0){1.0}
\end{picture} \
\end{center}}
\caption{Diagrams for neutralino-neutralino annihilation to the $h^0h^0$, $H^0H^0$, $h^0H^0$ and $A^0A^0$ final states.}\label{hhdiagrams}
\end{figure}

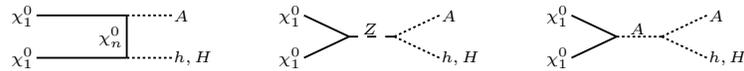
\begin{figure}[h!]
{\begin{center}
\unitlength=1.0 pt
\SetScale{1.0}
\SetWidth{0.7}      % line    size control
\tiny    %  letter  size control
{} \qquad\allowbreak
%  diagram # 1
\begin{picture}(79,40)(0,30)
\Text(13.0,57.0)[r]{$\chi^0_1$}
\Line(14.0,57.0)(48.0,57.0)
\Text(66.0,57.0)[l]{$A$}
\DashLine(48.0,57.0)(65.0,57.0){1.0}
\Text(47.0,49.0)[r]{$\chi^0_n$}
\Line(48.0,57.0)(48.0,41.0)
\Text(13.0,41.0)[r]{$\chi^0_1$}
\Line(14.0,41.0)(48.0,41.0)
\Text(66.0,41.0)[l]{$h,H$}
\DashLine(48.0,41.0)(65.0,41.0){1.0}
\end{picture} \
{} \qquad\allowbreak
\begin{picture}(79,40)(0,30)
\Text(13.0,57.0)[r]{$\chi^0_1$}
\Line(14.0,57.0)(31.0,49.0)
\Text(13.0,41.0)[r]{$\chi^0_1$}
\Line(14.0,41.0)(31.0,49.0)
\Text(39.0,50.0)[b]{$Z$}
\DashLine(31.0,49.0)(48.0,49.0){3.0}
\Text(66.0,57.0)[l]{$A$}
\DashLine(48.0,49.0)(65.0,57.0){1.0}
\Text(66.0,41.0)[l]{$h,H$}
\DashLine(48.0,49.0)(65.0,41.0){1.0}
\end{picture} \
{} \qquad\allowbreak
\begin{picture}(79,40)(0,30)
\Text(13.0,57.0)[r]{$\chi^0_1$}
\Line(14.0,57.0)(31.0,49.0)
\Text(13.0,41.0)[r]{$\chi^0_1$}
\Line(14.0,41.0)(31.0,49.0)
\Text(39.0,50.0)[b]{$A$}
\DashLine(31.0,49.0)(48.0,49.0){1.0}
\Text(66.0,57.0)[l]{$A$}
\DashLine(48.0,49.0)(65.0,57.0){1.0}
\Text(66.0,41.0)[l]{$h,H$}
\DashLine(48.0,49.0)(65.0,41.0){1.0}
\end{picture} \
\end{center}}
\caption{Diagrams for the processes containing one pseudoscalar Higgs boson and one scalar Higgs boson in the final state.}\label{hAdiagrams}
\end{figure}

\begin{figure}[h!]
{\begin{center}
\unitlength=1.0 pt
\SetScale{1.0}
\SetWidth{0.7}      % line    size control
\tiny    %  letter  size control
{} \qquad\allowbreak
%  diagram # 1
\begin{picture}(79,40)(0,30)
\Text(13.0,57.0)[r]{$\chi^0_1$}
\Line(14.0,57.0)(48.0,57.0)
\Text(66.0,57.0)[l]{$H^+$}
\DashArrowLine(65.0,57.0)(48.0,57.0){1.0}
\Text(44.0,49.0)[r]{$\chi^+_n$}
\ArrowLine(48.0,57.0)(48.0,41.0)
\Text(13.0,41.0)[r]{$\chi^0_1$}
\Line(14.0,41.0)(48.0,41.0)
\Text(66.0,41.0)[l]{$H^-$}
\DashArrowLine(48.0,41.0)(65.0,41.0){1.0}
\end{picture} \
{} \qquad\allowbreak
%  diagram # 5
\begin{picture}(79,40)(0,30)
\Text(13.0,57.0)[r]{$\chi^0_1$}
\Line(14.0,57.0)(31.0,49.0)
\Text(13.0,41.0)[r]{$\chi^0_1$}
\Line(14.0,41.0)(31.0,49.0)
\Text(39.0,50.0)[b]{$Z$}
\DashLine(31.0,49.0)(48.0,49.0){3.0}
\Text(66.0,57.0)[l]{$H^+$}
\DashArrowLine(48.0,49.0)(65.0,57.0){1.0}
\Text(66.0,41.0)[l]{$H^-$}
\DashArrowLine(65.0,41.0)(48.0,49.0){1.0}
\end{picture} \
{} \qquad\allowbreak
%  diagram # 6
\begin{picture}(79,40)(0,30)
\Text(13.0,57.0)[r]{$\chi^0_1$}
\Line(14.0,57.0)(31.0,49.0)
\Text(13.0,41.0)[r]{$\chi^0_1$}
\Line(14.0,41.0)(31.0,49.0)
\Text(39.0,50.0)[b]{$h,H$}
\DashLine(31.0,49.0)(48.0,49.0){1.0}
\Text(66.0,57.0)[l]{$H^+$}
\DashArrowLine(48.0,49.0)(65.0,57.0){1.0}
\Text(66.0,41.0)[l]{$H^-$}
\DashArrowLine(65.0,41.0)(48.0,49.0){1.0}
\end{picture} \
\end{center}}
\caption{Diagrams for the process $\chi\chi \rightarrow H^+H^-$.}\label{cHdiagrams}
\end{figure}
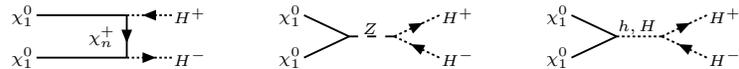

The annihilation of neutralino pairs to Higgs bosons may also be important when such channels are open. However, there are some cases where the s-wave amplitude vanishes because of the inability of the Higgs bosons to produce the appropriate final state quantum number $CP=-1$. These cases are $\chi\chi \rightarrow h^0h^0$, $\chi\chi \rightarrow H^0H^0$, $\chi\chi \rightarrow A^0A^0$, $\chi\chi \rightarrow H^+H^-$ and the mixed final state $\chi\chi \rightarrow Z^0A^0$. The processes which have non-vanishing s-wave contributions are $\chi\chi \rightarrow h^0A^0$, $\chi\chi \rightarrow H^0A^0$, $\chi\chi \rightarrow Z^0h^0$, $\chi\chi \rightarrow Z^0H^0$ and $\chi\chi \rightarrow W^\pm H^\mp$.\\
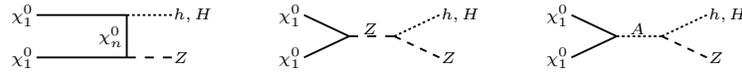
\begin{figure}[h!]
{\begin{center}
\unitlength=1.0 pt
\SetScale{1.0}
\SetWidth{0.7}      % line    size control
\tiny    %  letter  size control
{} \qquad\allowbreak
%  diagram # 3
\begin{picture}(79,40)(0,30)
\Text(13.0,57.0)[r]{$\chi^0_1$}
\Line(14.0,57.0)(48.0,57.0)
\Text(66.0,57.0)[l]{$h,H$}
\DashLine(48.0,57.0)(65.0,57.0){1.0}
\Text(47.0,49.0)[r]{$\chi^0_n$}
\Line(48.0,57.0)(48.0,41.0)
\Text(13.0,41.0)[r]{$\chi^0_1$}
\Line(14.0,41.0)(48.0,41.0)
\Text(66.0,41.0)[l]{$Z$}
\DashLine(48.0,41.0)(65.0,41.0){3.0}
\end{picture} \
{} \qquad\allowbreak
%  diagram # 2
\begin{picture}(79,40)(0,30)
\Text(13.0,57.0)[r]{$\chi^0_1$}
\Line(14.0,57.0)(31.0,49.0)
\Text(13.0,41.0)[r]{$\chi^0_1$}
\Line(14.0,41.0)(31.0,49.0)
\Text(39.0,50.0)[b]{$Z$}
\DashLine(31.0,49.0)(48.0,49.0){3.0}
\Text(66.0,57.0)[l]{$h,H$}
\DashLine(48.0,49.0)(65.0,57.0){1.0}
\Text(66.0,41.0)[l]{$Z$}
\DashLine(48.0,49.0)(65.0,41.0){3.0}
\end{picture} \
{} \qquad\allowbreak
%  diagram # 4
\begin{picture}(79,40)(0,30)
\Text(13.0,57.0)[r]{$\chi^0_1$}
\Line(14.0,57.0)(31.0,49.0)
\Text(13.0,41.0)[r]{$\chi^0_1$}
\Line(14.0,41.0)(31.0,49.0)
\Text(39.0,50.0)[b]{$A$}
\DashLine(31.0,49.0)(48.0,49.0){1.0}
\Text(66.0,57.0)[l]{$h,H$}
\DashLine(48.0,49.0)(65.0,57.0){1.0}
\Text(66.0,41.0)[l]{$Z$}
\DashLine(48.0,49.0)(65.0,41.0){3.0}
\end{picture} \
\end{center}}
\caption{Diagrams for the processes $\chi\chi \rightarrow Z^0 h^0$ and $\chi\chi \rightarrow Z^0 H^0$.}\label{ZhHdiagrams}
\end{figure}
\begin{figure}[h!]
{\begin{center}
\unitlength=1.0 pt
\SetScale{1.0}
\SetWidth{0.7}      % line    size control
\tiny    %  letter  size control
{} \qquad\allowbreak
%  diagram # 4
\begin{picture}(79,40)(0,30)
\Text(13.0,57.0)[r]{$\chi^0_1$}
\Line(14.0,57.0)(48.0,57.0)
\Text(66.0,57.0)[l]{$A$}
\DashLine(48.0,57.0)(65.0,57.0){1.0}
\Text(47.0,49.0)[r]{$\chi^0_n$}
\Line(48.0,57.0)(48.0,41.0)
\Text(13.0,41.0)[r]{$\chi^0_1$}
\Line(14.0,41.0)(48.0,41.0)
\Text(66.0,41.0)[l]{$Z$}
\DashLine(48.0,41.0)(65.0,41.0){3.0}
\end{picture} \
{} \qquad\allowbreak
%  diagram # 2
\begin{picture}(79,40)(0,30)
\Text(13.0,57.0)[r]{$\chi^0_1$}
\Line(14.0,57.0)(31.0,49.0)
\Text(13.0,41.0)[r]{$\chi^0_1$}
\Line(14.0,41.0)(31.0,49.0)
\Text(39.0,50.0)[b]{$h$}
\DashLine(31.0,49.0)(48.0,49.0){1.0}
\Text(66.0,57.0)[l]{$A$}
\DashLine(48.0,49.0)(65.0,57.0){1.0}
\Text(66.0,41.0)[l]{$Z$}
\DashLine(48.0,49.0)(65.0,41.0){3.0}
\end{picture} \
{} \qquad\allowbreak
%  diagram # 3
\begin{picture}(79,40)(0,30)
\Text(13.0,57.0)[r]{$\chi^0_1$}
\Line(14.0,57.0)(31.0,49.0)
\Text(13.0,41.0)[r]{$\chi^0_1$}
\Line(14.0,41.0)(31.0,49.0)
\Text(39.0,50.0)[b]{$H$}
\DashLine(31.0,49.0)(48.0,49.0){1.0}
\Text(66.0,57.0)[l]{$A$}
\DashLine(48.0,49.0)(65.0,57.0){1.0}
\Text(66.0,41.0)[l]{$Z$}
\DashLine(48.0,49.0)(65.0,41.0){3.0}
\end{picture} \
\end{center}}
\caption{Diagrams for the process $\chi \chi \rightarrow Z^0A^0$.}\label{ZAdiagrams}
\end{figure}
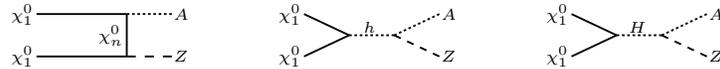\\
The diagrams for the processes $\chi\chi \rightarrow Z^0 h^0$ and $\chi\chi \rightarrow Z^0 H^0$ and the process $\chi\chi \rightarrow Z^0 A^0$ are shown in Fig. \ref{ZhHdiagrams} and \ref{ZAdiagrams}, respectively. All these processes receive contributions from t- and u-channel exchange of all four neutralinos. In the processes containing a neutral CP-even Higgs boson, $h^0$, $H^0$, in the final state s-channel exchange of $Z^0$ and $A^0$ contribute, while the process $\chi\chi\rightarrow Z^0A^0$ contains contributions from s-channel exchange of $h^0$ and $H^0$.\\
The diagrams of the process $\chi\chi \rightarrow W^+H^-$ (and its charge conjugate) are given in Fig. \ref{HWdiagrams}. It receives contributions from chargino exchange in t- and u-channel and all three neutral Higgs bosons in the s-channel.
\\
Annihilation to the two Higgs-boson final states $h^0h^0$, $H^0H^0$, $h^0H^0$ and $A^0A^0$ occurs via the exchange of neutralinos in the t- and u-channel and the exchange of $h^0$ and $H^0$ in the s-channel, as shown in Fig. \ref{hhdiagrams}.
\\
Next we consider the process with one CP-even scalar and one CP-odd scalar $A^0$ in the final state, i.e. $\chi\chi \rightarrow h^0A^0$ and $\chi\chi \rightarrow H^0A^0$. The diagrams for these channels are shown in Fig. \ref{hAdiagrams} and receive contributions from neutralino exchange in the t- and u-channel and from $Z^0$ and $A^0$ exchange in the s-channel.
\\
Finally, we consider the process $\chi\chi \rightarrow H^+H^-$. This proceeds by exchange of the two charginos in the t- and u-channel and by s-channel exchange of the $Z^0$, $h^0$ and $H^0$ bosons, as shown in Fig. \ref{cHdiagrams}.\\
\subsection{Calculating cross sections at tree-level: MadGraph and CompHEP}\label{sec_tree2}
For all tree-level cross-section calculations we use both {\scshape{MadGraph/MadEvent}} \cite{MadGraph} and {\scshape{CompHEP}} \cite{CompHEP}.\\
{\scshape{MadEvent}} is a multi-purpose, tree-level event generator which is powered by the matrix element creator {\scshape{MadGraph}}. It is a widely used tool in high energy physics to simulate events for collider experiments. Given a process and a specified model (SM, MSSM, 2HDM, etc), MadGraph automatically creates the amplitudes for all the relevant subprocesses and produces the mappings for the integration over the phase space. This information is passed to MadEvent, which calculates the cross section and may supply the user with unweighted events. These unweighted events are stored in the ``Les Houches Event Files'' (LHEF) format and may be passed directly to a Monte Carlo program (interface available for {\scshape{Pythia}}). However, for our purposes it is sufficient only to calculate the cross section.\\
{\scshape{CompHEP}} is a package for automatic calculations of elementary particle decay and collision properties at tree-level. The main idea prescribed in {\scshape{CompHEP}} is to make available passing on from the Lagrangian to the final distribution effectively with a high level of automation. {\scshape{CompHEP}} is a menu-driven system which makes it very user-friendly.  It is divided into two parts: a symbolical and a numerical calculation. During the symbolical part, the user is able to specify and exclude diagrams in order to find out, how the different diagrams interfere. The numerical calculation may provide amongst others an energy dependent cross-section plot.
\subsection{Neutralino annihilation cross section at mSUGRA point SPS1a}\label{sec_tree3}
In this section we study the cross section of the possible tree level neutralino annihilation processes at the mSUGRA benchmark point SPS1a. This first study has mainly two purposes: We want to see (i) how the cross section behaves with the energy and (ii) which channels are dominant. Moreover, this study was used to cross check the results from both MadGraph and CompHEP in order to find out how they work and which assumptions and approximations are used for the calculation.
\subsubsection{mSUGRA benchmark point SPS1a}
The mSUGRA parameters of the benchmark point SPSa1, which is defined in \cite{SPS}, are given in Table \ref{paraSPS1a}. In Table \ref{massSPS1atable} the mass values of some important particles are shown. Depending on the used SUSY spectrum calculator, the masses in MadGraph and CompHEP differ slightly but not significantly. The masses of the Higgs bosons $A^0$ and $H^0$ are almost degenerate.
\begin{table}[h!]
\centering
\begin{tabular}{|l|c|}
\hline
$m_0$		&	$100$ GeV\\
$m_{12}$	&	$250$ GeV\\
$\tan\beta$	&	$10$\\
$\mathrm{sign}(\mu)$	&	$+$\\
$A_0$	 	&      $-100$ GeV\\
\hline
\end{tabular}
\caption{mSUGRA parameters of benchmark point SPS1a.}\label{paraSPS1a}
\end{table}
\begin{table}[h!]
\centering
\begin{tabular}{|l|c|}
\hline
Particle		&	Mass /GeV		\\
\hline
$\chi_1^0$		&	96.23\\
$\chi_2^0$		&	176.97\\
$\chi_3^0$		&	363.76\\
$\chi_4^0$		&	381.73\\
$\chi_1^+$		&	181.70\\
$\chi_2^+$		&	379.94\\
$h^0$			&	118.37\\
$H^0$			&	394.23\\
$A^0$			&	394.49\\
$H^\pm$			&	402.53\\
\hline
\end{tabular}
\caption{Mass spectrum of mSUGRA benchmark point SPS1a.}\label{massSPS1atable}
\end{table}
\subsubsection{Energy dependence}\label{sec_treeenergy}
In the following, the energy dependence of the cross section of the most interesting tree level processes is presented. The $\sigma(E)$-plots are created by {\scshape{CompHEP}} in a zero-width-approximation of the exchanged Higgs bosons. Thus, it is possible to distinguish the resonance effects due to the s-channel exchange of $A^0$ and $H^0$, although they are almost mass-degenerate. Once we insert finite widths for these Higgs bosons, the distinguishablility of these resonances vanishes.\\
The cross section has been calculated for random energy values with both {\scshape{CompHEP}} and {\scshape{MadGraph}} to do a cross check. The results are in reasonable agreement.
\paragraph{Fermionic final states}
The energy dependence of the cross-section for the process $\chi\chi \rightarrow \tau^+\tau^-$ in the range from 200 GeV ($\approx 2m_\chi$) up to 1 TeV is shown in Fig. \ref{xsectau}. We see a clear resonance in the spectrum. A more detailed view on this resonance reveals two peaks at 394.2 GeV and 394.5 GeV. Comparing with the mass spectrum of SPS1a in Table \ref{massSPS1atable} we note that these peaks are due to the s-channel exchange of the Higgs bosons $A^0$ and $H^0$. In the region $2m_\chi < \sqrt s < 2m_A$ the cross section is of order $(0.5 - 1.0)\,\mathrm{pb}$ whereas, beyond the resonance, the cross section decreases as usual.
\begin{figure}[h!]
\centering
\begin{minipage}{7.2cm}
\scalebox{0.7}{
\unitlength=1.0 pt
\SetScale{1.0}
\SetWidth{0.7}      % line    size control
\normalsize    %  letter  size control
\begin{picture}(300,200)(0,0)
%\Text(168.2,199.4)[t]{$~o1,~o1 ->l,L$}
% ====================   X-axis =============
\LinAxis(45.90,36.72)(290.82,36.72)(4.000,2,-4,0.000,1.5)
\Text(45.9,29.9)[t]{$200$}
\Text(107.2,29.9)[t]{$400$}
\Text(168.6,29.9)[t]{$600$}
\Text(229.5,29.9)[t]{$800$}
\Text(290.8,29.9)[t]{$10^3$}
\Text(290.8,20.3)[rt]{$\sqrt(s)$}
% ====================   Y-axis =============
\LogAxis(45.90,36.72)(45.90,176.27)(1.777,4,2.061,1.5)
\Text(39.2,90.4)[r]{$1$}
\Text(39.2,168.9)[r]{$10$}
\rText(26.3,176.3)[tr][l]{Cross Section [pb]}
% ============== end of axis ============
\Line(48.0,80.2)(45.9,70.1) 
\Line(50.5,85.3)(48.0,80.2) 
\Line(53.0,88.1)(50.5,85.3) 
\Line(55.5,90.4)(53.0,88.1) 
\Line(58.0,91.5)(55.5,90.4) 
\Line(60.5,92.1)(58.0,91.5) 
\Line(63.0,92.7)(60.5,92.1) 
\Line(65.1,93.2)(63.0,92.7) 
\Line(67.6,93.2)(65.1,93.2) 
\Line(67.6,93.2)(70.1,93.2) 
\Line(70.1,93.2)(72.6,92.7) 
\Line(72.6,92.7)(75.1,92.7) 
\Line(75.1,92.7)(77.6,92.1) 
\Line(77.6,92.1)(80.1,91.5) 
\Line(80.1,91.5)(82.6,91.0) 
\Line(82.6,91.0)(84.7,90.4) 
\Line(84.7,90.4)(87.2,90.4) 
\Line(87.2,90.4)(89.7,89.8) 
\Line(89.7,89.8)(92.2,89.3) 
\Line(92.2,89.3)(94.7,89.3) 
\Line(97.2,89.8)(94.7,89.3) 
\Line(99.7,92.7)(97.2,89.8) 
\Line(102.2,102.8)(99.7,92.7) 
\Line(104.3,176.8)(102.2,102.8) 
\Line(104.3,176.8)(106.8,128.2) 
\Line(106.8,128.2)(109.3,96.6) 
\Line(109.3,96.6)(111.8,88.7) 
\Line(111.8,88.7)(114.3,85.9) 
\Line(114.3,85.9)(116.8,83.6) 
\Line(116.8,83.6)(119.3,82.5) 
\Line(119.3,82.5)(121.4,80.8) 
\Line(121.4,80.8)(123.9,80.2) 
\Line(123.9,80.2)(126.4,79.1) 
\Line(126.4,79.1)(128.9,78.0) 
\Line(128.9,78.0)(131.4,77.4) 
\Line(131.4,77.4)(133.9,76.3) 
\Line(133.9,76.3)(136.4,75.7) 
\Line(136.4,75.7)(138.9,75.1) 
\Line(138.9,75.1)(141.0,74.0) 
\Line(141.0,74.0)(143.5,73.4) 
\Line(143.5,73.4)(146.0,72.3) 
\Line(146.0,72.3)(148.5,71.8) 
\Line(148.5,71.8)(151.0,71.2) 
\Line(151.0,71.2)(153.5,70.6) 
\Line(153.5,70.6)(156.1,69.5) 
\Line(156.1,69.5)(158.6,68.9) 
\Line(158.6,68.9)(160.6,68.4) 
\Line(160.6,68.4)(163.1,67.2) 
\Line(163.1,67.2)(165.6,66.7) 
\Line(165.6,66.7)(168.2,66.1) 
\Line(168.2,66.1)(170.7,65.5) 
\Line(170.7,65.5)(173.2,64.4) 
\Line(173.2,64.4)(175.7,63.8) 
\Line(175.7,63.8)(177.7,63.3) 
\Line(177.7,63.3)(180.3,62.7) 
\Line(180.3,62.7)(182.8,62.1) 
\Line(182.8,62.1)(185.3,61.0) 
\Line(185.3,61.0)(187.8,60.5) 
\Line(187.8,60.5)(190.3,59.9) 
\Line(190.3,59.9)(192.8,59.3) 
\Line(192.8,59.3)(195.3,58.8) 
\Line(195.3,58.8)(197.4,58.2) 
\Line(197.4,58.2)(199.9,57.6) 
\Line(199.9,57.6)(202.4,57.1) 
\Line(202.4,57.1)(204.9,55.9) 
\Line(204.9,55.9)(207.4,55.4) 
\Line(207.4,55.4)(209.9,54.8) 
\Line(209.9,54.8)(212.4,54.2) 
\Line(212.4,54.2)(214.9,53.7) 
\Line(214.9,53.7)(217.0,53.1) 
\Line(217.0,53.1)(219.5,52.5) 
\Line(219.5,52.5)(222.0,52.0) 
\Line(222.0,52.0)(224.5,51.4) 
\Line(224.5,51.4)(227.0,50.8) 
\Line(227.0,50.8)(229.5,50.3) 
\Line(229.5,50.3)(232.0,49.7) 
\Line(232.0,49.7)(234.1,49.2) 
\Line(234.1,49.2)(236.6,48.6) 
\Line(236.6,48.6)(239.1,48.0) 
\Line(239.1,48.0)(241.6,47.5) 
\Line(241.6,47.5)(244.1,46.9) 
\Line(244.1,46.9)(246.6,46.3) 
\Line(246.6,46.3)(249.1,45.8) 
\Line(249.1,45.8)(251.6,45.2) 
\Line(251.6,45.2)(253.7,44.6) 
\Line(253.7,44.6)(256.2,44.1) 
\Line(256.2,44.1)(258.7,43.5) 
\Line(258.7,43.5)(261.2,42.9) 
\Line(261.2,42.9)(263.7,42.4) 
\Line(263.7,42.4)(266.2,41.8) 
\Line(266.2,41.8)(268.7,41.2) 
\Line(268.7,41.2)(271.2,41.2) 
\Line(271.2,41.2)(273.3,40.7)
\Line(273.3,40.7)(275.8,40.1) 
\Line(275.8,40.1)(278.3,39.5) 
\Line(278.3,39.5)(280.8,39.0) 
\Line(280.8,39.0)(283.3,38.4) 
\Line(283.3,38.4)(285.8,37.9) 
\Line(285.8,37.9)(288.3,37.3) 
\Line(288.3,37.3)(290.8,36.7) 
\end{picture}
}
\caption{Cross section energy dependence of the process $\chi\chi \rightarrow \tau^+\tau^-$.}\label{xsectau}
\end{minipage}
\hspace{1cm}
\begin{minipage}{7.2cm}
\begin{center}
\scalebox{0.7}{
\unitlength=1.0 pt
\SetScale{1.0}
\SetWidth{0.7}      % line    size control
\normalsize    %  letter  size control
\begin{picture}(300,200)(0,0)
%\Text(168.2,199.4)[t]{$~o1,~o1 ->b,B$}
% ====================   X-axis =============
\LinAxis(45.90,36.72)(290.82,36.72)(4.025,2,-4,-0.050,1.5)
\Text(47.6,29.9)[t]{$200$}
\Text(108.1,29.9)[t]{$400$}
\Text(169.0,29.9)[t]{$600$}
\Text(229.9,29.9)[t]{$800$}
\Text(290.8,29.9)[t]{$10^3$}
\Text(290.8,20.3)[rt]{$\sqrt(s)$}
% ====================   Y-axis =============
\LogAxis(45.90,36.72)(45.90,176.27)(4.841,4,5.641,1.5)
\Text(39.2,44.1)[r]{$10^-2$}
\Text(39.2,72.9)[r]{$0.1$}
\Text(39.2,101.7)[r]{$1$}
\Text(39.2,130.5)[r]{$10$}
\Text(39.2,159.3)[r]{$100$}
\rText(7.9,176.3)[tr][l]{Cross Section [pb]}
% ============== end of axis ============
\Line(45.9,71.2)(48.0,65.0) 
\Line(48.0,65.0)(50.5,63.8) 
\Line(50.5,63.8)(53.0,63.3) 
\Line(55.5,63.8)(53.0,63.3) 
\Line(58.0,65.0)(55.5,63.8) 
\Line(60.5,66.1)(58.0,65.0) 
\Line(63.0,67.2)(60.5,66.1) 
\Line(65.1,68.9)(63.0,67.2) 
\Line(67.6,70.1)(65.1,68.9) 
\Line(70.1,71.8)(67.6,70.1) 
\Line(72.6,74.0)(70.1,71.8) 
\Line(75.1,75.7)(72.6,74.0) 
\Line(77.6,78.0)(75.1,75.7) 
\Line(80.1,80.2)(77.6,78.0) 
\Line(82.6,82.5)(80.1,80.2) 
\Line(84.7,85.3)(82.6,82.5) 
\Line(87.2,88.1)(84.7,85.3) 
\Line(89.7,92.1)(87.2,88.1) 
\Line(92.2,96.0)(89.7,92.1) 
\Line(94.7,100.6)(92.2,96.0) 
\Line(97.2,106.8)(94.7,100.6) 
\Line(99.7,114.1)(97.2,106.8) 
\Line(102.2,125.4)(99.7,114.1) 
\Line(104.3,146.3)(102.2,125.4) 
\Line(106.8,176.3)(104.3,146.3) 
\Line(106.8,176.3)(109.3,134.5) 
\Line(109.3,134.5)(111.8,119.8) 
\Line(111.8,119.8)(114.3,110.2) 
\Line(114.3,110.2)(116.8,103.4) 
\Line(116.8,103.4)(119.3,98.3) 
\Line(119.3,98.3)(121.4,93.8) 
\Line(121.4,93.8)(123.9,89.8) 
\Line(123.9,89.8)(126.4,87.0) 
\Line(126.4,87.0)(128.9,83.6) 
\Line(128.9,83.6)(131.4,81.4) 
\Line(131.4,81.4)(133.9,79.1) 
\Line(133.9,79.1)(136.4,76.8) 
\Line(136.4,76.8)(138.9,74.6) 
\Line(138.9,74.6)(141.0,72.9) 
\Line(141.0,72.9)(143.5,71.2) 
\Line(143.5,71.2)(146.0,69.5) 
\Line(146.0,69.5)(148.5,68.4) 
\Line(148.5,68.4)(151.0,66.7) 
\Line(151.0,66.7)(153.5,65.5) 
\Line(153.5,65.5)(156.1,63.8) 
\Line(156.1,63.8)(158.6,62.7) 
\Line(158.6,62.7)(160.6,61.6) 
\Line(160.6,61.6)(163.1,60.5) 
\Line(163.1,60.5)(165.6,59.9) 
\Line(165.6,59.9)(168.2,58.8) 
\Line(168.2,58.8)(170.7,57.6) 
\Line(170.7,57.6)(173.2,57.1) 
\Line(173.2,57.1)(175.7,55.9) 
\Line(175.7,55.9)(177.7,55.4) 
\Line(177.7,55.4)(180.3,54.2) 
\Line(180.3,54.2)(182.8,53.7) 
\Line(182.8,53.7)(185.3,53.1) 
\Line(185.3,53.1)(187.8,52.0) 
\Line(187.8,52.0)(190.3,51.4) 
\Line(190.3,51.4)(192.8,50.8) 
\Line(192.8,50.8)(195.3,50.3) 
\Line(195.3,50.3)(197.4,49.7) 
\Line(197.4,49.7)(199.9,49.2) 
\Line(199.9,49.2)(202.4,48.6) 
\Line(202.4,48.6)(204.9,48.0) 
\Line(204.9,48.0)(207.4,47.5) 
\Line(207.4,47.5)(209.9,46.9) 
\Line(209.9,46.9)(212.4,46.3) 
\Line(212.4,46.3)(214.9,46.3) 
\Line(214.9,46.3)(217.0,45.8) 
\Line(217.0,45.8)(219.5,45.2) 
\Line(219.5,45.2)(222.0,44.6) 
\Line(222.0,44.6)(224.5,44.6) 
\Line(224.5,44.6)(227.0,44.1) 
\Line(227.0,44.1)(229.5,43.5) 
\Line(229.5,43.5)(232.0,43.5) 
\Line(232.0,43.5)(234.1,42.9) 
\Line(234.1,42.9)(236.6,42.4) 
\Line(236.6,42.4)(239.1,42.4) 
\Line(239.1,42.4)(241.6,41.8) 
\Line(241.6,41.8)(244.1,41.8) 
\Line(244.1,41.8)(246.6,41.2) 
\Line(246.6,41.2)(249.1,41.2) 
\Line(249.1,41.2)(251.6,40.7) 
\Line(251.6,40.7)(253.7,40.7) 
\Line(253.7,40.7)(256.2,40.1) 
\Line(256.2,40.1)(258.7,40.1) 
\Line(258.7,40.1)(261.2,39.5) 
\Line(261.2,39.5)(263.7,39.5) 
\Line(263.7,39.5)(266.2,39.0) 
\Line(266.2,39.0)(268.7,39.0) 
\Line(268.7,39.0)(271.2,38.4) 
\Line(271.2,38.4)(273.3,38.4) 
\Line(273.3,38.4)(275.8,38.4) 
\Line(275.8,38.4)(278.3,37.9) 
\Line(278.3,37.9)(280.8,37.9) 
\Line(280.8,37.9)(283.3,37.9) 
\Line(283.3,37.9)(285.8,37.3) 
\Line(285.8,37.3)(288.3,37.3) 
\Line(288.3,37.3)(290.8,36.7) 
\end{picture}
}
\caption{Cross section energy dependence of the process $\chi\chi \rightarrow b \bar b$.}\label{xsecbb}
\end{center}
\end{minipage}
\end{figure}\\
Another important fermionic channel is the process $\chi\chi \rightarrow b \bar b$, whose cross section-energy dependence is shown in Fig. \ref{xsecbb}. At the resonance the cross section increases significantly up to order $\mathcal{O}(10^2 \,\mathrm{pb})$ and this process is therefore the dominant channel in this energy region.
\paragraph{Final states containing gauge bosons}
Next we look at neutralino annihilation processes with gauge bosons in the final state, i.e. $\chi\chi \rightarrow W^+W^-$ and $\chi\chi\rightarrow Z^0Z^0$. The cross sections of these processes are shown in Fig. \ref{xsecww} and \ref{xseczz}, respectively. Since there is no s-channel exchange of the pseudoscalar Higgs $A^0$ there is only one resonance from the $H^0$ in the spectrum. In addition, there is a destructive interference between the s-channel diagrams of $h^0$ and $H^0$ which we study in detail in Appendix \ref{sec_treefeyn} for the process $\chi\chi \rightarrow Z^0Z^0$. The neutralino annihilation to gauge bosons is dominated by the process $\chi\chi \rightarrow W^+W^-$.
\begin{figure}[h!]
\centering
\begin{minipage}{7.2cm}
\begin{center}
\scalebox{0.7}{
\unitlength=1.0 pt
\SetScale{1.0}
\SetWidth{0.7}      % line    size control
\normalsize    %  letter  size control
\begin{picture}(300,200)(0,0)
%\Text(168.2,199.4)[t]{$~o1,~o1 ->W+,W-$}
% ====================   X-axis =============
\LinAxis(45.90,36.72)(290.82,36.72)(4.000,2,-4,0.000,1.5)
\Text(45.9,29.9)[t]{$200$}
\Text(107.2,29.9)[t]{$400$}
\Text(168.6,29.9)[t]{$600$}
\Text(229.5,29.9)[t]{$800$}
\Text(290.8,29.9)[t]{$10^3$}
\Text(290.8,20.3)[rt]{$\sqrt(s)$}
% ====================   Y-axis =============
\LogAxis(45.90,36.72)(45.90,176.27)(1.240,4,9.186,1.5)
\Text(39.2,40.7)[r]{$10^-2$}
\Text(39.2,153.7)[r]{$0.1$}
\rText(7.9,176.3)[tr][l]{Cross Section [pb]}
% ============== end of axis ============
\Line(48.0,50.3)(45.9,36.7)
\Line(50.5,58.2)(48.0,50.3)
\Line(53.0,63.8)(50.5,58.2)
\Line(55.5,68.4)(53.0,63.8)
\Line(58.0,71.2)(55.5,68.4)
\Line(60.5,74.0)(58.0,71.2)
\Line(63.0,75.7)(60.5,74.0)
\Line(65.1,77.4)(63.0,75.7)
\Line(67.6,78.5)(65.1,77.4)
\Line(70.1,79.7)(67.6,78.5)
\Line(72.6,80.8)(70.1,79.7)
\Line(75.1,81.4)(72.6,80.8)
\Line(77.6,81.9)(75.1,81.4)
\Line(80.1,82.5)(77.6,81.9)
\Line(82.6,83.1)(80.1,82.5)
\Line(84.7,83.6)(82.6,83.1)
\Line(87.2,84.2)(84.7,83.6)
\Line(89.7,84.7)(87.2,84.2)
\Line(92.2,85.3)(89.7,84.7)
\Line(94.7,87.0)(92.2,85.3)
\Line(97.2,88.7)(94.7,87.0)
\Line(99.7,92.7)(97.2,88.7)
\Line(102.2,104.0)(99.7,92.7)
\Line(104.3,176.8)(102.2,104.0)
\Line(104.3,176.8)(106.8,70.6)
\Line(106.8,70.6)(109.3,66.1)
\Line(111.8,68.9)(109.3,66.1)
\Line(114.3,70.6)(111.8,68.9)
\Line(116.8,71.8)(114.3,70.6)
\Line(119.3,72.3)(116.8,71.8)
\Line(121.4,72.3)(119.3,72.3)
\Line(123.9,72.9)(121.4,72.3)
\Line(123.9,72.9)(126.4,72.9)
\Line(126.4,72.9)(128.9,72.3)
\Line(128.9,72.3)(131.4,72.3)
\Line(131.4,72.3)(133.9,72.3)
\Line(133.9,72.3)(136.4,71.8)
\Line(136.4,71.8)(138.9,71.8)
\Line(138.9,71.8)(141.0,71.2)
\Line(141.0,71.2)(143.5,71.2)
\Line(143.5,71.2)(146.0,70.6)
\Line(146.0,70.6)(148.5,70.6)
\Line(148.5,70.6)(151.0,70.1)
\Line(151.0,70.1)(153.5,69.5)
\Line(153.5,69.5)(156.1,69.5)
\Line(156.1,69.5)(158.6,68.9)
\Line(158.6,68.9)(160.6,68.4)
\Line(160.6,68.4)(163.1,68.4)
\Line(163.1,68.4)(165.6,67.8)
\Line(165.6,67.8)(168.2,67.2)
\Line(168.2,67.2)(170.7,66.7)
\Line(170.7,66.7)(173.2,66.7)
\Line(173.2,66.7)(175.7,66.1)
\Line(175.7,66.1)(177.7,65.5)
\Line(177.7,65.5)(180.3,65.5)
\Line(180.3,65.5)(182.8,65.0)
\Line(182.8,65.0)(185.3,64.4)
\Line(185.3,64.4)(187.8,63.8)
\Line(187.8,63.8)(190.3,63.8)
\Line(190.3,63.8)(192.8,63.3)
\Line(192.8,63.3)(195.3,62.7)
\Line(195.3,62.7)(197.4,62.7)
\Line(197.4,62.7)(199.9,62.1)
\Line(199.9,62.1)(202.4,61.6)
\Line(202.4,61.6)(204.9,61.6)
\Line(204.9,61.6)(207.4,61.0)
\Line(207.4,61.0)(209.9,60.5)
\Line(209.9,60.5)(212.4,59.9)
\Line(212.4,59.9)(214.9,59.9)
\Line(214.9,59.9)(217.0,59.3)
\Line(217.0,59.3)(219.5,58.8)
\Line(219.5,58.8)(222.0,58.8)
\Line(222.0,58.8)(224.5,58.2)
\Line(224.5,58.2)(227.0,57.6)
\Line(227.0,57.6)(229.5,57.6)
\Line(229.5,57.6)(232.0,57.1)
\Line(232.0,57.1)(234.1,56.5)
\Line(234.1,56.5)(236.6,56.5)
\Line(236.6,56.5)(239.1,55.9)
\Line(239.1,55.9)(241.6,55.4)
\Line(241.6,55.4)(244.1,54.8)
\Line(244.1,54.8)(246.6,54.8)
\Line(246.6,54.8)(249.1,54.2)
\Line(249.1,54.2)(251.6,53.7)
\Line(251.6,53.7)(253.7,53.7)
\Line(253.7,53.7)(256.2,53.1)
\Line(256.2,53.1)(258.7,53.1)
\Line(258.7,53.1)(261.2,52.5)
\Line(261.2,52.5)(263.7,52.0)
\Line(263.7,52.0)(266.2,52.0)
\Line(266.2,52.0)(268.7,51.4)
\Line(268.7,51.4)(271.2,50.8)
\Line(271.2,50.8)(273.3,50.8)
\Line(273.3,50.8)(275.8,50.3)
\Line(275.8,50.3)(278.3,49.7)
\Line(278.3,49.7)(280.8,49.7)
\Line(280.8,49.7)(283.3,49.2)
\Line(283.3,49.2)(285.8,48.6)
\Line(285.8,48.6)(288.3,48.6)
\Line(288.3,48.6)(290.8,48.0)
\end{picture}
}
\caption{Energy dependence of the cross section of the process $\chi\chi \rightarrow W^+W^-$.}\label{xsecww}
\end{center}
\end{minipage}
\hspace{1cm}
\begin{minipage}{7.2cm}
\scalebox{0.7}{\unitlength=1.0 pt
\SetScale{1.0}
\SetWidth{0.7}      % line    size control
\normalsize    %  letter  size control
\begin{picture}(300,200)(0,0)
%\Text(168.2,199.4)[t]{$~o1,~o1 ->Z,Z$}
% ====================   X-axis =============
\LinAxis(45.90,36.72)(290.82,36.72)(4.000,2,-4,0.000,1.5)
\Text(45.9,29.9)[t]{$200$}
\Text(107.2,29.9)[t]{$400$}
\Text(168.6,29.9)[t]{$600$}
\Text(229.5,29.9)[t]{$800$}
\Text(290.8,29.9)[t]{$10^3$}
\Text(290.8,20.3)[rt]{$\sqrt(s)$}
% ====================   Y-axis =============
\LogAxis(45.90,36.72)(45.90,176.27)(1.873,4,1.123,1.5)
\Text(39.2,107.3)[r]{$10^-2$}
\rText(7.9,176.3)[tr][l]{Cross Section [pb]}
% ============== end of axis ============
\Line(48.0,82.5)(45.9,70.1) 
\Line(50.5,88.7)(48.0,82.5) 
\Line(53.0,92.7)(50.5,88.7) 
\Line(55.5,94.9)(53.0,92.7) 
\Line(58.0,97.2)(55.5,94.9) 
\Line(60.5,98.9)(58.0,97.2) 
\Line(63.0,100.0)(60.5,98.9) 
\Line(65.1,101.1)(63.0,100.0) 
\Line(67.6,101.7)(65.1,101.1) 
\Line(70.1,102.3)(67.6,101.7) 
\Line(72.6,102.8)(70.1,102.3) 
\Line(75.1,103.4)(72.6,102.8) 
\Line(77.6,103.4)(75.1,103.4) 
\Line(80.1,104.0)(77.6,103.4) 
\Line(82.6,104.5)(80.1,104.0) 
\Line(84.7,105.1)(82.6,104.5) 
\Line(87.2,105.1)(84.7,105.1) 
\Line(89.7,105.6)(87.2,105.1) 
\Line(92.2,106.8)(89.7,105.6) 
\Line(94.7,107.9)(92.2,106.8) 
\Line(97.2,110.2)(94.7,107.9) 
\Line(99.7,114.7)(97.2,110.2) 
\Line(102.2,124.9)(99.7,114.7) 
\Line(104.3,176.8)(102.2,124.9) 
\Line(104.3,176.8)(106.8,36.7) 
\Line(109.3,70.1)(106.8,36.7) 
\Line(111.8,80.8)(109.3,70.1) 
\Line(114.3,84.7)(111.8,80.8) 
\Line(116.8,87.0)(114.3,84.7) 
\Line(119.3,88.1)(116.8,87.0) 
\Line(121.4,88.7)(119.3,88.1) 
\Line(123.9,89.3)(121.4,88.7) 
\Line(126.4,89.8)(123.9,89.3) 
\Line(128.9,89.8)(126.4,89.8) 
\Line(128.9,89.8)(131.4,89.8) 
\Line(131.4,89.8)(133.9,89.3) 
\Line(133.9,89.3)(136.4,89.3) 
\Line(136.4,89.3)(138.9,89.3) 
\Line(138.9,89.3)(141.0,88.7) 
\Line(141.0,88.7)(143.5,88.7) 
\Line(143.5,88.7)(146.0,88.1) 
\Line(146.0,88.1)(148.5,88.1) 
\Line(148.5,88.1)(151.0,87.6) 
\Line(151.0,87.6)(153.5,87.6) 
\Line(153.5,87.6)(156.1,87.0) 
\Line(156.1,87.0)(158.6,87.0) 
\Line(158.6,87.0)(160.6,86.4) 
\Line(160.6,86.4)(163.1,85.9) 
\Line(163.1,85.9)(165.6,85.9) 
\Line(165.6,85.9)(168.2,85.3) 
\Line(168.2,85.3)(170.7,85.3) 
\Line(170.7,85.3)(173.2,84.7) 
\Line(173.2,84.7)(175.7,84.7) 
\Line(175.7,84.7)(177.7,84.2) 
\Line(177.7,84.2)(180.3,83.6) 
\Line(180.3,83.6)(182.8,83.6) 
\Line(182.8,83.6)(185.3,83.1) 
\Line(185.3,83.1)(187.8,83.1) 
\Line(187.8,83.1)(190.3,82.5) 
\Line(190.3,82.5)(192.8,82.5) 
\Line(192.8,82.5)(195.3,81.9) 
\Line(195.3,81.9)(197.4,81.9) 
\Line(197.4,81.9)(199.9,81.4) 
\Line(199.9,81.4)(202.4,80.8) 
\Line(202.4,80.8)(204.9,80.8) 
\Line(204.9,80.8)(207.4,80.2) 
\Line(207.4,80.2)(209.9,80.2) 
\Line(209.9,80.2)(212.4,79.7) 
\Line(212.4,79.7)(214.9,79.7) 
\Line(214.9,79.7)(217.0,79.1) 
\Line(217.0,79.1)(219.5,79.1) 
\Line(219.5,79.1)(222.0,78.5) 
\Line(222.0,78.5)(224.5,78.5) 
\Line(224.5,78.5)(227.0,78.5) 
\Line(227.0,78.5)(229.5,78.0) 
\Line(229.5,78.0)(232.0,78.0) 
\Line(232.0,78.0)(234.1,77.4) 
\Line(234.1,77.4)(236.6,77.4) 
\Line(236.6,77.4)(239.1,76.8) 
\Line(239.1,76.8)(241.6,76.8) 
\Line(241.6,76.8)(244.1,76.3) 
\Line(244.1,76.3)(246.6,76.3) 
\Line(246.6,76.3)(249.1,76.3) 
\Line(249.1,76.3)(251.6,75.7) 
\Line(251.6,75.7)(253.7,75.7) 
\Line(253.7,75.7)(256.2,75.1) 
\Line(256.2,75.1)(258.7,75.1) 
\Line(258.7,75.1)(261.2,75.1) 
\Line(261.2,75.1)(263.7,74.6) 
\Line(263.7,74.6)(266.2,74.6) 
\Line(266.2,74.6)(268.7,74.0) 
\Line(268.7,74.0)(271.2,74.0) 
\Line(271.2,74.0)(273.3,74.0) 
\Line(273.3,74.0)(275.8,73.4) 
\Line(275.8,73.4)(278.3,73.4) 
\Line(278.3,73.4)(280.8,73.4) 
\Line(280.8,73.4)(283.3,72.9) 
\Line(283.3,72.9)(285.8,72.9) 
\Line(285.8,72.9)(288.3,72.9) 
\Line(288.3,72.9)(290.8,72.3) 
\end{picture}
}
\caption{Energy dependence of the cross section of the process $\chi\chi \rightarrow Z^0Z^0$.}\label{xseczz}
\end{minipage}
\end{figure}
\paragraph{Final states containing Higgs bosons}
Neutralino annihilation processes which contain Higgs bosons in the final state are only available at higher energies since the masses are much larger than $m_\chi$. An exception may be the processes $\chi \chi \rightarrow Z^0 h^0$ and $\chi \chi \rightarrow h^0 h^0$, which contain the lightest Higgs in the final state. These channels are already open at lower energies and therefore contain resonances as well. The process $\chi \chi \rightarrow h^0 h^0$ shows similar behavior as the process $\chi\chi \rightarrow Z^0Z^0$ with a significant interference between the s-channel diagrams of $h^0$ and $H^0$.
\subsubsection{Comparing cross sections}
The cross sections for all tree-level neutralino annihilation processes have been calculated and are given in Table \ref{xsectotal} for five different energies, i.e. 200 GeV, 360 GeV, 600 GeV, 1000 GeV and 10000 GeV. The neutralino annihilation is dominated by processes with the leptons $e$, $\mu$ and $\tau$ in the final state. The total cross section is about 1.7 pb at 200 GeV where the neutralinos are nearly at rest. At 360 GeV where the energy is close to the mass of $A^0$ and $H^0$, the cross section increases up to around 4.3 pb. In the resonance region, the neutralino annihilation is dominated by the process $\chi\chi \rightarrow b \bar b$.
\begin{table}[h!]
\centering
\scalebox{0.9}{
\begin{tabular}{|l|c|c|c|c|c|}
\hline
$\quad\sigma \,[\mathrm{pb}]$ 			& \multicolumn{5}{c|}{Center-of-mass energy $\sqrt{s}$/GeV}\\
\hline
Final state	&	200	&	360	&	600	&	1000	&	10000\\
\hline
$e^+e^-$		&	4.89E-01&	8.79E-01&	4.72E-01&	2.04E-01&	2.37E-03\\
$\mu^+\mu^-$	&	4.89E-01&	8.79E-01&	4.72E-01&	2.04E-01&	2.37E-03\\
$\tau^+\tau^-$	&	5.44E-01&	9.57E-01&	4.82E-01&	2.06E-01&	2.38E-03\\
$\nu_e\bar\nu_e$&	3.14E-02&	6.69E-02&	4.03E-02&	1.86E-02&	2.30E-04\\
$\nu_\mu\bar\nu_\mu$&	3.14E-02&	6.69E-02&	4.03E-02&	1.86E-02&	2.30E-04\\
$\nu_\tau\bar\nu_\tau$&	3.18E-02&	6.73E-02&	4.05E-02&	1.84E-02&	2.30E-04\\
$u \bar u$	&		8.13E-03&	4.32E-02&	6.10E-02&	5.27E-02&	1.35E-03\\
$d \bar d$	&		6.41E-04&	2.08E-03&	3.46E-03&	3.28E-03&	9.10E-05\\
$c \bar c$	&		8.13E-03&	4.32E-02&	6.10E-02&	5.27E-02&	1.35E-03\\
$s \bar s$	&		6.41E-04&	2.08E-03&	3.46E-03&	3.28E-03&	9.10E-05\\
$t \bar t$		&			&	5.56E-02&	9.93E-02&	7.49E-02&	1.45E-03\\
$b \bar b$		&	6.76E-02&	1.13E+00&	3.11E-02&		5.64E-03&	1.05E-04\\
$W^+W^-$	&	9.19E-03&	2.54E-02&	1.71E-02&	1.15E-02&	5.46E-04\\
$ZZ$			&	3.12E-03&	1.01E-02&	5.02E-03&	3.36E-03&	2.44E-04\\
$Z h$		&			&	2.19E-02&	2.17E-02& 	1.58E-02&	6.32E-04\\
$Z H$		&				&		&	2.66E-02&	1.40E-02&	1.94E-04\\
$Z A$		&				&		&	2.06E-02&	1.20E-02&	1.58E-04\\
$W^+ H^-$	&				&		&	2.37E-02&	1.28E-02&	2.25E-04\\
$W^- H^+$	&				&		&	2.37E-02&	1.28E-02&	2.25E-04\\
$hh$		&				&	2.88E-02&	4.91E-03&	2.52E-03&	2.11E-04\\
$HH$		&					&		&		&	5.69E-04&	2.50E-04\\
$hH$		&				&		&	3.06E-02&	1.37E-02&	1.05E-04\\
$H^+H^-$	&						&		&		&	4.45E-03&	6.83E-04\\
$Ah$		&					&		&	3.65E-02&	1.58E-02&	1.49E-04\\
$AH$		&					&		&		&	6.41E-03&	7.13E-04\\
\hline
$\sigma_{tot}$	& 1.71E+00	&4.28E+00	&2.02E+00	&	0.99E+00&	1.66E-02	\\
\hline
\end{tabular}}
\caption{Cross section in pb of all tree-level neutralino annihilation channels at different energies $\sqrt{s}$ for the mSUGRA-parameter point SPSa1.}\label{xsectotal}
\end{table}
\subsection{Parameter dependence}\label{sec_tree4}
\subsubsection{Neutralino mass}
The dependence of the lightest neutralino mass $m_\chi$ on $m_0$ and $m_{1/2}$ is sketched in Fig. \ref{neutralinomass1} and \ref{neutralinomass2}, respectively, for $\tan\beta = 10$, $A_0=-100$ GeV and $\mu > 0$. We see, that $m_{\chi}$ increases slightly with $m_0$ whereas significantly with $m_{1/2}$. The dependence on $\tan\beta$, $A_0$ and $\mathrm{sign}(\mu)$ is negligible.
\begin{figure}[h!]
\centering
\begin{minipage}{7.2cm}
{
\unitlength=0.6 pt
\SetScale{0.6}
\SetWidth{0.42}     % line    size control
\scriptsize    %  letter  size control
\begin{picture}(300,200)(0,0)
% ====================   X-axis =============
\LinAxis(37.55,31.11)(292.49,31.11)(3.500,10,-4,0.000,1.5)
\Text(37.6,25.6)[t]{$100$}
\Text(110.6,25.6)[t]{$200$}
\Text(183.2,25.6)[t]{$300$}
\Text(256.2,25.6)[t]{$400$}
\Text(300.5,17.8)[rt]{$m_{1/2}$}
% ====================   Y-axis =============
\LinAxis(37.55,31.11)(37.55,180.56)(3.022,5,4,-1.902,1.5)
\Text(31.7,50.0)[r]{$50$}
\Text(31.7,99.4)[r]{$100$}
\Text(31.7,148.9)[r]{$150$}
\rText(7.1,180.6)[tr][l]{$m_\chi$}
% ============== end of axis ============
\Line(50.9,40.0)(37.6,31.1) 
\Line(64.3,48.3)(50.9,40.0) 
\Line(77.6,56.1)(64.3,48.3) 
\Line(91.0,63.9)(77.6,56.1) 
\Line(104.3,72.2)(91.0,63.9) 
\Line(117.7,79.4)(104.3,72.2) 
\Line(131.4,87.2)(117.7,79.4) 
\Line(144.8,95.0)(131.4,87.2) 
\Line(158.1,102.2)(144.8,95.0) 
\Line(171.5,110.6)(158.1,102.2) 
\Line(184.8,118.3)(171.5,110.6) 
\Line(198.2,126.1)(184.8,118.3) 
\Line(212.0,133.9)(198.2,126.1) 
\Line(225.3,141.1)(212.0,133.9) 
\Line(238.7,148.9)(225.3,141.1) 
\Line(252.0,157.8)(238.7,148.9) 
\Line(265.4,165.6)(252.0,157.8) 
\Line(278.7,172.8)(265.4,165.6) 
\Line(292.5,180.6)(278.7,172.8) 
\end{picture}
}
\caption{Neutralino mass dependence on $m_0$ for $m_{1/2} = 250$ GeV, $\tan\beta = 10$, $A_0=-100$ GeV and $\mu > 0$.}\label{neutralinomass1}
\end{minipage}
\hspace{1cm}
\begin{minipage}{7.2cm}
{
\unitlength=0.6 pt
\SetScale{0.6}
\SetWidth{0.42}      % line    size control
\scriptsize    %  letter  size control
\begin{picture}(300,200)(0,0)
% ====================   X-axis =============
\LinAxis(37.55,30.17)(292.49,30.17)(4.000,10,-4,0.000,1.5)
\Text(37.6,24.6)[t]{$100$}
\Text(101.4,24.6)[t]{$200$}
\Text(165.2,24.6)[t]{$300$}
\Text(228.7,24.6)[t]{$400$}
\Text(292.5,24.6)[t]{$500$}
\Text(275.5,16.8)[rt]{$m_0$}
% ====================   Y-axis =============
\LinAxis(37.55,30.17)(37.55,180.45)(3.948,2,4,0.284,1.5)
\Text(31.7,62.6)[r]{$96.4$}
\Text(31.7,101.1)[r]{$96.6$}
\Text(31.7,139.1)[r]{$96.8$}
\Text(31.7,177.1)[r]{$97.0$}
\rText(2.9,172.4)[tr][l]{$m_\chi$}
% ============== end of axis ============
\Line(65.5,41.3)(37.6,30.2) 
\Line(93.9,59.8)(65.5,41.3) 
\Line(122.3,77.7)(93.9,59.8) 
\Line(150.6,98.3)(122.3,77.7) 
\Line(179.0,120.1)(150.6,98.3) 
\Line(207.4,136.9)(179.0,120.1) 
\Line(235.7,147.5)(207.4,136.9) 
\Line(264.1,167.6)(235.7,147.5) 
\Line(292.5,180.4)(264.1,167.6) 
\end{picture}
}
\caption{Neutralino mass dependence on $m_{1/2}$ for $m_0 = 100$ GeV, $\tan\beta = 10$, $A_0=-100$ GeV and $\mu > 0$.}\label{neutralinomass2}
\end{minipage}
\end{figure}
\subsubsection{Total neutralino annihilation cross section}
\begin{figure}[p]
\centering
\begin{minipage}{7.8cm}
\scalebox{0.26}{\includegraphics{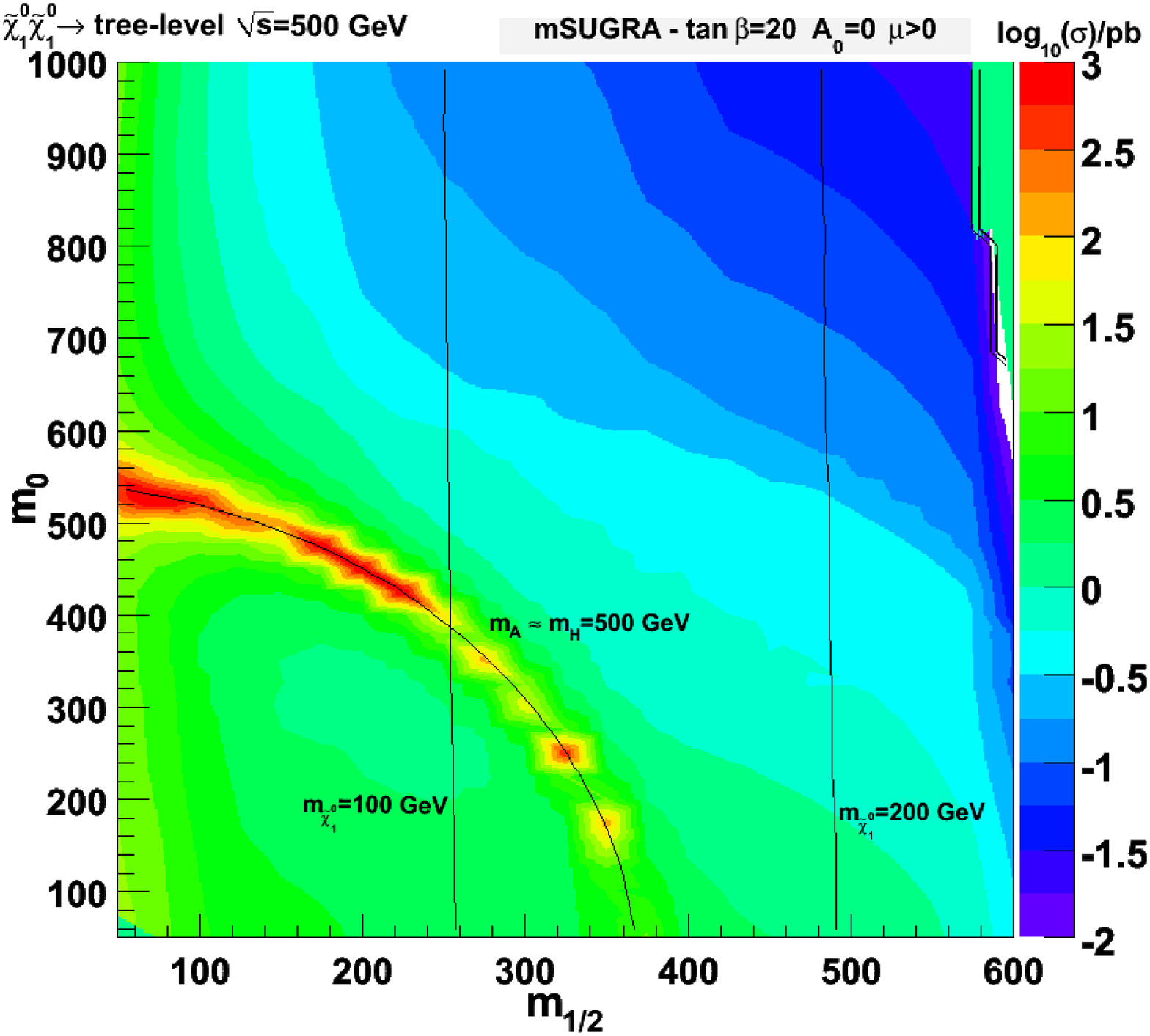}}
\end{minipage}
\hspace{1cm}
\begin{minipage}{7.8cm}
\scalebox{0.26}{\includegraphics{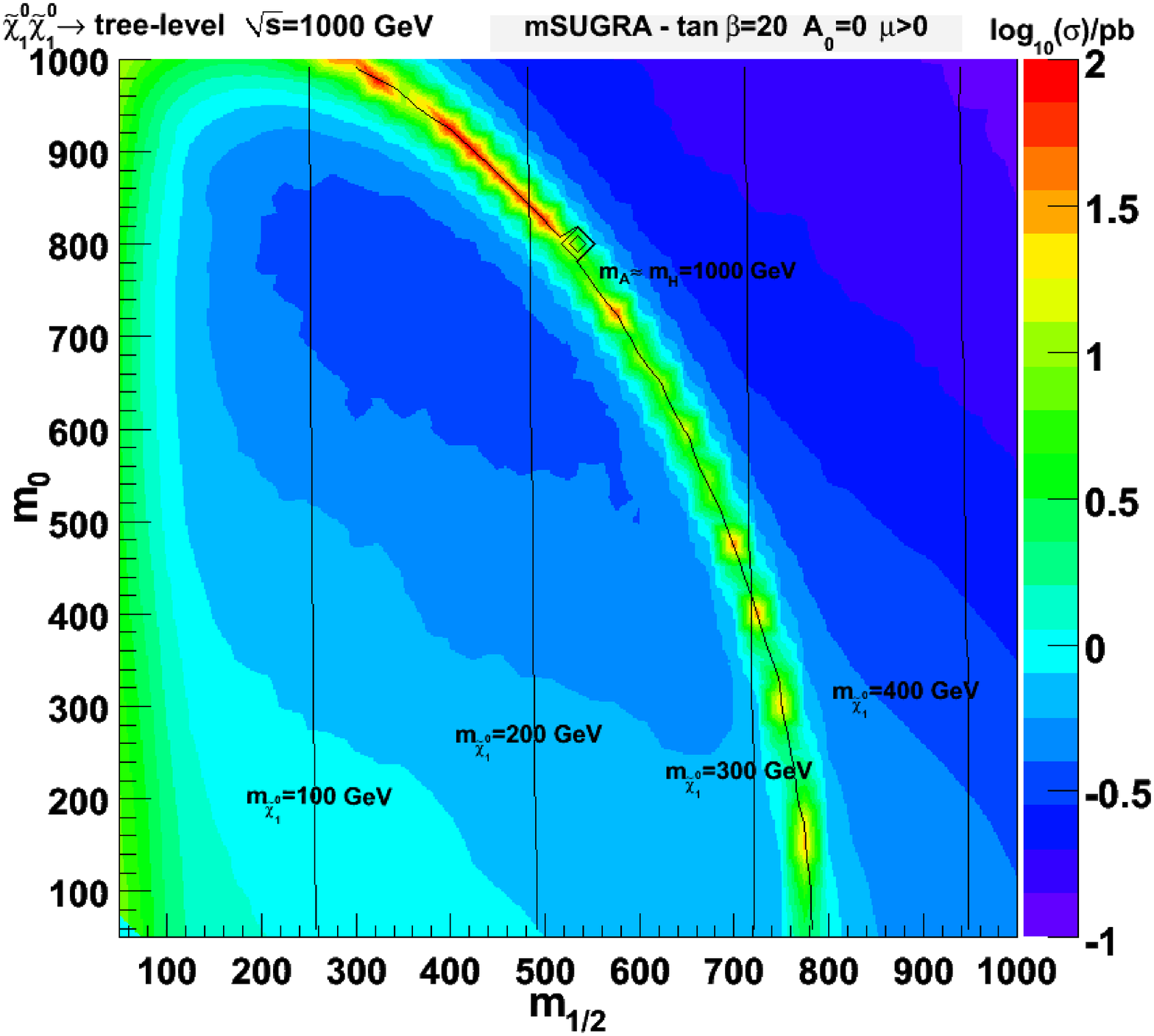}}
\end{minipage}
\caption{Total tree-level neutralino annihilation cross section in $m_{1/2}$-$m_{0}$ plane for $\tan\beta=20$, $A_0=0$ and $\mu>0$ and center-of-mass energies 500 GeV (left) and 1000 GeV (right). Contour lines for neutralino masses of $m_\chi=$ 100 GeV, 200 GeV, 300 GeV and 400 GeV as well as for Higgs masses $m_A\approx m_H =$ 500 GeV (left) and 1000 GeV (right) were added.}\label{tanb20A0}
\end{figure}

\begin{figure}[p]
\centering
\begin{minipage}{7.8cm}
\centering
\scalebox{0.26}{\includegraphics{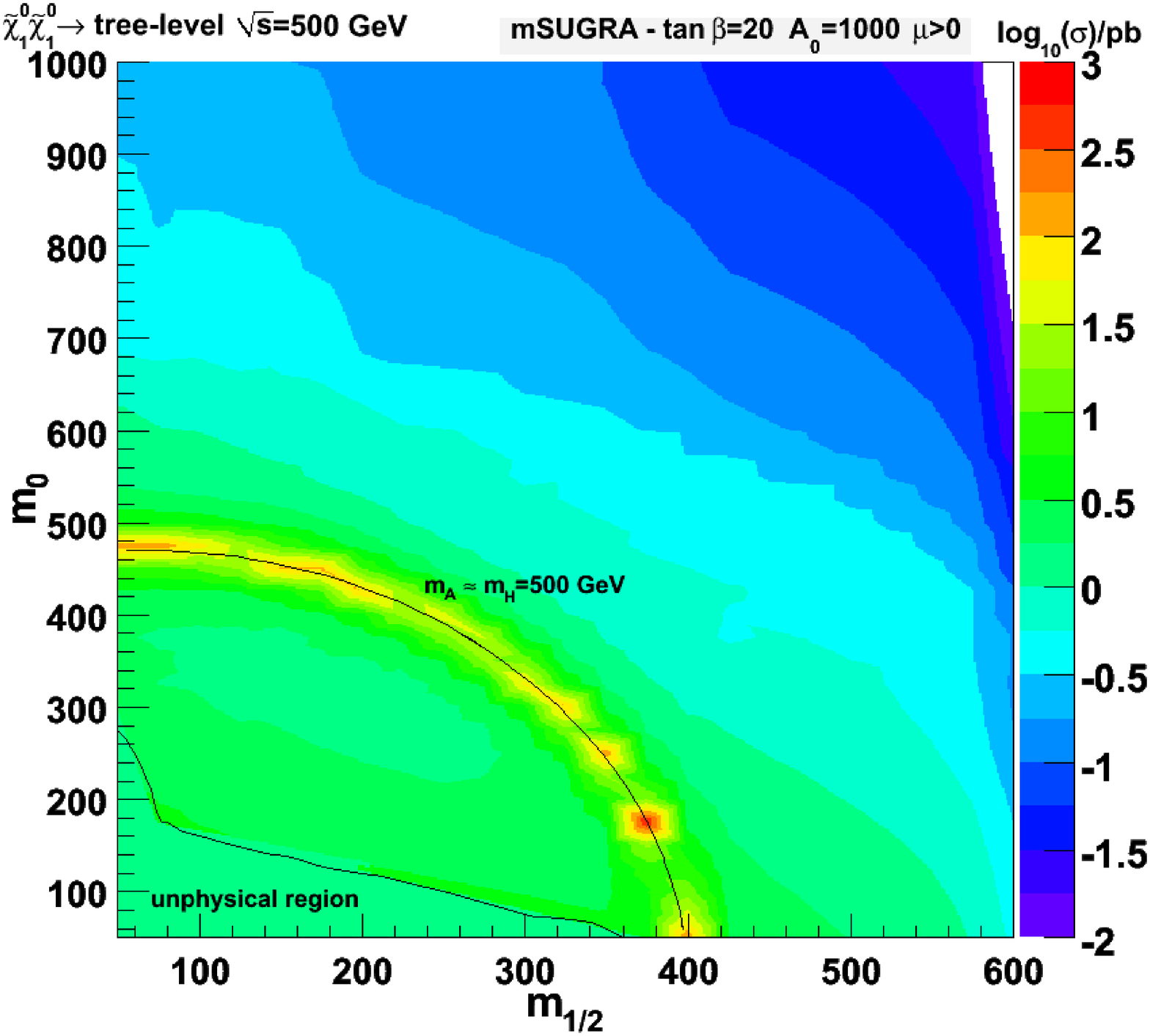}}
\end{minipage}
\hspace{1cm}
\begin{minipage}{7.8cm}
\centering
\scalebox{0.26}{\includegraphics{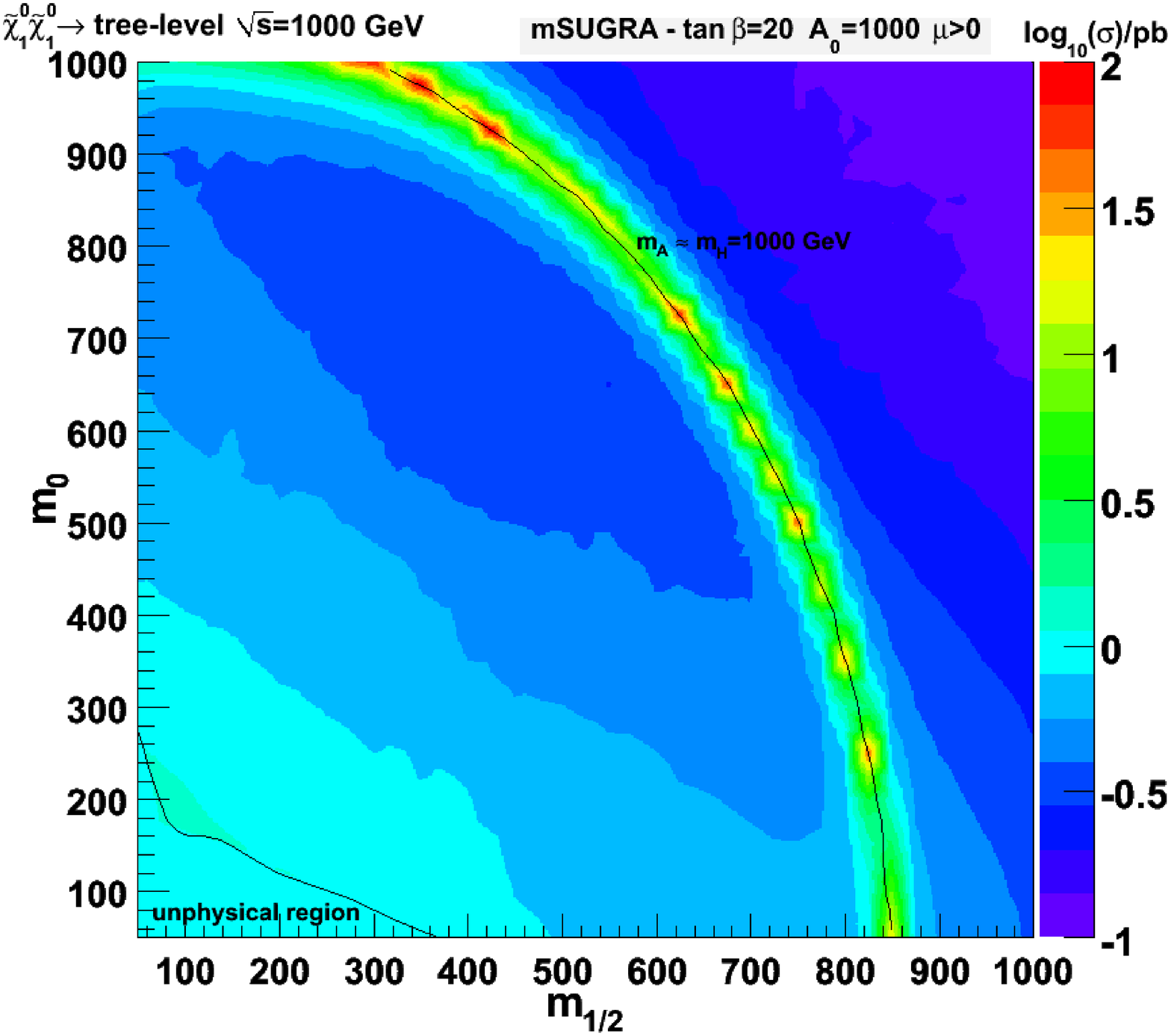}}
\end{minipage}
\caption{Total tree-level neutralino annihilation cross section in $m_{1/2}$-$m_{0}$ plane for $\tan\beta=20$, $A_0=1000$ GeV and $\mu>0$ and center-of-mass energies 500 GeV (left) and 1000 GeV (right).}\label{tanb20A1000}
\end{figure}

\begin{figure}[p]
\centering
\begin{minipage}{7.8cm}
\centering
\scalebox{0.26}{\includegraphics{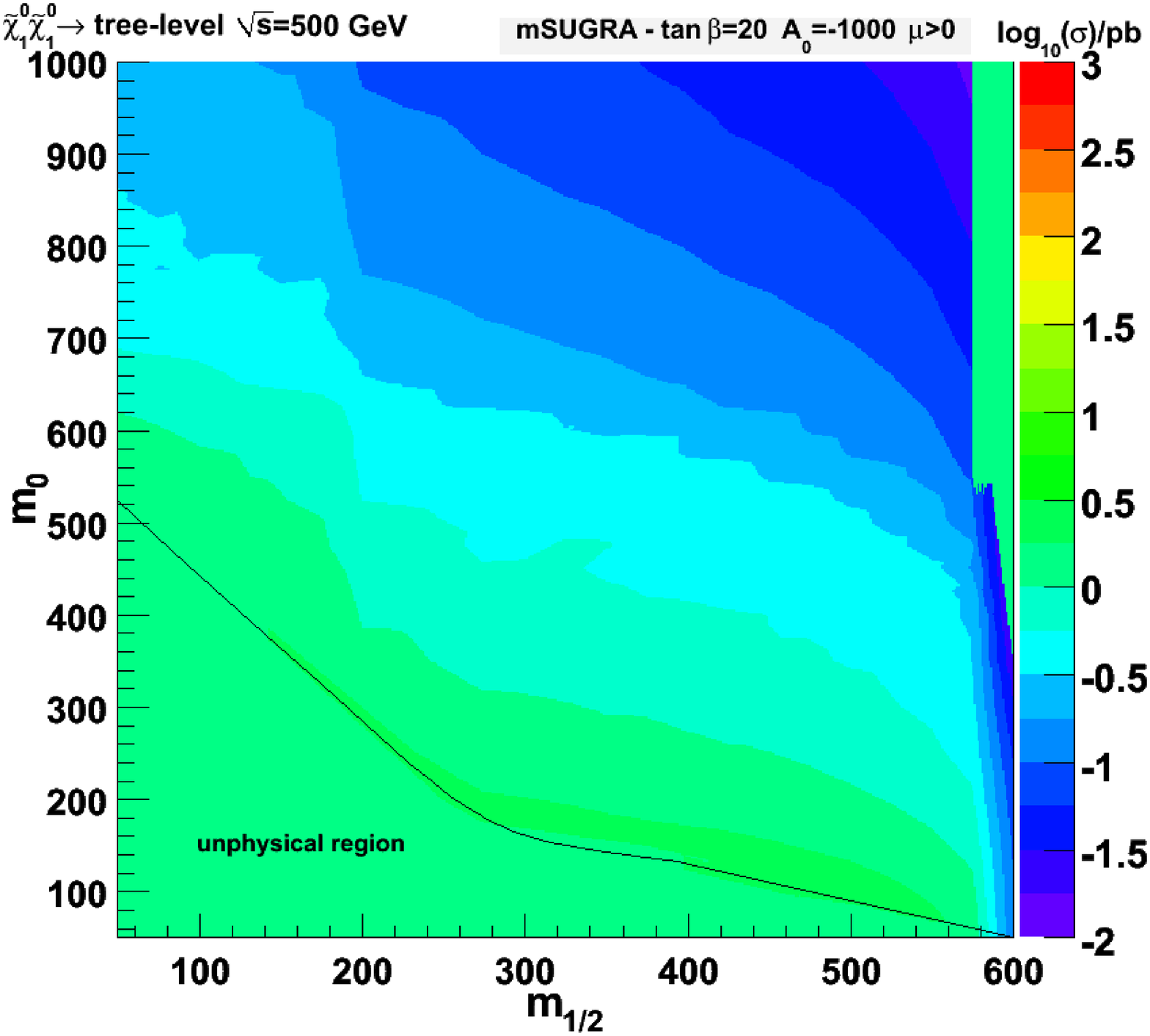}}
\end{minipage}
\hspace{1cm}
\begin{minipage}{7.8cm}
\centering
\scalebox{0.26}{\includegraphics{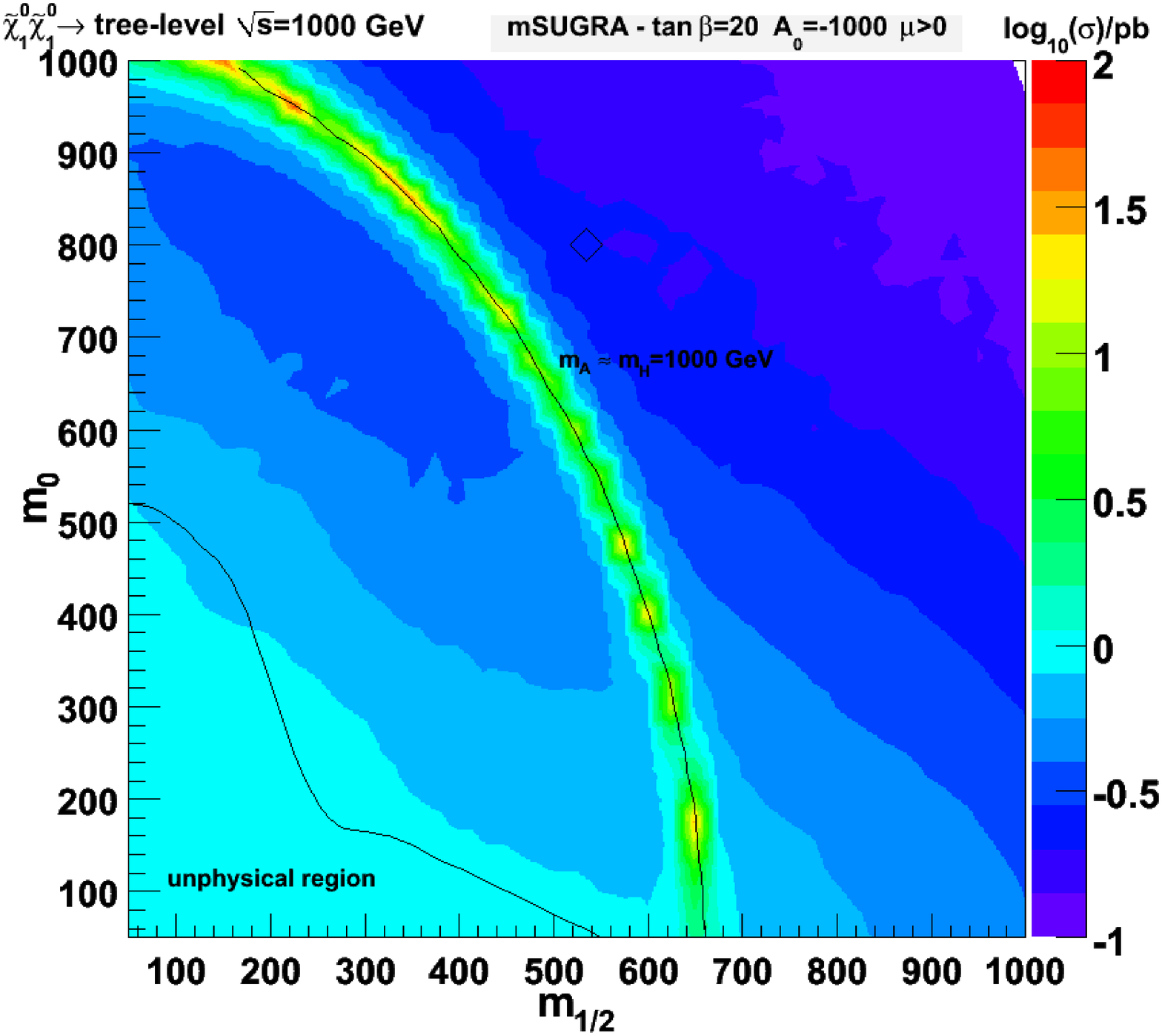}}
\end{minipage}
\caption{Total tree-level neutralino annihilation cross section in $m_{1/2}$-$m_{0}$ plane for $\tan\beta=20$, $A_0=-1000$ GeV and $\mu>0$ and center-of-mass energies 500 GeV (left) and 1000 GeV (right).}\label{tanb20A-1000}
\end{figure}

\begin{figure}[p]
\centering
\begin{minipage}{7.8cm}
\centering
\scalebox{0.26}{\includegraphics{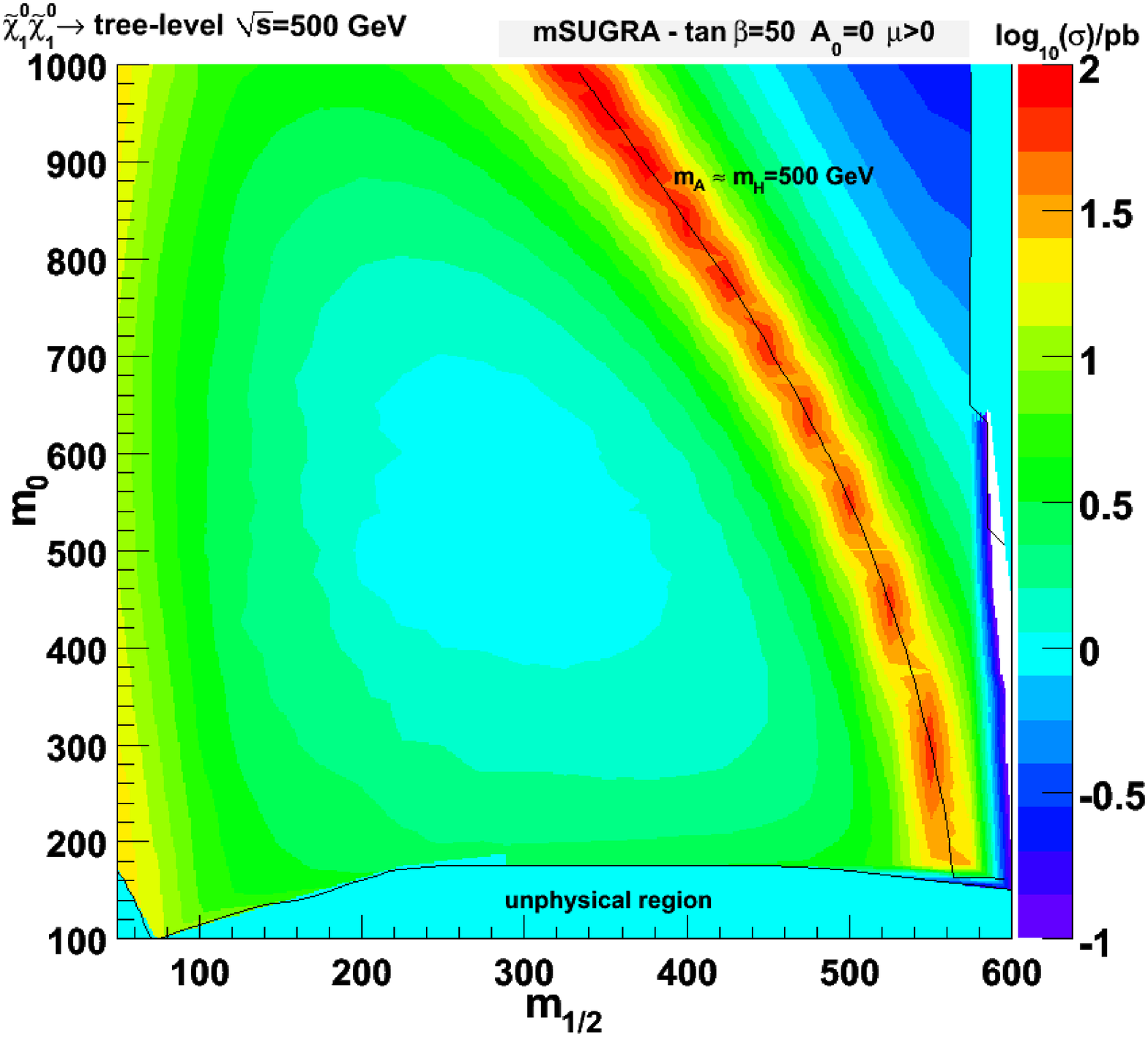}}
\end{minipage}
\hspace{1cm}
\begin{minipage}{7.8cm}
\centering
\scalebox{0.26}{\includegraphics{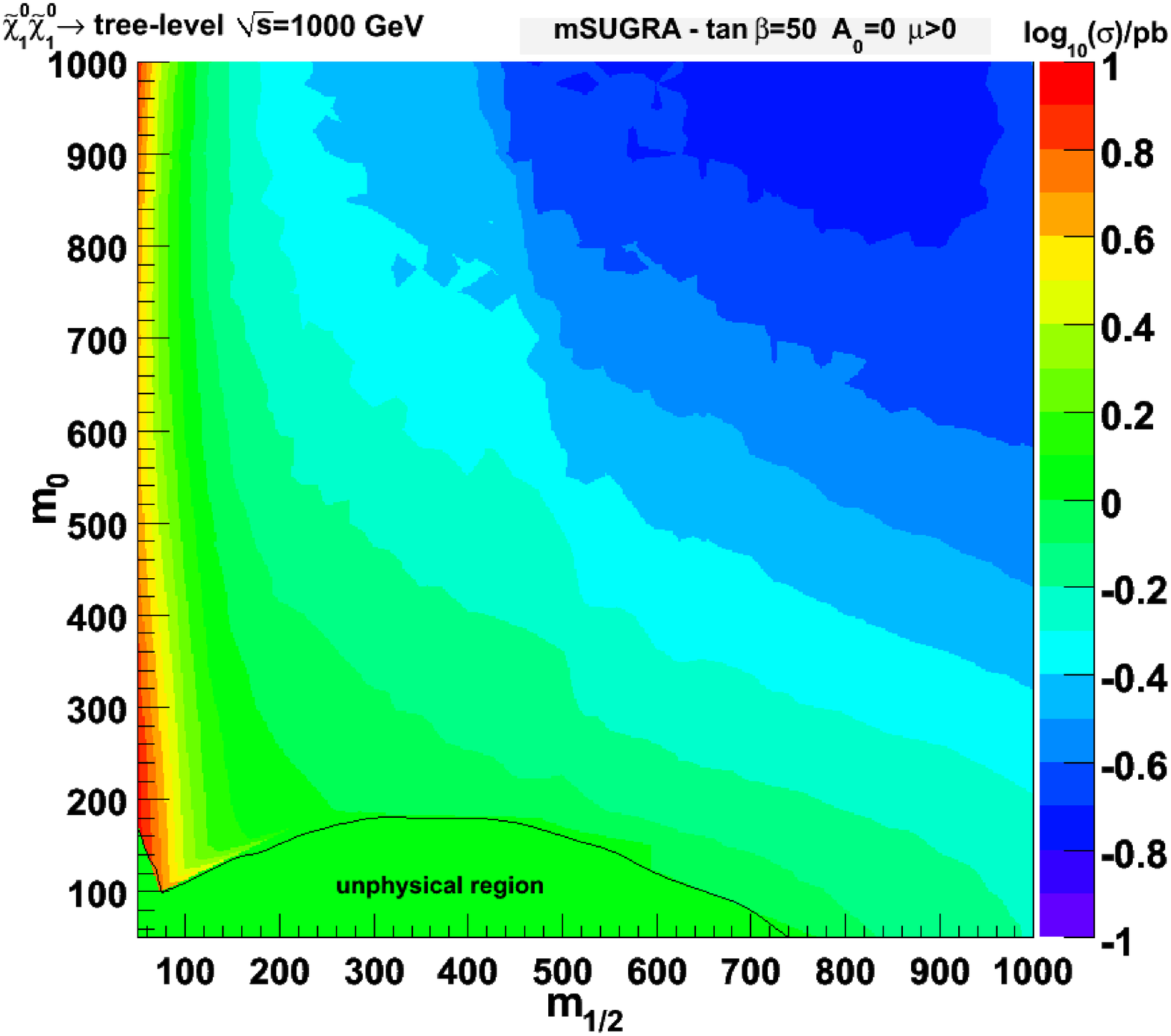}}
\end{minipage}
\caption{Total tree-level neutralino annihilation cross section in $m_{1/2}$-$m_{0}$ plane for $\tan\beta=50$, $A_0=0$ and $\mu>0$ and center-of-mass energies 500 GeV (left) and 1000 GeV (right).}\label{tanb50A0}
\end{figure}

\begin{figure}[p]
\centering
\begin{minipage}{7.8cm}
\centering
\scalebox{0.26}{\includegraphics{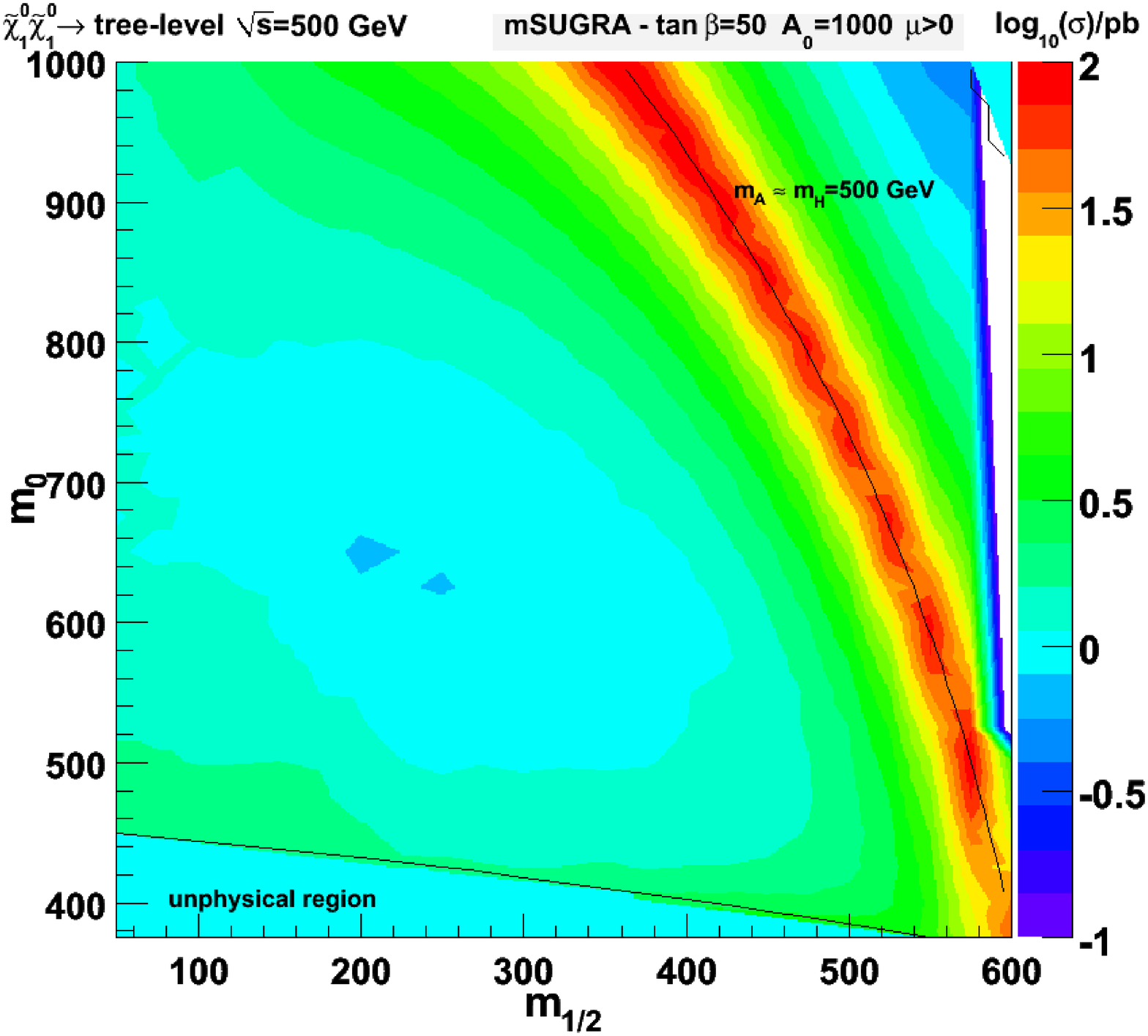}}
\end{minipage}
\hspace{1cm}
\begin{minipage}{7.8cm}
\centering
\scalebox{0.26}{\includegraphics{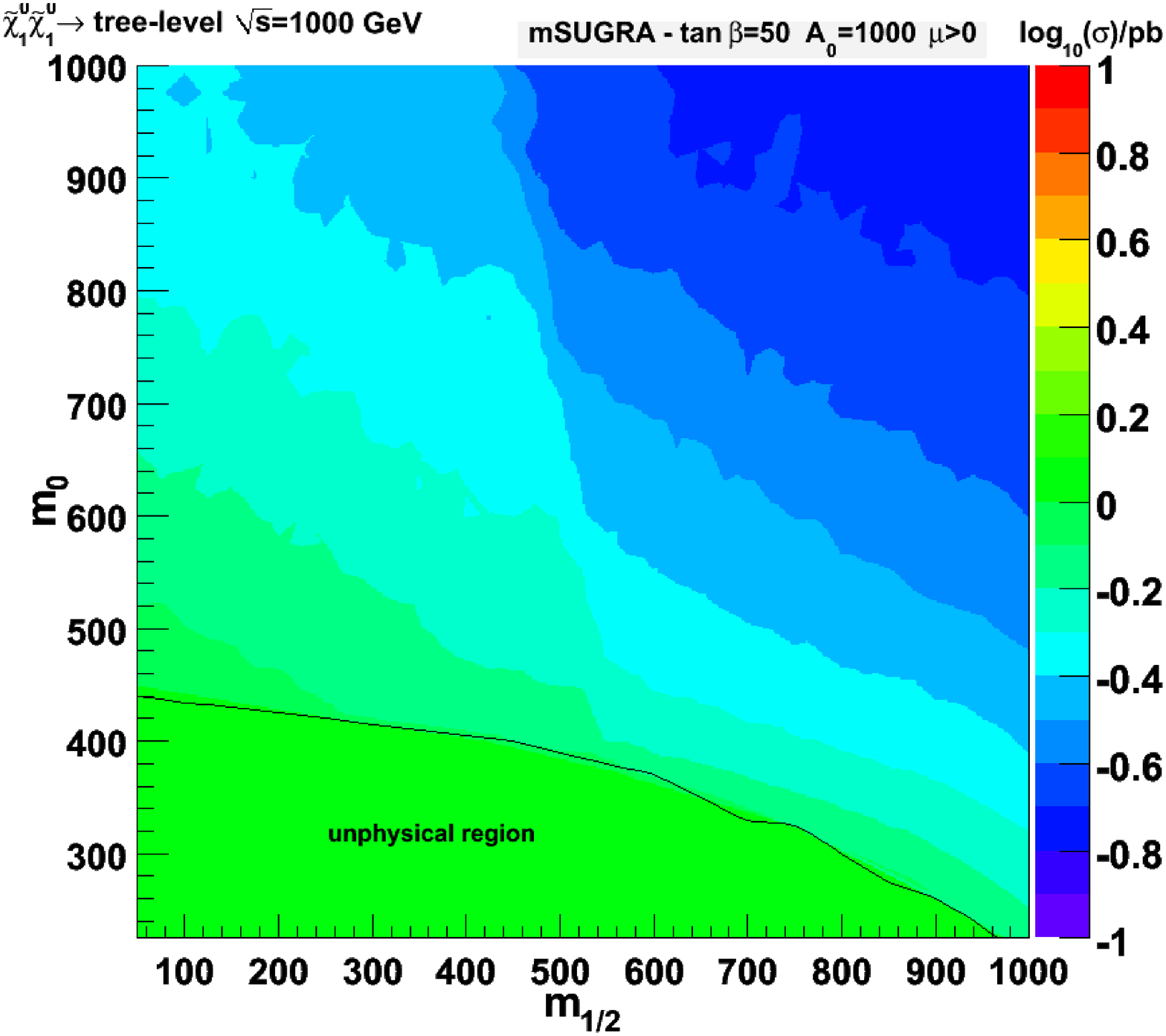}}
\end{minipage}
\caption{Total tree-level neutralino annihilation cross section in $m_{1/2}$-$m_{0}$ plane for $\tan\beta=50$, $A_0=1000$ GeV and $\mu>0$ and center-of-mass energies 500 GeV (left) and 1000 GeV (right).}\label{tanb50A1000}
\end{figure}

\begin{figure}[p]
\centering
\begin{minipage}{7.8cm}
\centering
\scalebox{0.26}{\includegraphics{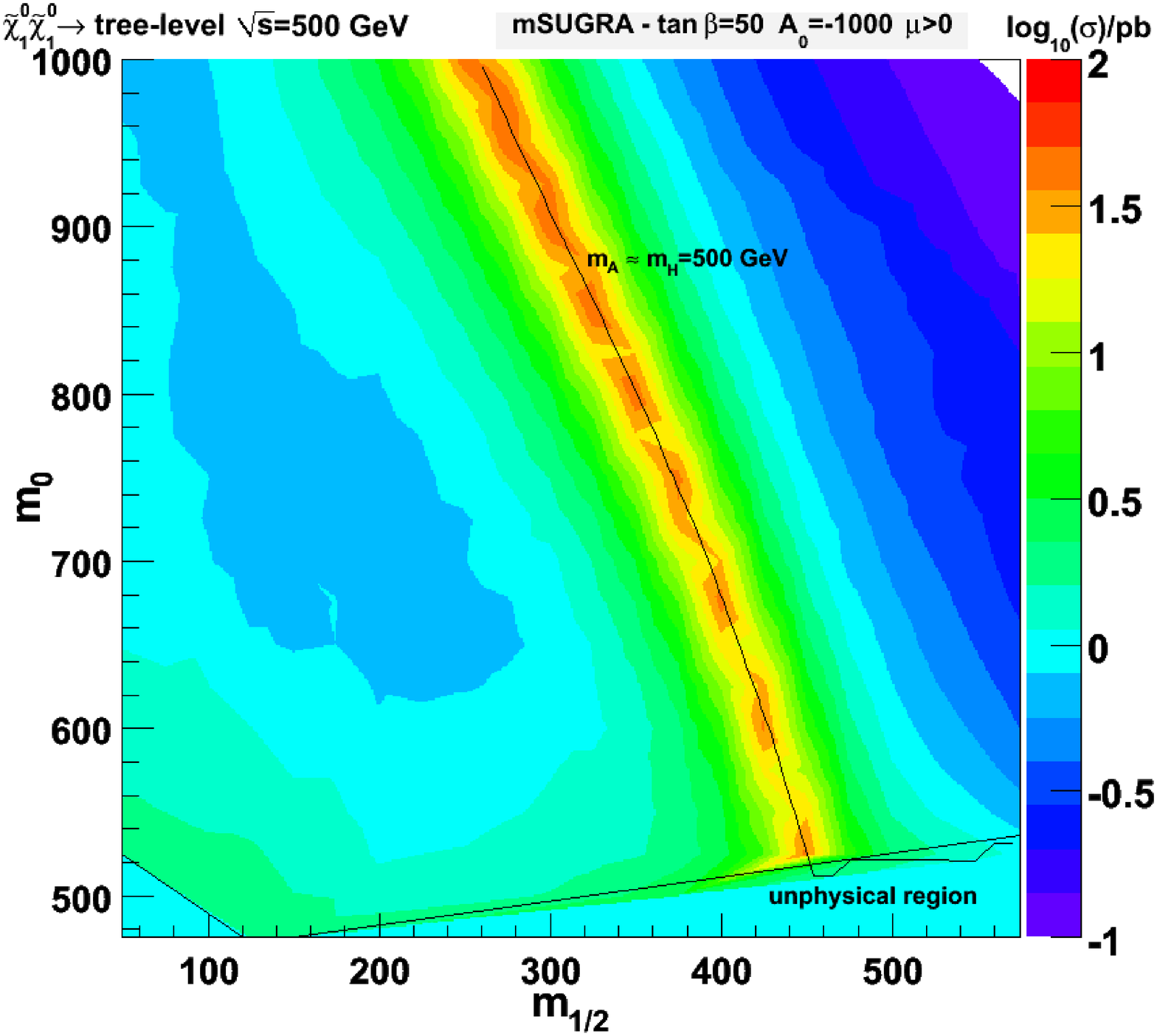}}
\end{minipage}
\hspace{1cm}
\begin{minipage}{7.8cm}
\centering
\scalebox{0.26}{\includegraphics{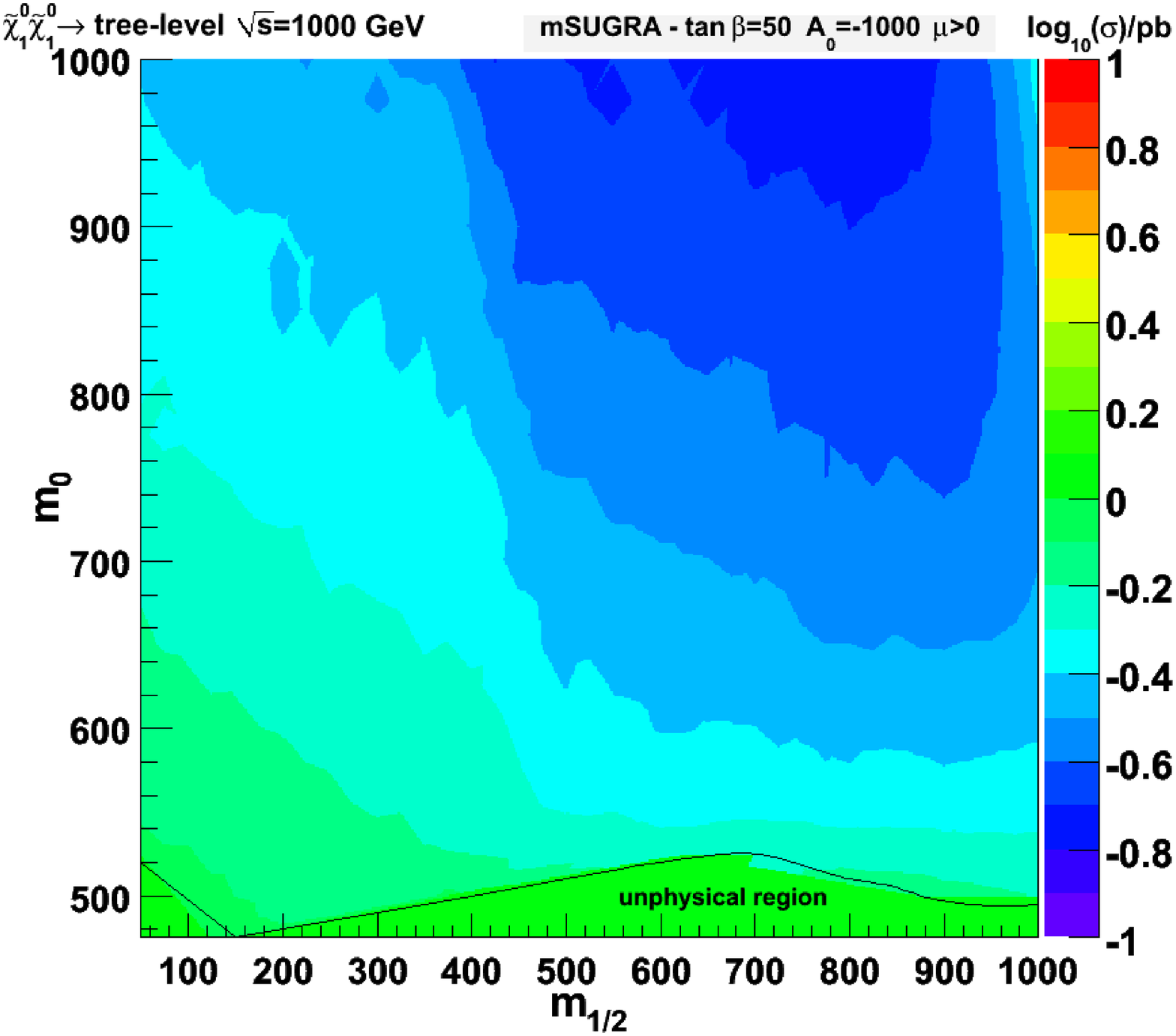}}
\end{minipage}
\caption{Total tree-level neutralino annihilation cross section in $m_{1/2}$-$m_{0}$ plane for $\tan\beta=50$, $A_0=-1000$ GeV and $\mu>0$ and center-of-mass energies 500 GeV (left) and 1000 GeV (right).}\label{tanb50A-1000}
\end{figure}

In order to study the dependence of the neutralino annihilation cross section on the mSUGRA parameter space scans have been performed in the $m_{1/2}$-$m_{0}$-plane for fixed values of $\tan\beta=20,50$, $A_0=-1000,0,1000$ GeV and $\mu>0$ for two different center-of-mass energies 500 GeV and 1000 GeV.\\
\\
For every parameter point, the particle spectrum was calculated by {\scshape{SoftSUSY}} \cite{SoftSUSY} and decay widths were added by {\scshape{MSSMCalc}} \cite{MSSMCalc}. Then, the cross-section calculation was performed by {\scshape{MadGraph}}. The $m_{1/2}$-$m_{0}$-plane was scanned in steps of 25 in each direction from 50 up to 1000. For the interpolation and plotting we used {\scshape{Root}} \cite{root}.\\
\\
A parameter scan with incoming neutralinos nearly at rest would also have been interesting. However, {\scshape{MadGraph}} failed to calculate the cross section for a process with initial particles nearly at rest. But in fact, in the threshold regions, instead of looking at the cross sections one should consider the mean value of the annihilation cross section times the relative velocity, which is also the quantity to use in order to calculate the neutralino relic density.\\
\\
In Fig. \ref{tanb20A0} the total neutralino annihilation cross section at tree-level is shown for $\tan\beta=20$, $A_0=0$ and $\mu>0$ and center-of-mass energies 500 GeV (left) and 1000 GeV (right). In addition, contour lines for neutralino masses of $m_\chi=$ 100 GeV, 200 GeV, 300 GeV and 400 GeV as well as for Higgs masses $m_A\approx m_H =$ 500 GeV (left) and 1000 GeV (right) are included. The masses of the heavy neutral Higgs $H^0$ and of the pseudoscalar Higgs $A^0$ are almost degenerate nearly in the whole mSUGRA parameter space. Since the neutralino mass is not very sensitive to $\tan \beta$ and $A_0$, these contour lines are omitted in the following diagrams.\\
For very small values of $m_{1/2}$, i.e. for light neutralino masses $m_\chi$, the cross section is of order $\mathcal{O}(1\,\mathrm{pb})$ for $\sqrt{s} = 1000\,\mathrm{GeV}$ and of order $\mathcal{O}(10\,\mathrm{pb})$ for $\sqrt{s} = 500\,\mathrm{GeV}$. In this region the process $\chi\chi \rightarrow W^+W^-$ is dominant, but there are also considerable contributions from $\chi\chi\rightarrow Z^0Z^0$. If $m_0$ and $m_{1/2}$ are small annihilation processes with Higgs particles in the final state cannot be neglected, but with increase of $m_0$ and $m_{1/2}$ these contributions decrease rapidly. If we approach a parameter region where the center-of-mass energy matches the mass of the Higgs particles $A^0$ and $H^0$, the cross section increases significantly due to the resonance of these particles in the s-channel, which we discussed earlier in section \ref{sec_treeenergy} for the parameter point SPS1a. Here, the process $\chi\chi \rightarrow b\bar b$ is dominant ($\approx 75 - 80 \%$), but we get also a large contribution from $\chi\chi \rightarrow \tau^+\tau^-$ ($\approx 16- 18 \%$) and considerable contributions from $\chi\chi \rightarrow t\bar t$ ($\approx 2 \%$).\\
\\
Next we look at the parameter scan for $\tan \beta = 20$, $A_0 = 1000$ GeV and $\mu > 0$ for both energies 500 GeV and 1000 GeV which are shown in Fig. \ref{tanb20A1000}. Here, no specific increase of the cross section due to the process $\chi\chi\rightarrow W^+W^-$ appears for very low values of $m_{1/2}$. In fact, this process does not contribute significantly over all the scan region. 
In the resonance-region, the main contribution comes again from $\chi\chi \rightarrow b\bar b$ ($\approx 80 - 82 \%$), $\chi\chi \rightarrow \tau^+\tau^-$ ($\approx 15 - 17 \%$) and $\chi\chi \rightarrow t \bar t$ ($\approx 2 \%$). For small values of $m_0$ and $m_{1/2}$, {\scshape{SoftSUSY}} failed to generate a new parameter point due to theoretical inconsistencies. Thus, we denote this area as \textit{unphysical region}.\\
\\
In Fig. \ref{tanb20A-1000} the total neutralino annihilation cross section at tree-level is shown for $\tan\beta=20$, $A_0=-1000$ GeV and $\mu>0$ and center-of-mass energies 500 GeV (left) and 1000 GeV (right). Apart from the resonance, the main contributions come from processes with the leptonic final states $e^+e^-$, $\mu^+\mu^-$ and $\tau^+\tau-$ and from the process $\chi\chi \rightarrow t \bar t$. At the resonance we find the same behavior as before with a dominating $b\bar b$ production ($\approx 79 -81\%$) and contributions from the channels $\chi\chi \rightarrow \tau^+\tau^-$ ($\approx 16 - 17 \%$) and $\chi\chi \rightarrow t \bar t$ ($\approx 2 - 2.5\%$). In the case $\sqrt{s} =$ 500 GeV the resonance region disappears into the \textit{unphysical region}, i.e. the masses of $A^0$ and $H^0$ are larger than 500 GeV for all plotted points.\\
\\
Comparing the parameter scans in Fig. \ref{tanb20A0}, \ref{tanb20A1000} and \ref{tanb20A-1000} for the three values $A_0 =$ 0, 1000, -1000 GeV, respectively, we may conclude that the neutralino annihilation cross section at tree-level is not very sensitive to the parameter $A_0$ except for the slight shifting of the masses of $A^0$ and $H^0$  and therewith of the resonance regions.\\
\\
Next, we look at the parameter plots for $\tan\beta=50$, $\mu>0$ and $A_0= 0, 1000, -1000$ GeV in Fig. \ref{tanb50A0}, \ref{tanb50A1000} and \ref{tanb50A-1000}, respectively, for center-of-mass energies of 500 GeV (left) and 1 TeV (right). The masses of $A_0$ and $H_0$ decrease with $\tan\beta$, so that the resonance region shifts to higher values of $m_0$ and $m_{1/2}$. Therefore, the resonance region disappears from the scanned region for the plots with $\sqrt{s}=$ 1 TeV.\\
\\
For fixed parameters $\tan\beta = 50$, $A_0 = 0$ and $\mu>0$ (Fig. \ref{tanb50A0}) the process $\chi\chi \rightarrow W^+W^-$ contributes significantly for $m_{1/2}$ smaller than 100 GeV as seen before for $\tan\beta = 20$, $A_0 = 0$ and $\mu>0$ (Fig. \ref{tanb20A0}). But in addition, there is also a large contribution from $\chi\chi\rightarrow b\bar b$ in this region which we did not see before in Fig. \ref{tanb20A0}.
The resonance region for $\sqrt{s}=$ 500 GeV is again dominated by $\chi\chi \rightarrow b \bar b$ ($\approx 79 - 82 \%$) and $\chi\chi \rightarrow \tau^+\tau^-$ ($\approx 18 - 20 \%$).\\
\\
The total tree-level neutralino annihilation cross section for $\tan\beta = 50$, $A_0 = 1000$ GeV and $\mu>0$ is shown in Fig. \ref{tanb50A1000}. At the resonance at center-of-mass energy $\sqrt{s}=$ 500 GeV the main contributions come from $\chi \chi \rightarrow b \bar b$ ($\approx 80 \%$) and $\chi \chi \rightarrow \tau^+ \tau^-$ ($\approx 20 \%$). Apart from the resonance, the most important channels are those with final states $e^+e^-$, $\mu^+\mu^-$, $\tau^+\tau^-$, $b\bar b$ and $t \bar t$, the latter especially in the case of $\sqrt{s}=$ 1 TeV.\\
\\
In Fig. \ref{tanb50A-1000} the parameter scans for $\tan\beta = 50$, $A_0 = -1000$ GeV and $\mu>0$ for center-of-mass energies 500 GeV and 1 TeV are shown. At the resonance, contributions come from $\chi \chi \rightarrow b \bar b$ ($\approx 77-80 \%$) and $\chi \chi \rightarrow \tau^+ \tau^-$ ($\approx 20-22 \%$). In the remaining region, main contributions come from channels with final states $e^+e^-$, $\mu^+\mu^-$, $\tau^+\tau^-$, $b\bar b$ and $t \bar t$.\\
\\
Comparing the magnitute of the cross-section in Fig. \ref{tanb50A0}, \ref{tanb50A1000} and \ref{tanb50A-1000} in the resonance region, we see that the cross section increases slightly with $A_0$.
\subsubsection{Conclusions of parameter dependence study}
We have seen that the resonances due to the s-channel exchange of $A^0$ and $H^0$ in the channels with massive fermionic final states lead to a significant increase of the total neutralino annihilation cross-section. In these resonance regions, the process $\chi\chi \rightarrow b \bar b$ dominates ($\approx 75 - 82 \%$). Other important contributions come from $\chi\chi \rightarrow \tau^+ \tau^-$ ($\approx 15 -22 \%$) and $\chi\chi \rightarrow t \bar t$ ($\approx 2 \%$). Since the masses of the Higgs bosons increase with $m_0$ and $m_{12}$, channels with Higgs bosons in the final state only play a role for very low values of $m_0$ and $m_{1/2}$. Apart from the resonances, the most important channels are those with final states $e^+e^-$, $\mu^+\mu^-$, $\tau^+\tau^-$, $b\bar b$ and $t \bar t$.\\
\\
The neutralino annhilation cross section seems not to be very sensitive to the parameter $A_0$. However, for $A_0 = 0$, there appears a region of increased cross section for very small $m_{1/2}$ due to the process $\chi\chi \rightarrow W^+W^-$. Since the masses of $A^0$ and $H^0$ depend on $A_0$ there is a slight shifting of the resonance regions.\\
\\
The dependence on $\tan\beta$ seems to be more delicate, since the masses of the Higgs bosons $A^0$ and $H^0$ are more sensitive to this parameter. Thus, for a higher $\tan \beta$, the resonance region appears for a given center-of-mass energy at higher values of $m_{0}$  and $m_{1/2}$.\\
\\
We want to note that, for cosmology, those parameter regions where $2 m_\chi \approx m_{A}$ are of greater importance. Large regions of the mSUGRA parameter space predict much larger values for the neutralino relic density than the values given by WMAP. To produce the observed dark matter relic density an enhanced neutralino annihilation is needed. One possibility would be an accidental degeneracy $2 m_\chi \approx m_{A}$. Assuming CP-invariance, the s-channel exchange of the $A^0$ boson would still be open at threshold, while the exchange of the CP-even $H^0$ would be suppressed.
\section{Summary and outlook}\label{summary}
We studied the cross section of all tree-level neutralino annihilation processes concerning energy dependence and channel dominance for the mSUGRA benchmark point SPS1a. In processes which contain contributions from s-channel exchange of a neutral heavy Higgs $H^0$ and/or a pseudoscalar Higgs boson $A^0$, these diagrams lead to resonances when the exchanged Higgs bosons are produced on-shell. Since the Higgs bosons couple to the mass, those channels with third generation fermions in the final state, particularly the process $\chi\chi \rightarrow b\bar b$, dominate the total cross section in this resonance region. Apart from the resonance region, the neutralino annihilation cross section receives the largest contributions from processes with $e^+e^-$, $\mu^+\mu^-$ and $\tau^+\tau^-$ final states.\\
\\
We studied the mSUGRA parameter dependence of the neutralino annihilation cross section by scanning the $m_{1/2}$-$m_{0}$ plane for fixed values $\tan \beta = 20,\, 50$, $A_0 = -1000,\,0,\,1000$ GeV and $\mu>0$ for two different center-of-mass energies $\sqrt{s}=$ 500 GeV, 1000 GeV. The parameter scans have shown that the dominance of the process $\chi\chi \rightarrow b\bar b$ in the resonance region is not sensitive to the mSUGRA parameters. The behavior apart from the resonance, i.e. that the main contributions come from channels $\chi\chi\rightarrow e^+e^-,\mu^+\mu^-, \tau^+\tau^-$, does not vary significantly with the mSUGRA parameters either. However, there appear some regions in the parameter space where the processes $\chi\chi \rightarrow W^+W^-$ and $\chi\chi \rightarrow t\bar t$ provide considerable contributions. For $A_0=0$, there is even a small region at low values of $m_{1/2}$, where the process  $\chi\chi \rightarrow W^+W^-$ is dominating. Since the masses of the Higgs bosons $A^0$ and $H^0$ depend on the mSUGRA parameter, the resonance region shifts in the $m_0$-$m_{1/2}$-plane for a given center-of-mass energy.\\
\\
After this study on neutralino annihilation processes at tree-level we started to investigate higher-order corrections by calculating the cross section up to one-loop-level with {\scshape{FormCalc}} \cite{FormCalc}. We also studied neutralino annihilation to the two photon final state where the leading order happens to be at loop-level. The tree-level results, obtained by {\scshape{FormCalc}}, are in good agreement with those derived with {\scshape{CompHEP}}. At loop-level, the overall form of the cross section plots seems reasonable. However, since counter-terms have not yet been included for the MSSM in {\scshape{FormCalc}}, we are not totally certain about our results and thus decided not to present them here.
\appendix
\section*{Appendix A\\ Detailed view on the Feynman diagrams contributing to $\chi\chi\rightarrow Z^0Z^0$}\label{sec_treefeyn}\label{appendix}
\begin{figure}[h!]
{\begin{center}
\unitlength=1.0 pt
\SetScale{1.0}
\SetWidth{0.7}      % line    size control
\tiny    %  letter  size control
{} \qquad\allowbreak
%  diagram # 1
\begin{picture}(79,40)(0,30)
\Text(13.0,57.0)[r]{$\chi^0_1$}
\Line(14.0,57.0)(48.0,57.0)
\Text(66.0,57.0)[l]{$Z$}
\DashLine(48.0,57.0)(65.0,57.0){3.0}
\Text(47.0,49.0)[r]{$\chi^0_{n}$}
\Line(48.0,57.0)(48.0,41.0)
\Text(13.0,41.0)[r]{$\chi^0_1$}
\Line(14.0,41.0)(48.0,41.0)
\Text(66.0,41.0)[l]{$Z$}
\DashLine(48.0,41.0)(65.0,41.0){3.0}
\end{picture} \
{} \qquad\allowbreak
%  diagram # 2
\begin{picture}(79,40)(0,30)
\Text(13.0,57.0)[r]{$\chi^0_1$}
\Line(14.0,57.0)(31.0,49.0)
\Text(13.0,41.0)[r]{$\chi^0_1$}
\Line(14.0,41.0)(31.0,49.0)
\Text(39.0,50.0)[b]{$h$}
\DashLine(31.0,49.0)(48.0,49.0){1.0}
\Text(66.0,57.0)[l]{$Z$}
\DashLine(48.0,49.0)(65.0,57.0){3.0}
\Text(66.0,41.0)[l]{$Z$}
\DashLine(48.0,49.0)(65.0,41.0){3.0}
\end{picture} \
{} \qquad\allowbreak
%  diagram # 3
\begin{picture}(79,40)(0,30)
\Text(13.0,57.0)[r]{$\chi^0_1$}
\Line(14.0,57.0)(31.0,49.0)
\Text(13.0,41.0)[r]{$\chi^0_1$}
\Line(14.0,41.0)(31.0,49.0)
\Text(39.0,50.0)[b]{$H$}
\DashLine(31.0,49.0)(48.0,49.0){1.0}
\Text(66.0,57.0)[l]{$Z$}
\DashLine(48.0,49.0)(65.0,57.0){3.0}
\Text(66.0,41.0)[l]{$Z$}
\DashLine(48.0,49.0)(65.0,41.0){3.0}
\end{picture}
\end{center}}
\caption{Six Feynman diagrams are contributing to the process $\chi\chi \rightarrow Z^0Z^0$: exchange of all 4 neutralinos in the t-channel (n=1,2,3,4) and s-channel exchange of the neutral Higgs bosons $h^0$ and $H^0$.}\label{zzdiagrams2}
\end{figure}
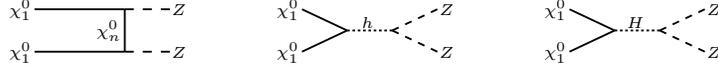

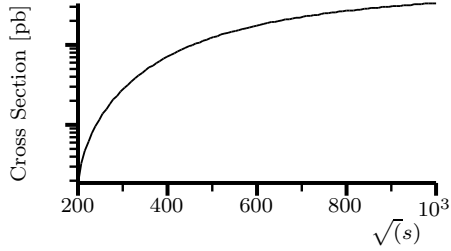
\begin{figure}[h!]
\centering
\scalebox{1}
{
\unitlength=1.0 pt
\SetScale{1.0}
\SetWidth{0.7}      % line    size control
\scriptsize    %  letter  size control
\begin{picture}(180,120)(0,0)
%\Text(104.9,119.7)[t]{$\chi\chi \rightarrow $}
% ====================   X-axis =============
\LinAxis(37.55,33.18)(172.49,33.18)(4.000,2,-4,0.000,1.5)
\Text(37.6,27.5)[t]{$200$}
\Text(71.3,27.5)[t]{$400$}
\Text(105.1,27.5)[t]{$600$}
\Text(138.7,27.5)[t]{$800$}
\Text(172.5,27.5)[t]{$10^3$}
\Text(167.5,19.8)[rt]{$\sqrt(s)$}
% ====================   Y-axis =============
\LogAxis(37.55,33.18)(37.55,100.89)(2.246,4,1.874,1.5)
%\Text(31.8,55.0)[r]{$10^-3$}
%\Text(31.8,85.1)[r]{$0.01$}
\rText(19.8,100.9)[tr][l]{Cross Section [pb]}
% ============== end of axis ============
\Line(38.8,40.6)(37.6,33.2)
\Line(40.1,45.3)(38.8,40.6)
\Line(41.6,48.9)(40.1,45.3)
\Line(42.8,52.0)(41.6,48.9)
\Line(44.1,54.6)(42.8,52.0)
\Line(45.6,57.0)(44.1,54.6)
\Line(46.8,59.3)(45.6,57.0)
\Line(48.3,61.0)(46.8,59.3)
\Line(49.6,63.0)(48.3,61.0)
\Line(50.8,64.7)(49.6,63.0)
\Line(52.3,66.0)(50.8,64.7)
\Line(53.6,67.7)(52.3,66.0)
\Line(55.1,69.1)(53.6,67.7)
\Line(56.3,70.1)(55.1,69.1)
\Line(57.6,71.4)(56.3,70.1)
\Line(59.1,72.7)(57.6,71.4)
\Line(60.3,73.7)(59.1,72.7)
\Line(61.8,74.7)(60.3,73.7)
\Line(63.1,75.8)(61.8,74.7)
\Line(64.3,76.8)(63.1,75.8)
\Line(65.8,77.4)(64.3,76.8)
\Line(67.1,78.4)(65.8,77.4)
\Line(68.3,79.1)(67.1,78.4)
\Line(69.8,80.1)(68.3,79.1)
\Line(71.1,80.8)(69.8,80.1)
\Line(72.6,81.5)(71.1,80.8)
\Line(73.9,82.1)(72.6,81.5)
\Line(75.1,82.8)(73.9,82.1)
\Line(76.6,83.5)(75.1,82.8)
\Line(77.9,84.1)(76.6,83.5)
\Line(79.4,84.8)(77.9,84.1)
\Line(80.6,85.1)(79.4,84.8)
\Line(81.9,85.8)(80.6,85.1)
\Line(83.4,86.1)(81.9,85.8)
\Line(84.6,86.8)(83.4,86.1)
\Line(86.1,87.2)(84.6,86.8)
\Line(87.4,87.8)(86.1,87.2)
\Line(88.6,88.2)(87.4,87.8)
\Line(90.1,88.5)(88.6,88.2)
\Line(91.4,89.2)(90.1,88.5)
\Line(92.6,89.5)(91.4,89.2)
\Line(94.1,89.8)(92.6,89.5)
\Line(95.4,90.2)(94.1,89.8)
\Line(96.9,90.5)(95.4,90.2)
\Line(98.1,90.8)(96.9,90.5)
\Line(99.4,91.2)(98.1,90.8)
\Line(100.9,91.5)(99.4,91.2)
\Line(102.1,91.8)(100.9,91.5)
\Line(103.6,92.2)(102.1,91.8)
\Line(104.9,92.5)(103.6,92.2)
\Line(106.1,92.8)(104.9,92.5)
\Line(107.6,93.2)(106.1,92.8)
\Line(108.9,93.5)(107.6,93.2)
\Line(110.4,93.9)(108.9,93.5)
\Line(111.7,93.9)(110.4,93.9)
\Line(112.9,94.2)(111.7,93.9)
\Line(114.4,94.5)(112.9,94.2)
\Line(115.7,94.9)(114.4,94.5)
\Line(117.2,94.9)(115.7,94.9)
\Line(118.4,95.2)(117.2,94.9)
\Line(119.7,95.5)(118.4,95.2)
\Line(121.2,95.5)(119.7,95.5)
\Line(122.4,95.9)(121.2,95.5)
\Line(123.7,96.2)(122.4,95.9)
\Line(125.2,96.2)(123.7,96.2)
\Line(126.4,96.5)(125.2,96.2)
\Line(127.9,96.5)(126.4,96.5)
\Line(129.2,96.9)(127.9,96.5) 
\Line(130.4,97.2)(129.2,96.9)
\Line(131.9,97.2)(130.4,97.2) 
\Line(133.2,97.5)(131.9,97.2) 
\Line(134.7,97.5)(133.2,97.5)
\Line(135.9,97.9)(134.7,97.5) 
\Line(137.2,97.9)(135.9,97.9) 
\Line(138.7,98.2)(137.2,97.9) 
\Line(139.9,98.2)(138.7,98.2) 
\Line(141.4,98.2)(139.9,98.2) 
\Line(142.7,98.5)(141.4,98.2)
\Line(143.9,98.5)(142.7,98.5) 
\Line(145.5,98.9)(143.9,98.5)
\Line(146.7,98.9)(145.5,98.9) 
\Line(148.0,99.2)(146.7,98.9)
\Line(149.5,99.2)(148.0,99.2)
\Line(150.7,99.2)(149.5,99.2) 
\Line(152.2,99.6)(150.7,99.2) 
\Line(153.5,99.6)(152.2,99.6) 
\Line(154.7,99.6)(153.5,99.6)
\Line(156.2,99.9)(154.7,99.6)
\Line(157.5,99.9)(156.2,99.9)
\Line(159.0,99.9)(157.5,99.9) 
\Line(160.2,100.2)(159.0,99.9)
\Line(161.5,100.2)(160.2,100.2)
\Line(163.0,100.2)(161.5,100.2)
\Line(164.2,100.6)(163.0,100.2)
\Line(165.7,100.6)(164.2,100.6)
\Line(167.0,100.6)(165.7,100.6)
\Line(168.2,100.9)(167.0,100.6) 
\Line(169.7,100.9)(168.2,100.9)
\Line(171.0,100.9)(169.7,100.9)
\Line(172.5,100.9)(171.0,100.9)
\end{picture}}
\caption{Contribution of single t-channel diagram $\chi\chi\rightarrow (\chi_n^0) \rightarrow Z^0Z^0$. The slope is for all four neutralinos similar while the magnitude differs significantly (see text).}\label{singlet}
\end{figure}

In this appendix, we study the contributions and interferences of the six different Feynman diagrams for the process $\chi\chi \rightarrow Z^0Z^0$ given in Fig. \ref{zzdiagrams2} at the mSUGRA point SPS1a, as an example. 
For this, we use the possibility to exclude specific diagrams in {\scshape{CompHEP}} in order to extract the contributions of those diagrams we are interested in. Then, we plot the cross section energy dependence in the region from 200 GeV up to 1 TeV. In general, the cross section is proportional to the scattering amplitude squared, i.e. the sum of all diagrams squared:
\begin{eqnarray}
\sigma \propto {\left|\mathcal{M}\right|^2} = {\left|\sum{diagrams}\right|^2}
\end{eqnarray}
Hence, we have all in all 21 terms contributing, where 6 are squared diagrams and 15 are interference terms. In the following, we show the most important contributions.\\
\\
In Fig. \ref{xseczz2} we show again the whole cross-section of the process $\chi\chi \rightarrow Z^0Z^0$ with all diagrams contributing. As mentioned before, there is a resonance at 394.5 GeV due to the s-channel exchange of $H^0$. In addition, there happens to be a destructive interference, we want to find out about.
\begin{figure}[h!]
\centering
\scalebox{0.7}{
\unitlength=1.0 pt
\SetScale{1.0}
\SetWidth{0.7}      % line    size control
\normalsize    %  letter  size control
\begin{picture}(300,200)(0,0)
%\Text(168.2,199.4)[t]{$~o1,~o1 ->Z,Z$}
% ====================   X-axis =============
\LinAxis(45.90,36.72)(290.82,36.72)(4.000,2,-4,0.000,1.5)
\Text(45.9,29.9)[t]{$200$}
\Text(107.2,29.9)[t]{$400$}
\Text(168.6,29.9)[t]{$600$}
\Text(229.5,29.9)[t]{$800$}
\Text(290.8,29.9)[t]{$10^3$}
\Text(280.8,20.3)[rt]{$\sqrt(s)$}
% ====================   Y-axis =============
\LogAxis(45.90,36.72)(45.90,176.27)(1.873,4,1.123,1.5)
\Text(39.2,107.3)[r]{$10^-2$}
\rText(7.9,176.3)[tr][l]{Cross Section [pb]}
% ============== end of axis ============
\Line(48.0,82.5)(45.9,70.1)
\Line(50.5,88.7)(48.0,82.5)
\Line(53.0,92.7)(50.5,88.7) 
\Line(55.5,94.9)(53.0,92.7)
\Line(58.0,97.2)(55.5,94.9)
\Line(60.5,98.9)(58.0,97.2)
\Line(63.0,100.0)(60.5,98.9) 
\Line(65.1,101.1)(63.0,100.0)
\Line(67.6,101.7)(65.1,101.1)
\Line(70.1,102.3)(67.6,101.7)
\Line(72.6,102.8)(70.1,102.3)
\Line(75.1,103.4)(72.6,102.8)
\Line(77.6,103.4)(75.1,103.4) 
\Line(80.1,104.0)(77.6,103.4)
\Line(82.6,104.5)(80.1,104.0)
\Line(84.7,105.1)(82.6,104.5) 
\Line(87.2,105.1)(84.7,105.1)
\Line(89.7,105.6)(87.2,105.1)
\Line(92.2,106.8)(89.7,105.6) 
\Line(94.7,107.9)(92.2,106.8)
\Line(97.2,110.2)(94.7,107.9)
\Line(99.7,114.7)(97.2,110.2)
\Line(102.2,124.9)(99.7,114.7)
\Line(104.3,176.8)(102.2,124.9) 
\Line(104.3,176.8)(106.8,36.7)
\Line(109.3,70.1)(106.8,36.7)
\Line(111.8,80.8)(109.3,70.1)
\Line(114.3,84.7)(111.8,80.8)
\Line(116.8,87.0)(114.3,84.7)
\Line(119.3,88.1)(116.8,87.0) 
\Line(121.4,88.7)(119.3,88.1)
\Line(123.9,89.3)(121.4,88.7)
\Line(126.4,89.8)(123.9,89.3) 
\Line(128.9,89.8)(126.4,89.8)
\Line(128.9,89.8)(131.4,89.8)
\Line(131.4,89.8)(133.9,89.3)
\Line(133.9,89.3)(136.4,89.3) 
\Line(136.4,89.3)(138.9,89.3)
\Line(138.9,89.3)(141.0,88.7) 
\Line(141.0,88.7)(143.5,88.7)
\Line(143.5,88.7)(146.0,88.1) 
\Line(146.0,88.1)(148.5,88.1)
\Line(148.5,88.1)(151.0,87.6)
\Line(151.0,87.6)(153.5,87.6) 
\Line(153.5,87.6)(156.1,87.0)
\Line(156.1,87.0)(158.6,87.0)
\Line(158.6,87.0)(160.6,86.4) 
\Line(160.6,86.4)(163.1,85.9)
\Line(163.1,85.9)(165.6,85.9)
\Line(165.6,85.9)(168.2,85.3)
\Line(168.2,85.3)(170.7,85.3) 
\Line(170.7,85.3)(173.2,84.7)
\Line(173.2,84.7)(175.7,84.7) 
\Line(175.7,84.7)(177.7,84.2)
\Line(177.7,84.2)(180.3,83.6) 
\Line(180.3,83.6)(182.8,83.6)
\Line(182.8,83.6)(185.3,83.1) 
\Line(185.3,83.1)(187.8,83.1)
\Line(187.8,83.1)(190.3,82.5) 
\Line(190.3,82.5)(192.8,82.5)
\Line(192.8,82.5)(195.3,81.9)
\Line(195.3,81.9)(197.4,81.9)
\Line(197.4,81.9)(199.9,81.4) 
\Line(199.9,81.4)(202.4,80.8)
\Line(202.4,80.8)(204.9,80.8) 
\Line(204.9,80.8)(207.4,80.2)
\Line(207.4,80.2)(209.9,80.2)
\Line(209.9,80.2)(212.4,79.7)
\Line(212.4,79.7)(214.9,79.7) 
\Line(214.9,79.7)(217.0,79.1)
\Line(217.0,79.1)(219.5,79.1)
\Line(219.5,79.1)(222.0,78.5)
\Line(222.0,78.5)(224.5,78.5) 
\Line(224.5,78.5)(227.0,78.5)
\Line(227.0,78.5)(229.5,78.0) 
\Line(229.5,78.0)(232.0,78.0)
\Line(232.0,78.0)(234.1,77.4)
\Line(234.1,77.4)(236.6,77.4)
\Line(236.6,77.4)(239.1,76.8) 
\Line(239.1,76.8)(241.6,76.8)
\Line(241.6,76.8)(244.1,76.3)
\Line(244.1,76.3)(246.6,76.3)
\Line(246.6,76.3)(249.1,76.3)
\Line(249.1,76.3)(251.6,75.7)
\Line(251.6,75.7)(253.7,75.7)
\Line(253.7,75.7)(256.2,75.1)
\Line(256.2,75.1)(258.7,75.1)
\Line(258.7,75.1)(261.2,75.1)
\Line(261.2,75.1)(263.7,74.6)
\Line(263.7,74.6)(266.2,74.6)
\Line(266.2,74.6)(268.7,74.0)
\Line(268.7,74.0)(271.2,74.0)
\Line(271.2,74.0)(273.3,74.0)
\Line(273.3,74.0)(275.8,73.4)
\Line(275.8,73.4)(278.3,73.4)
\Line(278.3,73.4)(280.8,73.4)
\Line(280.8,73.4)(283.3,72.9)
\Line(283.3,72.9)(285.8,72.9)
\Line(285.8,72.9)(288.3,72.9)
\Line(288.3,72.9)(290.8,72.3)
\end{picture}
}
\caption{Cross section energy dependence of the process $\chi\chi \rightarrow Z^0Z^0$ with all diagrams contributing.}\label{xseczz2}
\end{figure}
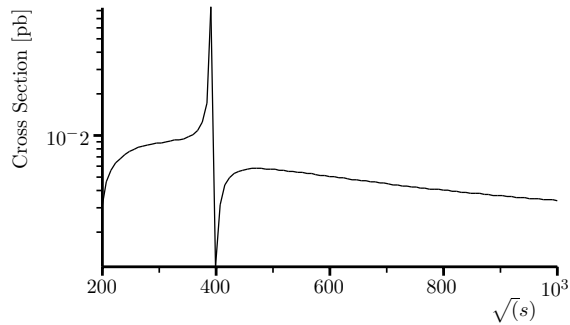\\
We start by looking at the six squared single diagram contributions. In Fig. \ref{singlet} the cross section energy dependence of one single t-channel diagram is shown. Since the slope is for all four t-channel diagrams similar, we omit here to show all four plots. But in contrary, the magnitude of these diagrams differs significantly. An overview of the magnitudes is given in Table \ref{tchannelmagnitudes}. Note, that these values are only given in order to get a feeling for the order of magnitudes. However, we see that the main t-channel diagram contributing is via the exchange of $\chi_3^0$.
\begin{table}[h!]
\centering
\begin{tabular}{|l|c|c|c|c|}
\hline
 			& \multicolumn{4}{c|}{exchanged neutralino}\\
\hline
Energy		&	$\chi_1^0$	&	$\chi_2^0$	&	$\chi_3^0$	&	$\chi_4^0$\\
\hline
200			&	3E-06		&	3E-06		&	2E-04		&	1E-05\\
300			&	7E-06		&	4E-05		&	2E-03		&	1E-04\\
1000			&	1E-05		&	2E-04		&	2E-02		&	3E-03\\
\hline
\end{tabular}
\caption{Cross section in pb for the four single t-channel diagram contributions with neutralino exchange for three different energy values $\sqrt{s}=$ 200 GeV, 300 GeV, 1 TeV.}\label{tchannelmagnitudes}
\end{table}\\
The cross section contribution of the single s-channel diagrams with exchanged $h^0$ and $H^0$ are shown in Fig. \ref{singleh0} and \ref{singleH0}, respectively. While the cross section contribution of the diagram with $h^0$-exchange increases monotonically, the $H^0$-exchange diagram contribution has a broad resonance and decreases for higher energies. 
\begin{figure}[h!]
\centering
\begin{minipage}{7.5cm}
{
\unitlength=1.0 pt
\SetScale{1.0}
\SetWidth{0.7}      % line    size control
\scriptsize    %  letter  size control
\begin{picture}(180,120)(0,0)
%\Text(104.9,119.7)[t]{$~o1,~o1 ->Z,Z$}
% ====================   X-axis =============
\LinAxis(37.55,33.18)(172.49,33.18)(4.000,2,-4,0.000,1.5)
\Text(37.6,27.5)[t]{$200$}
\Text(71.3,27.5)[t]{$400$}
\Text(105.1,27.5)[t]{$600$}
\Text(138.7,27.5)[t]{$800$}
\Text(172.5,27.5)[t]{$10^3$}
\Text(167.5,19.8)[rt]{$\sqrt(s)$}
% ====================   Y-axis =============
\LinAxis(37.55,33.18)(37.55,100.89)(3.598,5,4,-1.194,1.5)
\Text(31.8,37.5)[r]{$0.005$}
\Text(31.8,56.6)[r]{$10^-2$}
\Text(31.8,75.4)[r]{$0.015$}
\Text(31.8,94.2)[r]{$0.020$}
\rText(6.8,100.9)[tr][l]{Cross Section [pb]}
% ============== end of axis ============
\Line(38.8,40.6)(37.6,33.2)
\Line(40.1,45.6)(38.8,40.6)
\Line(41.6,49.6)(40.1,45.6)
\Line(42.8,53.3)(41.6,49.6)
\Line(44.1,56.3)(42.8,53.3)
\Line(45.6,58.7)(44.1,56.3)
\Line(46.8,61.0)(45.6,58.7)
\Line(48.3,63.4)(46.8,61.0)
\Line(49.6,65.4)(48.3,63.4)
\Line(50.8,67.0)(49.6,65.4)
\Line(52.3,68.7)(50.8,67.0)
\Line(53.6,70.4)(52.3,68.7)
\Line(55.1,71.7)(53.6,70.4)
\Line(56.3,73.1)(55.1,71.7)
\Line(57.6,74.4)(56.3,73.1)
\Line(59.1,75.8)(57.6,74.4)
\Line(60.3,77.1)(59.1,75.8)
\Line(61.8,78.1)(60.3,77.1)
\Line(63.1,79.1)(61.8,78.1)
\Line(64.3,80.1)(63.1,79.1)
\Line(65.8,81.1)(64.3,80.1)
\Line(67.1,81.8)(65.8,81.1)
\Line(68.3,82.8)(67.1,81.8)
\Line(69.8,83.5)(68.3,82.8)
\Line(71.1,84.1)(69.8,83.5)
\Line(72.6,84.8)(71.1,84.1)
\Line(73.9,85.5)(72.6,84.8)
\Line(75.1,86.1)(73.9,85.5)
\Line(76.6,86.8)(75.1,86.1)
\Line(77.9,87.5)(76.6,86.8)
\Line(79.4,87.8)(77.9,87.5)
\Line(80.6,88.5)(79.4,87.8)
\Line(81.9,89.2)(80.6,88.5)
\Line(83.4,89.5)(81.9,89.2)
\Line(84.6,89.8)(83.4,89.5)
\Line(86.1,90.5)(84.6,89.8)
\Line(87.4,90.8)(86.1,90.5)
\Line(88.6,91.2)(87.4,90.8)
\Line(90.1,91.5)(88.6,91.2)
\Line(91.4,91.8)(90.1,91.5)
\Line(92.6,92.5)(91.4,91.8)
\Line(94.1,92.8)(92.6,92.5)
\Line(95.4,93.2)(94.1,92.8)
\Line(96.9,93.2)(95.4,93.2)
\Line(98.1,93.5)(96.9,93.2)
\Line(99.4,93.9)(98.1,93.5)
\Line(100.9,94.2)(99.4,93.9)
\Line(102.1,94.5)(100.9,94.2)
\Line(103.6,94.9)(102.1,94.5)
\Line(104.9,94.9)(103.6,94.9)
\Line(106.1,95.2)(104.9,94.9)
\Line(107.6,95.5)(106.1,95.2)
\Line(108.9,95.5)(107.6,95.5)
\Line(110.4,95.9)(108.9,95.5)
\Line(111.7,96.2)(110.4,95.9)
\Line(112.9,96.2)(111.7,96.2)
\Line(114.4,96.5)(112.9,96.2)
\Line(115.7,96.5)(114.4,96.5)
\Line(117.2,96.9)(115.7,96.5)
\Line(118.4,97.2)(117.2,96.9)
\Line(119.7,97.2)(118.4,97.2)
\Line(121.2,97.5)(119.7,97.2)
\Line(122.4,97.5)(121.2,97.5)
\Line(123.7,97.5)(122.4,97.5)
\Line(125.2,97.9)(123.7,97.5)
\Line(126.4,97.9)(125.2,97.9)
\Line(127.9,98.2)(126.4,97.9)
\Line(129.2,98.2)(127.9,98.2) 
\Line(130.4,98.5)(129.2,98.2) 
\Line(131.9,98.5)(130.4,98.5)
\Line(133.2,98.5)(131.9,98.5) 
\Line(134.7,98.9)(133.2,98.5) 
\Line(135.9,98.9)(134.7,98.9) 
\Line(137.2,98.9)(135.9,98.9) 
\Line(138.7,99.2)(137.2,98.9) 
\Line(139.9,99.2)(138.7,99.2) 
\Line(141.4,99.2)(139.9,99.2) 
\Line(142.7,99.6)(141.4,99.2) 
\Line(143.9,99.6)(142.7,99.6) 
\Line(145.5,99.6)(143.9,99.6)
\Line(146.7,99.6)(145.5,99.6) 
\Line(148.0,99.9)(146.7,99.6) 
\Line(149.5,99.9)(148.0,99.9) 
\Line(150.7,99.9)(149.5,99.9)
\Line(152.2,99.9)(150.7,99.9) 
\Line(153.5,100.2)(152.2,99.9) 
\Line(154.7,100.2)(153.5,100.2) 
\Line(156.2,100.2)(154.7,100.2) 
\Line(157.5,100.2)(156.2,100.2)
\Line(159.0,100.6)(157.5,100.2) 
\Line(160.2,100.6)(159.0,100.6)
\Line(161.5,100.6)(160.2,100.6)
\Line(163.0,100.6)(161.5,100.6) 
\Line(164.2,100.6)(163.0,100.6)
\Line(165.7,100.9)(164.2,100.6)
\Line(167.0,100.9)(165.7,100.9)
\Line(168.2,100.9)(167.0,100.9)
\Line(169.7,100.9)(168.2,100.9)
\Line(171.0,100.9)(169.7,100.9)
\Line(172.5,100.9)(171.0,100.9)
\end{picture}
}
\caption{Contribution of the s-channel diagram $\chi\chi\rightarrow (h^0) \rightarrow Z^0Z^0$.}\label{singleh0}
\end{minipage}
\hspace{1cm}
\begin{minipage}{7.5cm}
{
\unitlength=1.0 pt
\SetScale{1.0}
\SetWidth{0.7}      % line    size control
\scriptsize    %  letter  size control
\begin{picture}(180,120)(0,0)
%\Text(104.9,119.7)[t]{$~o1,~o1 ->Z,Z$}
% ====================   X-axis =============
\LinAxis(37.55,33.18)(172.49,33.18)(4.000,2,-4,0.000,1.5)
\Text(37.6,27.5)[t]{$200$}
\Text(71.3,27.5)[t]{$400$}
\Text(105.1,27.5)[t]{$600$}
\Text(138.7,27.5)[t]{$800$}
\Text(172.5,27.5)[t]{$10^3$}
\Text(167.5,19.8)[rt]{$\sqrt(s)$}
% ====================   Y-axis =============
\LinAxis(37.55,33.18)(37.55,100.89)(2.608,2,4,-0.911,1.5)
\Text(31.8,44.9)[r]{$0.004$}
\Text(31.8,71.1)[r]{$0.006$}
\Text(31.8,96.9)[r]{$0.008$}
\rText(9.3,100.9)[tr][l]{Cross Section [pb]}
% ============== end of axis ============
\Line(38.8,51.6)(37.6,33.2) 
\Line(40.1,64.0)(38.8,51.6) 
\Line(41.6,73.1)(40.1,64.0) 
\Line(42.8,79.8)(41.6,73.1) 
\Line(44.1,85.1)(42.8,79.8) 
\Line(45.6,89.2)(44.1,85.1) 
\Line(46.8,92.5)(45.6,89.2) 
\Line(48.3,95.2)(46.8,92.5) 
\Line(49.6,97.2)(48.3,95.2) 
\Line(50.8,98.5)(49.6,97.2) 
\Line(52.3,99.6)(50.8,98.5) 
\Line(53.6,100.6)(52.3,99.6) 
\Line(55.1,100.9)(53.6,100.6) 
\Line(56.3,101.2)(55.1,100.9) 
\Line(56.3,101.2)(57.6,100.9) 
\Line(57.6,100.9)(59.1,100.9) 
\Line(59.1,100.9)(60.3,100.2) 
\Line(60.3,100.2)(61.8,99.9) 
\Line(61.8,99.9)(63.1,99.2) 
\Line(63.1,99.2)(64.3,98.2) 
\Line(64.3,98.2)(65.8,97.5) 
\Line(65.8,97.5)(67.1,96.5) 
\Line(67.1,96.5)(68.3,95.5) 
\Line(68.3,95.5)(69.8,94.5) 
\Line(69.8,94.5)(71.1,93.2) 
\Line(71.1,93.2)(72.6,92.2) 
\Line(72.6,92.2)(73.9,90.8) 
\Line(73.9,90.8)(75.1,89.8) 
\Line(75.1,89.8)(76.6,88.5) 
\Line(76.6,88.5)(77.9,87.2) 
\Line(77.9,87.2)(79.4,86.1) 
\Line(79.4,86.1)(80.6,84.8) 
\Line(80.6,84.8)(81.9,83.5) 
\Line(81.9,83.5)(83.4,82.5) 
\Line(83.4,82.5)(84.6,81.1) 
\Line(84.6,81.1)(86.1,80.1) 
\Line(86.1,80.1)(87.4,78.8) 
\Line(87.4,78.8)(88.6,77.8) 
\Line(88.6,77.8)(90.1,76.8) 
\Line(90.1,76.8)(91.4,75.4) 
\Line(91.4,75.4)(92.6,74.4) 
\Line(92.6,74.4)(94.1,73.4) 
\Line(94.1,73.4)(95.4,72.4) 
\Line(95.4,72.4)(96.9,71.4) 
\Line(96.9,71.4)(98.1,70.4) 
\Line(98.1,70.4)(99.4,69.4) 
\Line(99.4,69.4)(100.9,68.4) 
\Line(100.9,68.4)(102.1,67.4) 
\Line(102.1,67.4)(103.6,66.4) 
\Line(103.6,66.4)(104.9,65.7) 
\Line(104.9,65.7)(106.1,64.7) 
\Line(106.1,64.7)(107.6,64.0) 
\Line(107.6,64.0)(108.9,63.0) 
\Line(108.9,63.0)(110.4,62.3) 
\Line(110.4,62.3)(111.7,61.7) 
\Line(111.7,61.7)(112.9,60.7) 
\Line(112.9,60.7)(114.4,60.0) 
\Line(114.4,60.0)(115.7,59.3) 
\Line(115.7,59.3)(117.2,58.7) 
\Line(117.2,58.7)(118.4,58.0) 
\Line(118.4,58.0)(119.7,57.3) 
\Line(119.7,57.3)(121.2,56.6) 
\Line(121.2,56.6)(122.4,56.0) 
\Line(122.4,56.0)(123.7,55.3) 
\Line(123.7,55.3)(125.2,54.6) 
\Line(125.2,54.6)(126.4,54.0) 
\Line(126.4,54.0)(127.9,53.3) 
\Line(127.9,53.3)(129.2,53.0) 
\Line(129.2,53.0)(130.4,52.3) 
\Line(130.4,52.3)(131.9,51.6) 
\Line(131.9,51.6)(133.2,51.3) 
\Line(133.2,51.3)(134.7,50.6) 
\Line(134.7,50.6)(135.9,50.3) 
\Line(135.9,50.3)(137.2,49.6) 
\Line(137.2,49.6)(138.7,49.3) 
\Line(138.7,49.3)(139.9,48.6) 
\Line(139.9,48.6)(141.4,48.3) 
\Line(141.4,48.3)(142.7,47.9) 
\Line(142.7,47.9)(143.9,47.3) 
\Line(143.9,47.3)(145.5,46.9) 
\Line(145.5,46.9)(146.7,46.6) 
\Line(146.7,46.6)(148.0,46.3) 
\Line(148.0,46.3)(149.5,45.6) 
\Line(149.5,45.6)(150.7,45.3) 
\Line(150.7,45.3)(152.2,44.9) 
\Line(152.2,44.9)(153.5,44.6) 
\Line(153.5,44.6)(154.7,44.2) 
\Line(154.7,44.2)(156.2,43.9) 
\Line(156.2,43.9)(157.5,43.6) 
\Line(157.5,43.6)(159.0,42.9) 
\Line(159.0,42.9)(160.2,42.6) 
\Line(160.2,42.6)(161.5,42.2) 
\Line(161.5,42.2)(163.0,41.9) 
\Line(163.0,41.9)(164.2,41.6) 
\Line(164.2,41.6)(165.7,41.2) 
\Line(165.7,41.2)(167.0,40.9) 
\Line(167.0,40.9)(168.2,40.6) 
\Line(168.2,40.6)(169.7,40.6) 
\Line(169.7,40.6)(171.0,40.2) 
\Line(171.0,40.2)(172.5,39.9) 
\end{picture}}
\caption{Contribution of the s-channel diagram $\chi\chi\rightarrow (H^0) \rightarrow Z^0Z^0$.}\label{singleH0}
\end{minipage}
\end{figure}
\\
Now, we look at the 15 interference terms, that we separate into t-channel interferences, s-channel interference and s-t-channel interferences. Note, that although we discuss in the following the interference terms, the squared diagrams are also included in the plots. We select (two) specific diagrams and calculate the cross-section from these diagrams, which of course also includes the squared single diagram contributions.\\
The t-channel interference, i.e. the six interference terms between the t-channel diagrams exchanging neutralinos, is reasonably simple. All interferences are dominated by the diagram exchanging $\chi_3^0$, if it is involved. The slopes of the t-channel interferences are similar to those of the single t-channel diagram contributions in Fig. \ref{singlet}. The t-channel interference terms are almost negligible compared to the squared term of the diagram exchanging $\chi_3^0$. This can be seen in Fig. \ref{tchannel}, where the total t-channel contribution, i.e. all four diagrams exchanging neutralinos, is shown.
\begin{figure}[h!]
\centering
\begin{minipage}{7.5cm}
{
\unitlength=1.0 pt
\SetScale{1.0}
\SetWidth{0.7}      % line    size control
\scriptsize    %  letter  size control
\begin{picture}(180,120)(0,0)
%\Text(104.9,119.7)[t]{$~o1,~o1 ->Z,Z$}
% ====================   X-axis =============
\LinAxis(37.55,33.18)(172.49,33.18)(4.000,2,-4,0.000,1.5)
\Text(37.6,27.5)[t]{$200$}
\Text(71.3,27.5)[t]{$400$}
\Text(105.1,27.5)[t]{$600$}
\Text(138.7,27.5)[t]{$800$}
\Text(172.5,27.5)[t]{$10^3$}
\Text(167.5,19.8)[rt]{$\sqrt(s)$}
% ====================   Y-axis =============
\LogAxis(37.55,33.18)(37.55,100.89)(2.000,4,1.497,1.5)
\Text(31.8,61.0)[r]{$10^-3$}
\Text(31.8,94.9)[r]{$0.01$}
\rText(6.8,100.9)[tr][l]{Cross Section [pb]}
% ============== end of axis ============
\Line(38.8,40.2)(37.6,33.2) 
\Line(40.1,44.9)(38.8,40.2)
\Line(41.6,48.6)(40.1,44.9)
\Line(42.8,51.6)(41.6,48.6)
\Line(44.1,54.3)(42.8,51.6)
\Line(45.6,56.6)(44.1,54.3)
\Line(46.8,58.7)(45.6,56.6)
\Line(48.3,60.7)(46.8,58.7) 
\Line(49.6,62.3)(48.3,60.7)
\Line(50.8,64.0)(49.6,62.3)
\Line(52.3,65.7)(50.8,64.0)
\Line(53.6,67.0)(52.3,65.7) 
\Line(55.1,68.4)(53.6,67.0)
\Line(56.3,69.7)(55.1,68.4)
\Line(57.6,70.7)(56.3,69.7)
\Line(59.1,72.1)(57.6,70.7)
\Line(60.3,73.1)(59.1,72.1) 
\Line(61.8,74.1)(60.3,73.1)
\Line(63.1,75.1)(61.8,74.1)
\Line(64.3,76.1)(63.1,75.1)
\Line(65.8,77.1)(64.3,76.1) 
\Line(67.1,78.1)(65.8,77.1)
\Line(68.3,78.8)(67.1,78.1)
\Line(69.8,79.4)(68.3,78.8)
\Line(71.1,80.4)(69.8,79.4) 
\Line(72.6,81.1)(71.1,80.4)
\Line(73.9,81.8)(72.6,81.1)
\Line(75.1,82.5)(73.9,81.8)
\Line(76.6,83.1)(75.1,82.5)
\Line(77.9,83.8)(76.6,83.1)
\Line(79.4,84.5)(77.9,83.8) 
\Line(80.6,85.1)(79.4,84.5)
\Line(81.9,85.5)(80.6,85.1)
\Line(83.4,86.1)(81.9,85.5)
\Line(84.6,86.5)(83.4,86.1) 
\Line(86.1,87.2)(84.6,86.5)
\Line(87.4,87.5)(86.1,87.2)
\Line(88.6,88.2)(87.4,87.5)
\Line(90.1,88.5)(88.6,88.2) 
\Line(91.4,88.8)(90.1,88.5)
\Line(92.6,89.5)(91.4,88.8)
\Line(94.1,89.8)(92.6,89.5)
\Line(95.4,90.2)(94.1,89.8) 
\Line(96.9,90.5)(95.4,90.2)
\Line(98.1,90.8)(96.9,90.5)
\Line(99.4,91.2)(98.1,90.8) 
\Line(100.9,91.5)(99.4,91.2)
\Line(102.1,91.8)(100.9,91.5)
\Line(103.6,92.2)(102.1,91.8) 
\Line(104.9,92.5)(103.6,92.2)
\Line(106.1,92.8)(104.9,92.5) 
\Line(107.6,93.2)(106.1,92.8)
\Line(108.9,93.5)(107.6,93.2) 
\Line(110.4,93.9)(108.9,93.5)
\Line(111.7,94.2)(110.4,93.9) 
\Line(112.9,94.2)(111.7,94.2)
\Line(114.4,94.5)(112.9,94.2) 
\Line(115.7,94.9)(114.4,94.5)
\Line(117.2,95.2)(115.7,94.9)
\Line(118.4,95.2)(117.2,95.2) 
\Line(119.7,95.5)(118.4,95.2)
\Line(121.2,95.9)(119.7,95.5) 
\Line(122.4,95.9)(121.2,95.9)
\Line(123.7,96.2)(122.4,95.9)
\Line(125.2,96.2)(123.7,96.2) 
\Line(126.4,96.5)(125.2,96.2)
\Line(127.9,96.9)(126.4,96.5) 
\Line(129.2,96.9)(127.9,96.9)
\Line(130.4,97.2)(129.2,96.9) 
\Line(131.9,97.2)(130.4,97.2)
\Line(133.2,97.5)(131.9,97.2) 
\Line(134.7,97.5)(133.2,97.5)
\Line(135.9,97.9)(134.7,97.5) 
\Line(137.2,97.9)(135.9,97.9)
\Line(138.7,98.2)(137.2,97.9) 
\Line(139.9,98.2)(138.7,98.2)
\Line(141.4,98.5)(139.9,98.2) 
\Line(142.7,98.5)(141.4,98.5)
\Line(143.9,98.9)(142.7,98.5) 
\Line(145.5,98.9)(143.9,98.9)
\Line(146.7,98.9)(145.5,98.9) 
\Line(148.0,99.2)(146.7,98.9)
\Line(149.5,99.2)(148.0,99.2)
\Line(150.7,99.2)(149.5,99.2) 
\Line(152.2,99.6)(150.7,99.2)
\Line(153.5,99.6)(152.2,99.6) 
\Line(154.7,99.9)(153.5,99.6)
\Line(156.2,99.9)(154.7,99.9)
\Line(157.5,99.9)(156.2,99.9) 
\Line(159.0,100.2)(157.5,99.9)
\Line(160.2,100.2)(159.0,100.2) 
\Line(161.5,100.2)(160.2,100.2)
\Line(163.0,100.6)(161.5,100.2) 
\Line(164.2,100.6)(163.0,100.6)
\Line(165.7,100.6)(164.2,100.6) 
\Line(167.0,100.6)(165.7,100.6)
\Line(168.2,100.9)(167.0,100.6) 
\Line(169.7,100.9)(168.2,100.9)
\Line(171.0,100.9)(169.7,100.9)
\Line(172.5,101.2)(171.0,100.9) 
\end{picture}
}
\caption{Contribution from all four t-channel diagrams $\chi\chi\rightarrow (\chi^0_n) \rightarrow Z^0Z^0$.}\label{tchannel}
\end{minipage}
\hspace{1cm}
\begin{minipage}{7.5cm}
{
\unitlength=1.0 pt
\SetScale{1.0}
\SetWidth{0.7}      % line    size control
\scriptsize    %  letter  size control
\begin{picture}(180,120)(0,0)
%\Text(104.9,119.7)[t]{$~o1,~o1 ->Z,Z$}
% ====================   X-axis =============
\LinAxis(37.55,33.18)(172.49,33.18)(4.000,2,-4,0.000,1.5)
\Text(37.6,27.5)[t]{$200$}
\Text(71.3,27.5)[t]{$400$}
\Text(105.1,27.5)[t]{$600$}
\Text(138.7,27.5)[t]{$800$}
\Text(172.5,27.5)[t]{$10^3$}
\Text(167.5,19.8)[rt]{$\sqrt(s)$}
% ====================   Y-axis =============
\LogAxis(37.55,33.18)(37.55,100.89)(1.673,4,2.407,1.5)
\Text(31.8,58.3)[r]{$10^-2$}
\Text(31.8,98.5)[r]{$0.1$}
\rText(6.8,100.9)[tr][l]{Cross Section [pb]}
% ============== end of axis ============
\Line(38.8,48.6)(37.6,41.6)
\Line(40.1,52.6)(38.8,48.6) 
\Line(41.6,55.0)(40.1,52.6)
\Line(42.8,57.0)(41.6,55.0) 
\Line(44.1,58.3)(42.8,57.0)
\Line(45.6,59.7)(44.1,58.3) 
\Line(46.8,60.7)(45.6,59.7)
\Line(48.3,61.7)(46.8,60.7) 
\Line(49.6,62.7)(48.3,61.7)
\Line(50.8,63.4)(49.6,62.7) 
\Line(52.3,64.0)(50.8,63.4)
\Line(53.6,64.7)(52.3,64.0) 
\Line(55.1,65.4)(53.6,64.7)
\Line(56.3,65.7)(55.1,65.4) 
\Line(57.6,66.4)(56.3,65.7)
\Line(59.1,67.0)(57.6,66.4) 
\Line(60.3,67.4)(59.1,67.0)
\Line(61.8,68.0)(60.3,67.4) 
\Line(63.1,69.1)(61.8,68.0)
\Line(64.3,69.7)(63.1,69.1) 
\Line(65.8,71.1)(64.3,69.7)
\Line(67.1,73.1)(65.8,71.1) 
\Line(68.3,77.8)(67.1,73.1)
\Line(69.8,101.2)(68.3,77.8)
\Line(69.8,101.2)(71.1,33.2) 
\Line(72.6,57.0)(71.1,33.2)
\Line(73.9,61.7)(72.6,57.0)
\Line(75.1,63.7)(73.9,61.7)
\Line(76.6,64.7)(75.1,63.7) 
\Line(77.9,65.4)(76.6,64.7)
\Line(79.4,66.0)(77.9,65.4)
\Line(80.6,66.4)(79.4,66.0)
\Line(81.9,67.0)(80.6,66.4) 
\Line(83.4,67.0)(81.9,67.0) 
\Line(84.6,67.4)(83.4,67.0) 
\Line(86.1,67.7)(84.6,67.4) 
\Line(87.4,68.0)(86.1,67.7) 
\Line(88.6,68.0)(87.4,68.0) 
\Line(90.1,68.4)(88.6,68.0) 
\Line(91.4,68.4)(90.1,68.4) 
\Line(92.6,68.7)(91.4,68.4) 
\Line(94.1,68.7)(92.6,68.7) 
\Line(95.4,68.7)(94.1,68.7) 
\Line(96.9,69.1)(95.4,68.7) 
\Line(98.1,69.1)(96.9,69.1) 
\Line(99.4,69.1)(98.1,69.1) 
\Line(100.9,69.4)(99.4,69.1) 
\Line(102.1,69.4)(100.9,69.4) 
\Line(103.6,69.4)(102.1,69.4) 
\Line(104.9,69.4)(103.6,69.4) 
\Line(106.1,69.7)(104.9,69.4) 
\Line(107.6,69.7)(106.1,69.7) 
\Line(108.9,69.7)(107.6,69.7) 
\Line(110.4,69.7)(108.9,69.7) 
\Line(111.7,70.1)(110.4,69.7) 
\Line(112.9,70.1)(111.7,70.1) 
\Line(114.4,70.1)(112.9,70.1) 
\Line(115.7,70.1)(114.4,70.1) 
\Line(117.2,70.1)(115.7,70.1) 
\Line(118.4,70.1)(117.2,70.1) 
\Line(119.7,70.4)(118.4,70.1) 
\Line(121.2,70.4)(119.7,70.4) 
\Line(122.4,70.4)(121.2,70.4) 
\Line(123.7,70.4)(122.4,70.4) 
\Line(125.2,70.4)(123.7,70.4) 
\Line(126.4,70.4)(125.2,70.4) 
\Line(127.9,70.4)(126.4,70.4) 
\Line(129.2,70.4)(127.9,70.4) 
\Line(130.4,70.7)(129.2,70.4) 
\Line(131.9,70.7)(130.4,70.7) 
\Line(133.2,70.7)(131.9,70.7) 
\Line(134.7,70.7)(133.2,70.7) 
\Line(135.9,70.7)(134.7,70.7) 
\Line(137.2,70.7)(135.9,70.7) 
\Line(138.7,70.7)(137.2,70.7) 
\Line(139.9,70.7)(138.7,70.7) 
\Line(141.4,70.7)(139.9,70.7) 
\Line(142.7,70.7)(141.4,70.7) 
\Line(143.9,70.7)(142.7,70.7) 
\Line(145.5,70.7)(143.9,70.7) 
\Line(146.7,71.1)(145.5,70.7) 
\Line(148.0,71.1)(146.7,71.1) 
\Line(149.5,71.1)(148.0,71.1) 
\Line(150.7,71.1)(149.5,71.1) 
\Line(152.2,71.1)(150.7,71.1) 
\Line(153.5,71.1)(152.2,71.1) 
\Line(154.7,71.1)(153.5,71.1) 
\Line(156.2,71.1)(154.7,71.1) 
\Line(157.5,71.1)(156.2,71.1) 
\Line(159.0,71.1)(157.5,71.1) 
\Line(160.2,71.1)(159.0,71.1) 
\Line(161.5,71.1)(160.2,71.1) 
\Line(163.0,71.1)(161.5,71.1) 
\Line(164.2,71.1)(163.0,71.1) 
\Line(165.7,71.1)(164.2,71.1) 
\Line(167.0,71.1)(165.7,71.1) 
\Line(168.2,71.1)(167.0,71.1) 
\Line(169.7,71.4)(168.2,71.1) 
\Line(171.0,71.4)(169.7,71.4) 
\Line(172.5,71.4)(171.0,71.4) 
\end{picture}}
\caption{Interference of the s-channel diagrams $\chi\chi\rightarrow (h^0) \rightarrow Z^0Z^0$ and $\chi\chi\rightarrow (H^0) \rightarrow Z^0Z^0$.}\label{schannel}
\end{minipage}
\end{figure}\\
The interference between the s-channel diagrams with exchange of $h^0$ and $H^0$ is shown in Fig. \ref{schannel}. Here, we see a significant interference at about 400 GeV, i.e. near the resonance of the heavy neutral Higgs boson $H^0$. This gives a first hint, where the large interference in the cross section of the process $\chi\chi \rightarrow Z^0Z^0$ comes from. At higher energies, the s-channel interference contribution seems to be constant or increasing slowly. Thus, the decrease of the cross section at higher energies results from interferences between t- and s-channel.\\
\\
In Fig. \ref{h1} - \ref{h4} we study the interference between the s-channel diagram exchanging a light Higgs boson $h^0$ and one of the four neutralinos exchanging t-channel diagrams. The cross section for those interference terms involving t-channel diagrams with $\chi_1^0$, $\chi_2^0$ and $\chi_4^0$ are quite similar in slope and magnitude, while the cross section for the interference term involving $\chi_3^0$ is about one order of magnitude lower and features a clear maximum at about 270 GeV and a minimum at about 650 GeV.
\begin{figure}[h!]
\centering
\begin{minipage}{7.5cm}
{
\unitlength=1.0 pt
\SetScale{1.0}
\SetWidth{0.7}      % line    size control
\scriptsize    %  letter  size control
\begin{picture}(180,120)(0,0)
%\Text(104.9,119.7)[t]{$~o1,~o1 ->Z,Z$}
% ====================   X-axis =============
\LinAxis(37.55,33.18)(172.49,33.18)(4.000,2,-4,0.000,1.5)
\Text(37.6,27.5)[t]{$200$}
\Text(71.3,27.5)[t]{$400$}
\Text(105.1,27.5)[t]{$600$}
\Text(138.7,27.5)[t]{$800$}
\Text(172.5,27.5)[t]{$10^3$}
\Text(167.5,19.8)[rt]{$\sqrt(s)$}
% ====================   Y-axis =============
\LinAxis(37.55,33.18)(37.55,100.89)(3.746,5,4,-1.073,1.5)
\Text(31.8,37.2)[r]{$0.005$}
\Text(31.8,55.0)[r]{$10^-2$}
\Text(31.8,73.1)[r]{$0.015$}
\Text(31.8,91.2)[r]{$0.020$}
\rText(6.8,100.9)[tr][l]{Cross Section [pb]}
% ============== end of axis ============
\Line(38.8,40.2)(37.6,33.2) 
\Line(40.1,45.6)(38.8,40.2) 
\Line(41.6,49.6)(40.1,45.6) 
\Line(42.8,53.0)(41.6,49.6) 
\Line(44.1,56.0)(42.8,53.0) 
\Line(45.6,58.7)(44.1,56.0) 
\Line(46.8,61.0)(45.6,58.7) 
\Line(48.3,63.0)(46.8,61.0) 
\Line(49.6,65.0)(48.3,63.0) 
\Line(50.8,66.7)(49.6,65.0) 
\Line(52.3,68.4)(50.8,66.7) 
\Line(53.6,70.1)(52.3,68.4) 
\Line(55.1,71.7)(53.6,70.1) 
\Line(56.3,73.1)(55.1,71.7) 
\Line(57.6,74.4)(56.3,73.1) 
\Line(59.1,75.4)(57.6,74.4) 
\Line(60.3,76.8)(59.1,75.4) 
\Line(61.8,77.8)(60.3,76.8) 
\Line(63.1,78.8)(61.8,77.8) 
\Line(64.3,79.8)(63.1,78.8) 
\Line(65.8,80.8)(64.3,79.8) 
\Line(67.1,81.5)(65.8,80.8) 
\Line(68.3,82.5)(67.1,81.5) 
\Line(69.8,83.1)(68.3,82.5) 
\Line(71.1,84.1)(69.8,83.1) 
\Line(72.6,84.8)(71.1,84.1) 
\Line(73.9,85.5)(72.6,84.8) 
\Line(75.1,86.1)(73.9,85.5) 
\Line(76.6,86.5)(75.1,86.1) 
\Line(77.9,87.2)(76.6,86.5) 
\Line(79.4,87.8)(77.9,87.2) 
\Line(80.6,88.2)(79.4,87.8) 
\Line(81.9,88.8)(80.6,88.2) 
\Line(83.4,89.2)(81.9,88.8) 
\Line(84.6,89.8)(83.4,89.2) 
\Line(86.1,90.2)(84.6,89.8) 
\Line(87.4,90.5)(86.1,90.2) 
\Line(88.6,91.2)(87.4,90.5) 
\Line(90.1,91.5)(88.6,91.2) 
\Line(91.4,91.8)(90.1,91.5) 
\Line(92.6,92.2)(91.4,91.8) 
\Line(94.1,92.5)(92.6,92.2) 
\Line(95.4,92.8)(94.1,92.5) 
\Line(96.9,93.2)(95.4,92.8) 
\Line(98.1,93.5)(96.9,93.2) 
\Line(99.4,93.9)(98.1,93.5) 
\Line(100.9,94.2)(99.4,93.9) 
\Line(102.1,94.2)(100.9,94.2) 
\Line(103.6,94.5)(102.1,94.2) 
\Line(104.9,94.9)(103.6,94.5) 
\Line(106.1,95.2)(104.9,94.9) 
\Line(107.6,95.2)(106.1,95.2) 
\Line(108.9,95.5)(107.6,95.2) 
\Line(110.4,95.9)(108.9,95.5) 
\Line(111.7,95.9)(110.4,95.9) 
\Line(112.9,96.2)(111.7,95.9) 
\Line(114.4,96.5)(112.9,96.2) 
\Line(115.7,96.5)(114.4,96.5) 
\Line(117.2,96.9)(115.7,96.5) 
\Line(118.4,96.9)(117.2,96.9) 
\Line(119.7,97.2)(118.4,96.9) 
\Line(121.2,97.2)(119.7,97.2) 
\Line(122.4,97.5)(121.2,97.2) 
\Line(123.7,97.5)(122.4,97.5) 
\Line(125.2,97.9)(123.7,97.5) 
\Line(126.4,97.9)(125.2,97.9) 
\Line(127.9,98.2)(126.4,97.9) 
\Line(129.2,98.2)(127.9,98.2) 
\Line(130.4,98.2)(129.2,98.2) 
\Line(131.9,98.5)(130.4,98.2) 
\Line(133.2,98.5)(131.9,98.5) 
\Line(134.7,98.5)(133.2,98.5) 
\Line(135.9,98.9)(134.7,98.5) 
\Line(137.2,98.9)(135.9,98.9) 
\Line(138.7,98.9)(137.2,98.9) 
\Line(139.9,99.2)(138.7,98.9) 
\Line(141.4,99.2)(139.9,99.2) 
\Line(142.7,99.2)(141.4,99.2) 
\Line(143.9,99.6)(142.7,99.2) 
\Line(145.5,99.6)(143.9,99.6) 
\Line(146.7,99.6)(145.5,99.6) 
\Line(148.0,99.9)(146.7,99.6) 
\Line(149.5,99.9)(148.0,99.9) 
\Line(150.7,99.9)(149.5,99.9) 
\Line(152.2,99.9)(150.7,99.9) 
\Line(153.5,100.2)(152.2,99.9) 
\Line(154.7,100.2)(153.5,100.2) 
\Line(156.2,100.2)(154.7,100.2) 
\Line(157.5,100.2)(156.2,100.2) 
\Line(159.0,100.6)(157.5,100.2) 
\Line(160.2,100.6)(159.0,100.6) 
\Line(161.5,100.6)(160.2,100.6) 
\Line(163.0,100.6)(161.5,100.6) 
\Line(164.2,100.6)(163.0,100.6) 
\Line(165.7,100.9)(164.2,100.6) 
\Line(167.0,100.9)(165.7,100.9) 
\Line(168.2,100.9)(167.0,100.9) 
\Line(169.7,100.9)(168.2,100.9) 
\Line(171.0,100.9)(169.7,100.9) 
\Line(172.5,101.2)(171.0,100.9) 
\end{picture}}
\caption{Interference of the s-channel diagram $\chi\chi\rightarrow (h^0) \rightarrow Z^0Z^0$ and the t-channel diagram $\chi\chi\rightarrow (\chi^0_1) \rightarrow Z^0Z^0$.}\label{h1}
\end{minipage}
\hspace{1cm}
\begin{minipage}{7.5cm}
{
\unitlength=1.0 pt
\SetScale{1.0}
\SetWidth{0.7}      % line    size control
\scriptsize    %  letter  size control
\begin{picture}(180,120)(0,0)
%\Text(108.9,119.7)[t]{$~o1,~o1 ->Z,Z$}
% ====================   X-axis =============
\LinAxis(47.57,42.37)(170.49,42.37)(4.000,2,-4,0.000,1.5)
\Text(47.6,35.6)[t]{$200$}
\Text(78.4,35.6)[t]{$400$}
\Text(109.2,35.6)[t]{$600$}
\Text(139.7,35.6)[t]{$800$}
\Text(170.5,35.6)[t]{$10^3$}
\Text(167.5,19.8)[rt]{$\sqrt(s)$}
% ====================   Y-axis =============
\LinAxis(47.57,42.37)(47.57,95.59)(4.264,5,4,-0.982,1.5)
\Text(40.8,44.7)[r]{$0.005$}
\Text(40.8,57.3)[r]{$10^-2$}
\Text(40.8,69.8)[r]{$0.015$}
\Text(40.8,82.4)[r]{$0.020$}
\Text(40.8,94.9)[r]{$0.025$}
\rText(9.8,95.6)[tr][l]{Cross Section [pb]}
% ============== end of axis ============
\Line(48.6,47.5)(47.6,42.4)
\Line(49.8,51.2)(48.6,47.5)
\Line(51.1,54.2)(49.8,51.2)
\Line(52.3,56.9)(51.1,54.2)
\Line(53.6,59.0)(52.3,56.9)
\Line(54.8,61.0)(53.6,59.0)
\Line(56.1,62.7)(54.8,61.0)
\Line(57.3,64.4)(56.1,62.7)
\Line(58.6,65.8)(57.3,64.4)
\Line(59.8,67.1)(58.6,65.8)
\Line(61.1,68.5)(59.8,67.1)
\Line(62.1,69.8)(61.1,68.5)
\Line(63.3,70.8)(62.1,69.8)
\Line(64.6,71.9)(63.3,70.8)
\Line(65.8,72.9)(64.6,71.9)
\Line(67.1,73.9)(65.8,72.9)
\Line(68.3,74.9)(67.1,73.9)
\Line(69.6,75.9)(68.3,74.9)
\Line(70.8,76.6)(69.6,75.9)
\Line(72.1,77.3)(70.8,76.6)
\Line(73.4,78.3)(72.1,77.3)
\Line(74.6,79.0)(73.4,78.3)
\Line(75.6,79.7)(74.6,79.0)
\Line(76.9,80.3)(75.6,79.7)
\Line(78.1,80.7)(76.9,80.3)
\Line(79.4,81.4)(78.1,80.7)
\Line(80.6,82.0)(79.4,81.4)
\Line(81.9,82.4)(80.6,82.0)
\Line(83.1,83.1)(81.9,82.4)
\Line(84.4,83.4)(83.1,83.1)
\Line(85.6,84.1)(84.4,83.4)
\Line(86.9,84.4)(85.6,84.1)
\Line(88.1,84.7)(86.9,84.4)
\Line(89.1,85.1)(88.1,84.7)
\Line(90.4,85.8)(89.1,85.1)
\Line(91.6,86.1)(90.4,85.8)
\Line(92.9,86.4)(91.6,86.1)
\Line(94.1,86.8)(92.9,86.4)
\Line(95.4,87.1)(94.1,86.8)
\Line(96.6,87.5)(95.4,87.1)
\Line(97.9,87.8)(96.6,87.5)
\Line(99.1,88.1)(97.9,87.8)
\Line(100.4,88.1)(99.1,88.1)
\Line(101.6,88.5)(100.4,88.1)
\Line(102.6,88.8)(101.6,88.5)
\Line(103.9,89.2)(102.6,88.8)
\Line(105.1,89.5)(103.9,89.2)
\Line(106.4,89.5)(105.1,89.5)
\Line(107.6,89.8)(106.4,89.5)
\Line(108.9,90.2)(107.6,89.8)
\Line(110.2,90.2)(108.9,90.2)
\Line(111.4,90.5)(110.2,90.2)
\Line(112.7,90.5)(111.4,90.5)
\Line(113.9,90.8)(112.7,90.5)
\Line(115.2,91.2)(113.9,90.8)
\Line(116.2,91.2)(115.2,91.2)
\Line(117.4,91.5)(116.2,91.2)
\Line(118.7,91.5)(117.4,91.5)
\Line(119.9,91.9)(118.7,91.5)
\Line(121.2,91.9)(119.9,91.9)
\Line(122.4,92.2)(121.2,91.9)
\Line(123.7,92.2)(122.4,92.2)
\Line(124.9,92.2)(123.7,92.2)
\Line(126.2,92.5)(124.9,92.2)
\Line(127.4,92.5)(126.2,92.5)
\Line(128.7,92.9)(127.4,92.5)
\Line(129.7,92.9)(128.7,92.9) 
\Line(130.9,92.9)(129.7,92.9)
\Line(132.2,93.2)(130.9,92.9) 
\Line(133.4,93.2)(132.2,93.2)
\Line(134.7,93.6)(133.4,93.2)
\Line(135.9,93.6)(134.7,93.6)
\Line(137.2,93.6)(135.9,93.6)
\Line(138.4,93.9)(137.2,93.6)
\Line(139.7,93.9)(138.4,93.9)
\Line(140.9,93.9)(139.7,93.9)
\Line(142.2,93.9)(140.9,93.9)
\Line(143.2,94.2)(142.2,93.9)
\Line(144.5,94.2)(143.2,94.2) 
\Line(145.7,94.2)(144.5,94.2)
\Line(147.0,94.6)(145.7,94.2)
\Line(148.2,94.6)(147.0,94.6) 
\Line(149.5,94.6)(148.2,94.6)
\Line(150.7,94.6)(149.5,94.6)
\Line(152.0,94.9)(150.7,94.6)
\Line(153.2,94.9)(152.0,94.9) 
\Line(154.5,94.9)(153.2,94.9)
\Line(155.7,94.9)(154.5,94.9)
\Line(156.7,94.9)(155.7,94.9)
\Line(158.0,95.3)(156.7,94.9)
\Line(159.2,95.3)(158.0,95.3)
\Line(160.5,95.3)(159.2,95.3)
\Line(161.7,95.3)(160.5,95.3)
\Line(163.0,95.3)(161.7,95.3)
\Line(164.2,95.6)(163.0,95.3)
\Line(165.5,95.6)(164.2,95.6)
\Line(166.7,95.6)(165.5,95.6)
\Line(168.0,95.6)(166.7,95.6)
\Line(169.2,95.6)(168.0,95.6)
\Line(170.5,95.6)(169.2,95.6)
\end{picture}}
\caption{Interference of the s-channel diagram $\chi\chi\rightarrow (h^0) \rightarrow Z^0Z^0$ and the t-channel diagram $\chi\chi\rightarrow (\chi^0_2) \rightarrow Z^0Z^0$.}\label{h2}
\end{minipage}
\end{figure}
\begin{figure}[h!]
\centering
\begin{minipage}{7.5cm}
{
\unitlength=1.0 pt
\SetScale{1.0}
\SetWidth{0.7}      % line    size control
\scriptsize    %  letter  size control
\begin{picture}(180,120)(0,0)
%\Text(104.9,119.7)[t]{$~o1,~o1 ->Z,Z$}
% ====================   X-axis =============
\LinAxis(37.55,33.18)(172.49,33.18)(4.000,2,-4,0.000,1.5)
\Text(37.6,27.5)[t]{$200$}
\Text(71.3,27.5)[t]{$400$}
\Text(105.1,27.5)[t]{$600$}
\Text(138.7,27.5)[t]{$800$}
\Text(172.5,27.5)[t]{$10^3$}
\Text(167.5,19.8)[rt]{$\sqrt(s)$}
% ====================   Y-axis =============
\LinAxis(37.55,33.18)(37.55,100.89)(4.213,10,4,0.933,1.5)
\Text(31.8,47.6)[r]{$0.0020$}
\Text(31.8,63.7)[r]{$0.0030$}
\Text(31.8,79.8)[r]{$0.0040$}
\Text(31.8,95.9)[r]{$0.0050$}
\rText(5.3,100.9)[tr][l]{Cross Section [pb]}
% ============== end of axis ============
\Line(38.8,73.1)(37.6,56.0) 
\Line(40.1,83.8)(38.8,73.1) 
\Line(41.6,90.8)(40.1,83.8) 
\Line(42.8,95.5)(41.6,90.8) 
\Line(44.1,98.5)(42.8,95.5) 
\Line(45.6,100.2)(44.1,98.5) 
\Line(46.8,100.9)(45.6,100.2) 
\Line(48.3,100.9)(46.8,100.9) 
\Line(48.3,100.9)(49.6,100.6) 
\Line(49.6,100.6)(50.8,99.6) 
\Line(50.8,99.6)(52.3,98.2) 
\Line(52.3,98.2)(53.6,96.5) 
\Line(53.6,96.5)(55.1,94.9) 
\Line(55.1,94.9)(56.3,92.8) 
\Line(56.3,92.8)(57.6,90.5) 
\Line(57.6,90.5)(59.1,88.2) 
\Line(59.1,88.2)(60.3,85.8) 
\Line(60.3,85.8)(61.8,83.1) 
\Line(61.8,83.1)(63.1,80.8) 
\Line(63.1,80.8)(64.3,78.1) 
\Line(64.3,78.1)(65.8,75.8) 
\Line(65.8,75.8)(67.1,73.4) 
\Line(67.1,73.4)(68.3,70.7) 
\Line(68.3,70.7)(69.8,68.4) 
\Line(69.8,68.4)(71.1,66.0) 
\Line(71.1,66.0)(72.6,64.0) 
\Line(72.6,64.0)(73.9,61.7) 
\Line(73.9,61.7)(75.1,59.7) 
\Line(75.1,59.7)(76.6,57.7) 
\Line(76.6,57.7)(77.9,55.6) 
\Line(77.9,55.6)(79.4,53.6) 
\Line(79.4,53.6)(80.6,52.0) 
\Line(80.6,52.0)(81.9,50.3) 
\Line(81.9,50.3)(83.4,48.9) 
\Line(83.4,48.9)(84.6,47.3) 
\Line(84.6,47.3)(86.1,45.9) 
\Line(86.1,45.9)(87.4,44.6) 
\Line(87.4,44.6)(88.6,43.2) 
\Line(88.6,43.2)(90.1,42.2) 
\Line(90.1,42.2)(91.4,41.2) 
\Line(91.4,41.2)(92.6,40.2) 
\Line(92.6,40.2)(94.1,39.2) 
\Line(94.1,39.2)(95.4,38.2) 
\Line(95.4,38.2)(96.9,37.5) 
\Line(96.9,37.5)(98.1,36.9) 
\Line(98.1,36.9)(99.4,36.2) 
\Line(99.4,36.2)(100.9,35.9) 
\Line(100.9,35.9)(102.1,35.2) 
\Line(102.1,35.2)(103.6,34.9) 
\Line(103.6,34.9)(104.9,34.5) 
\Line(104.9,34.5)(106.1,34.2) 
\Line(106.1,34.2)(107.6,33.9) 
\Line(107.6,33.9)(108.9,33.9) 
\Line(108.9,33.9)(110.4,33.5) 
\Line(110.4,33.5)(111.7,33.5) 
\Line(111.7,33.5)(112.9,33.5) 
\Line(112.9,33.5)(114.4,33.2) 
\Line(115.7,33.5)(114.4,33.2) 
\Line(117.2,33.5)(115.7,33.5) 
\Line(118.4,33.5)(117.2,33.5) 
\Line(119.7,33.5)(118.4,33.5) 
\Line(121.2,33.9)(119.7,33.5) 
\Line(122.4,34.2)(121.2,33.9) 
\Line(123.7,34.2)(122.4,34.2) 
\Line(125.2,34.5)(123.7,34.2) 
\Line(126.4,34.9)(125.2,34.5) 
\Line(127.9,35.2)(126.4,34.9) 
\Line(129.2,35.5)(127.9,35.2) 
\Line(130.4,35.9)(129.2,35.5) 
\Line(131.9,36.2)(130.4,35.9) 
\Line(133.2,36.5)(131.9,36.2) 
\Line(134.7,36.9)(133.2,36.5) 
\Line(135.9,37.5)(134.7,36.9) 
\Line(137.2,37.9)(135.9,37.5) 
\Line(138.7,38.2)(137.2,37.9) 
\Line(139.9,38.9)(138.7,38.2) 
\Line(141.4,39.2)(139.9,38.9) 
\Line(142.7,39.9)(141.4,39.2) 
\Line(143.9,40.2)(142.7,39.9) 
\Line(145.5,40.9)(143.9,40.2) 
\Line(146.7,41.2)(145.5,40.9) 
\Line(148.0,41.9)(146.7,41.2) 
\Line(149.5,42.6)(148.0,41.9) 
\Line(150.7,42.9)(149.5,42.6) 
\Line(152.2,43.6)(150.7,42.9) 
\Line(153.5,44.2)(152.2,43.6) 
\Line(154.7,44.9)(153.5,44.2) 
\Line(156.2,45.3)(154.7,44.9) 
\Line(157.5,45.9)(156.2,45.3) 
\Line(159.0,46.6)(157.5,45.9) 
\Line(160.2,47.3)(159.0,46.6) 
\Line(161.5,47.6)(160.2,47.3) 
\Line(163.0,48.3)(161.5,47.6) 
\Line(164.2,48.9)(163.0,48.3) 
\Line(165.7,49.6)(164.2,48.9) 
\Line(167.0,50.3)(165.7,49.6) 
\Line(168.2,50.9)(167.0,50.3) 
\Line(169.7,51.3)(168.2,50.9) 
\Line(171.0,52.0)(169.7,51.3) 
\Line(172.5,52.6)(171.0,52.0) 
\end{picture}}
\caption{Interference of the s-channel diagram $\chi\chi\rightarrow (h^0) \rightarrow Z^0Z^0$ and the t-channel diagram $\chi\chi\rightarrow (\chi^0_3) \rightarrow Z^0Z^0$.}\label{h3}
\end{minipage}
\hspace{1cm}
\begin{minipage}{7.5cm}
{
\unitlength=1.0 pt
\SetScale{1.0}
\SetWidth{0.7}      % line    size control
\scriptsize    %  letter  size control
\begin{picture}(180,120)(0,0)
%\Text(104.9,119.7)[t]{$~o1,~o1 ->Z,Z$}
% ====================   X-axis =============
\LinAxis(37.55,33.18)(172.49,33.18)(4.000,2,-4,0.000,1.5)
\Text(37.6,27.5)[t]{$200$}
\Text(71.3,27.5)[t]{$400$}
\Text(105.1,27.5)[t]{$600$}
\Text(138.7,27.5)[t]{$800$}
\Text(172.5,27.5)[t]{$10^3$}
\Text(167.5,19.8)[rt]{$\sqrt(s)$}
% ====================   Y-axis =============
\LinAxis(37.55,33.18)(37.55,100.89)(3.563,10,4,4.121,1.5)
\Text(31.8,44.2)[r]{$10^-2$}
\Text(31.8,63.4)[r]{$0.020$}
\Text(31.8,82.5)[r]{$0.030$}
\rText(5.3,100.9)[tr][l]{Cross Section [pb]}
% ============== end of axis ============
\Line(38.8,37.2)(37.6,33.2) 
\Line(40.1,40.2)(38.8,37.2) 
\Line(41.6,42.9)(40.1,40.2) 
\Line(42.8,44.9)(41.6,42.9) 
\Line(44.1,46.6)(42.8,44.9) 
\Line(45.6,48.6)(44.1,46.6) 
\Line(46.8,49.9)(45.6,48.6) 
\Line(48.3,51.6)(46.8,49.9) 
\Line(49.6,53.0)(48.3,51.6) 
\Line(50.8,54.3)(49.6,53.0) 
\Line(52.3,55.6)(50.8,54.3) 
\Line(53.6,56.6)(52.3,55.6) 
\Line(55.1,58.0)(53.6,56.6) 
\Line(56.3,59.0)(55.1,58.0) 
\Line(57.6,60.0)(56.3,59.0) 
\Line(59.1,61.0)(57.6,60.0) 
\Line(60.3,62.3)(59.1,61.0) 
\Line(61.8,63.4)(60.3,62.3) 
\Line(63.1,64.0)(61.8,63.4) 
\Line(64.3,65.0)(63.1,64.0) 
\Line(65.8,66.0)(64.3,65.0) 
\Line(67.1,67.0)(65.8,66.0) 
\Line(68.3,67.7)(67.1,67.0) 
\Line(69.8,68.7)(68.3,67.7) 
\Line(71.1,69.4)(69.8,68.7) 
\Line(72.6,70.4)(71.1,69.4) 
\Line(73.9,71.1)(72.6,70.4) 
\Line(75.1,71.7)(73.9,71.1) 
\Line(76.6,72.4)(75.1,71.7) 
\Line(77.9,73.4)(76.6,72.4) 
\Line(79.4,74.1)(77.9,73.4) 
\Line(80.6,74.7)(79.4,74.1) 
\Line(81.9,75.4)(80.6,74.7) 
\Line(83.4,76.1)(81.9,75.4) 
\Line(84.6,76.8)(83.4,76.1) 
\Line(86.1,77.4)(84.6,76.8) 
\Line(87.4,78.1)(86.1,77.4) 
\Line(88.6,78.4)(87.4,78.1) 
\Line(90.1,79.1)(88.6,78.4) 
\Line(91.4,79.8)(90.1,79.1) 
\Line(92.6,80.4)(91.4,79.8) 
\Line(94.1,80.8)(92.6,80.4) 
\Line(95.4,81.5)(94.1,80.8) 
\Line(96.9,82.1)(95.4,81.5) 
\Line(98.1,82.5)(96.9,82.1) 
\Line(99.4,83.1)(98.1,82.5) 
\Line(100.9,83.5)(99.4,83.1) 
\Line(102.1,84.1)(100.9,83.5) 
\Line(103.6,84.5)(102.1,84.1) 
\Line(104.9,85.1)(103.6,84.5) 
\Line(106.1,85.5)(104.9,85.1) 
\Line(107.6,85.8)(106.1,85.5) 
\Line(108.9,86.5)(107.6,85.8) 
\Line(110.4,86.8)(108.9,86.5) 
\Line(111.7,87.2)(110.4,86.8) 
\Line(112.9,87.8)(111.7,87.2) 
\Line(114.4,88.2)(112.9,87.8) 
\Line(115.7,88.5)(114.4,88.2) 
\Line(117.2,88.8)(115.7,88.5) 
\Line(118.4,89.5)(117.2,88.8) 
\Line(119.7,89.8)(118.4,89.5) 
\Line(121.2,90.2)(119.7,89.8) 
\Line(122.4,90.5)(121.2,90.2) 
\Line(123.7,90.8)(122.4,90.5) 
\Line(125.2,91.2)(123.7,90.8) 
\Line(126.4,91.5)(125.2,91.2) 
\Line(127.9,91.8)(126.4,91.5) 
\Line(129.2,92.5)(127.9,91.8) 
\Line(130.4,92.8)(129.2,92.5) 
\Line(131.9,93.2)(130.4,92.8) 
\Line(133.2,93.5)(131.9,93.2) 
\Line(134.7,93.9)(133.2,93.5) 
\Line(135.9,94.2)(134.7,93.9) 
\Line(137.2,94.2)(135.9,94.2) 
\Line(138.7,94.5)(137.2,94.2) 
\Line(139.9,94.9)(138.7,94.5) 
\Line(141.4,95.2)(139.9,94.9) 
\Line(142.7,95.5)(141.4,95.2) 
\Line(143.9,95.9)(142.7,95.5) 
\Line(145.5,96.2)(143.9,95.9) 
\Line(146.7,96.5)(145.5,96.2) 
\Line(148.0,96.5)(146.7,96.5) 
\Line(149.5,96.9)(148.0,96.5) 
\Line(150.7,97.2)(149.5,96.9) 
\Line(152.2,97.5)(150.7,97.2) 
\Line(153.5,97.9)(152.2,97.5) 
\Line(154.7,97.9)(153.5,97.9) 
\Line(156.2,98.2)(154.7,97.9) 
\Line(157.5,98.5)(156.2,98.2) 
\Line(159.0,98.9)(157.5,98.5) 
\Line(160.2,98.9)(159.0,98.9) 
\Line(161.5,99.2)(160.2,98.9) 
\Line(163.0,99.6)(161.5,99.2) 
\Line(164.2,99.9)(163.0,99.6) 
\Line(165.7,99.9)(164.2,99.9) 
\Line(167.0,100.2)(165.7,99.9) 
\Line(168.2,100.6)(167.0,100.2) 
\Line(169.7,100.6)(168.2,100.6) 
\Line(171.0,100.9)(169.7,100.6) 
\Line(172.5,101.2)(171.0,100.9) 
\end{picture}}
\caption{Interference of the s-channel diagram $\chi\chi\rightarrow (h^0) \rightarrow Z^0Z^0$ and the t-channel diagram $\chi\chi\rightarrow (\chi^0_4) \rightarrow Z^0Z^0$.}\label{h4}
\end{minipage}
\end{figure}\\
Finally, we look at the interference term contributions between the s-channel diagram exchanging a heavy neutral Higgs boson $H^0$ and one of the four neutralinos exchanging t-channel diagrams, shown in Fig. \ref{H1} - \ref{H4}. In all plots we see a resonance at around 400 GeV, i.e. at the mass of the Higgs boson $H^0$. In the interferences containing $\chi_1^0$, $\chi_2^0$ and $\chi_4^0$ in Fig. \ref{H1}, \ref{H2} and \ref{H4}, respectively, we have first the maximum and then a minimum (or a monotonically decreasing slope in the case of Fig. \ref{H1}). However, for the interference between the diagrams exchanging $H^0$ and $\chi_3^0$ we have the opposite, first a minimum and then a maximum.
\begin{figure}[h!]
\centering
\begin{minipage}{7.5cm}
{
\unitlength=1.0 pt
\SetScale{1.0}
\SetWidth{0.7}      % line    size control
\scriptsize    %  letter  size control
\begin{picture}(180,120)(0,0)
%\Text(104.9,119.7)[t]{$~o1,~o1 ->Z,Z$}
% ====================   X-axis =============
\LinAxis(37.55,33.18)(172.49,33.18)(4.000,2,-4,0.000,1.5)
\Text(37.6,27.5)[t]{$200$}
\Text(71.3,27.5)[t]{$400$}
\Text(105.1,27.5)[t]{$600$}
\Text(138.7,27.5)[t]{$800$}
\Text(172.5,27.5)[t]{$10^3$}
\Text(167.5,19.8)[rt]{$\sqrt(s)$}
% ====================   Y-axis =============
\LogAxis(37.55,33.18)(37.55,100.89)(4.527,4,1.287,1.5)
\Text(31.8,46.6)[r]{$10^-5$}
\Text(31.8,61.3)[r]{$10^-4$}
\Text(31.8,76.4)[r]{$10^-3$}
\Text(31.8,91.5)[r]{$0.01$}
\rText(6.8,100.9)[tr][l]{Cross Section [pb]}
% ============== end of axis ============
\Line(38.8,41.6)(37.6,39.2) 
\Line(40.1,42.9)(38.8,41.6) 
\Line(41.6,43.9)(40.1,42.9) 
\Line(42.8,44.9)(41.6,43.9) 
\Line(44.1,45.9)(42.8,44.9) 
\Line(45.6,46.6)(44.1,45.9) 
\Line(46.8,47.6)(45.6,46.6) 
\Line(48.3,48.3)(46.8,47.6) 
\Line(49.6,49.3)(48.3,48.3) 
\Line(50.8,49.9)(49.6,49.3) 
\Line(52.3,50.9)(50.8,49.9) 
\Line(53.6,52.0)(52.3,50.9) 
\Line(55.1,53.0)(53.6,52.0) 
\Line(56.3,54.3)(55.1,53.0) 
\Line(57.6,55.6)(56.3,54.3) 
\Line(59.1,57.0)(57.6,55.6) 
\Line(60.3,58.7)(59.1,57.0) 
\Line(61.8,60.7)(60.3,58.7) 
\Line(63.1,63.0)(61.8,60.7) 
\Line(64.3,65.7)(63.1,63.0) 
\Line(65.8,69.1)(64.3,65.7) 
\Line(67.1,73.7)(65.8,69.1) 
\Line(68.3,81.1)(67.1,73.7) 
\Line(69.8,101.2)(68.3,81.1) 
\Line(69.8,101.2)(71.1,88.8) 
\Line(71.1,88.8)(72.6,77.4) 
\Line(72.6,77.4)(73.9,71.4) 
\Line(73.9,71.4)(75.1,67.4) 
\Line(75.1,67.4)(76.6,64.4) 
\Line(76.6,64.4)(77.9,62.0) 
\Line(77.9,62.0)(79.4,59.7) 
\Line(79.4,59.7)(80.6,58.0) 
\Line(80.6,58.0)(81.9,56.6) 
\Line(81.9,56.6)(83.4,55.3) 
\Line(83.4,55.3)(84.6,54.0) 
\Line(84.6,54.0)(86.1,53.0) 
\Line(86.1,53.0)(87.4,52.0) 
\Line(87.4,52.0)(88.6,50.9) 
\Line(88.6,50.9)(90.1,49.9) 
\Line(90.1,49.9)(91.4,49.3) 
\Line(91.4,49.3)(92.6,48.6) 
\Line(92.6,48.6)(94.1,47.9) 
\Line(94.1,47.9)(95.4,47.3) 
\Line(95.4,47.3)(96.9,46.6) 
\Line(96.9,46.6)(98.1,45.9) 
\Line(98.1,45.9)(99.4,45.6) 
\Line(99.4,45.6)(100.9,44.9) 
\Line(100.9,44.9)(102.1,44.6) 
\Line(102.1,44.6)(103.6,43.9) 
\Line(103.6,43.9)(104.9,43.6) 
\Line(104.9,43.6)(106.1,43.2) 
\Line(106.1,43.2)(107.6,42.9) 
\Line(107.6,42.9)(108.9,42.2) 
\Line(108.9,42.2)(110.4,41.9) 
\Line(110.4,41.9)(111.7,41.6) 
\Line(111.7,41.6)(112.9,41.2) 
\Line(112.9,41.2)(114.4,40.9) 
\Line(114.4,40.9)(115.7,40.6) 
\Line(115.7,40.6)(117.2,40.6) 
\Line(117.2,40.6)(118.4,40.2) 
\Line(118.4,40.2)(119.7,39.9) 
\Line(119.7,39.9)(121.2,39.6) 
\Line(121.2,39.6)(122.4,39.2) 
\Line(122.4,39.2)(123.7,39.2) 
\Line(123.7,39.2)(125.2,38.9) 
\Line(125.2,38.9)(126.4,38.5) 
\Line(126.4,38.5)(127.9,38.5) 
\Line(127.9,38.5)(129.2,38.2) 
\Line(129.2,38.2)(130.4,37.9) 
\Line(130.4,37.9)(131.9,37.9) 
\Line(131.9,37.9)(133.2,37.5) 
\Line(133.2,37.5)(134.7,37.2) 
\Line(134.7,37.2)(135.9,37.2) 
\Line(135.9,37.2)(137.2,36.9) 
\Line(137.2,36.9)(138.7,36.9) 
\Line(138.7,36.9)(139.9,36.5) 
\Line(139.9,36.5)(141.4,36.5) 
\Line(141.4,36.5)(142.7,36.2) 
\Line(142.7,36.2)(143.9,36.2) 
\Line(143.9,36.2)(145.5,35.9) 
\Line(145.5,35.9)(146.7,35.9) 
\Line(146.7,35.9)(148.0,35.5) 
\Line(148.0,35.5)(149.5,35.5) 
\Line(149.5,35.5)(150.7,35.2) 
\Line(150.7,35.2)(152.2,35.2) 
\Line(152.2,35.2)(153.5,35.2) 
\Line(153.5,35.2)(154.7,34.9) 
\Line(154.7,34.9)(156.2,34.9) 
\Line(156.2,34.9)(157.5,34.5) 
\Line(157.5,34.5)(159.0,34.5) 
\Line(159.0,34.5)(160.2,34.5) 
\Line(160.2,34.5)(161.5,34.2) 
\Line(161.5,34.2)(163.0,34.2) 
\Line(163.0,34.2)(164.2,34.2) 
\Line(164.2,34.2)(165.7,33.9) 
\Line(165.7,33.9)(167.0,33.9) 
\Line(167.0,33.9)(168.2,33.5) 
\Line(168.2,33.5)(169.7,33.5) 
\Line(169.7,33.5)(171.0,33.5) 
\Line(171.0,33.5)(172.5,33.2) 
\end{picture}}
\caption{Interference of the s-channel diagram $\chi\chi\rightarrow (H^0) \rightarrow Z^0Z^0$ and the t-channel diagram $\chi\chi\rightarrow (\chi^0_1) \rightarrow Z^0Z^0$.}\label{H1}
\end{minipage}
\hspace{1cm}
\begin{minipage}{7.5cm}
{
\unitlength=1.0 pt
\SetScale{1.0}
\SetWidth{0.7}      % line    size control
\scriptsize    %  letter  size control
\begin{picture}(180,120)(0,0)
%\Text(104.9,119.7)[t]{$~o1,~o1 ->Z,Z$}
% ====================   X-axis =============
\LinAxis(37.55,33.18)(172.49,33.18)(4.000,2,-4,0.000,1.5)
\Text(37.6,27.5)[t]{$200$}
\Text(71.3,27.5)[t]{$400$}
\Text(105.1,27.5)[t]{$600$}
\Text(138.7,27.5)[t]{$800$}
\Text(172.5,27.5)[t]{$10^3$}
\Text(167.5,19.8)[rt]{$\sqrt(s)$}
% ====================   Y-axis =============
\LogAxis(37.55,33.18)(37.55,100.89)(3.896,4,5.772,1.5)
\Text(31.8,37.2)[r]{$10^-5$}
\Text(31.8,54.6)[r]{$10^-4$}
\Text(31.8,72.1)[r]{$10^-3$}
\Text(31.8,89.5)[r]{$0.01$}
\rText(6.8,100.9)[tr][l]{Cross Section [pb]}
% ============== end of axis ============
\Line(38.8,37.2)(37.6,33.2) 
\Line(40.1,39.9)(38.8,37.2) 
\Line(41.6,41.6)(40.1,39.9) 
\Line(42.8,43.2)(41.6,41.6) 
\Line(44.1,44.9)(42.8,43.2) 
\Line(45.6,45.9)(44.1,44.9) 
\Line(46.8,47.3)(45.6,45.9) 
\Line(48.3,48.3)(46.8,47.3) 
\Line(49.6,49.3)(48.3,48.3) 
\Line(50.8,50.3)(49.6,49.3) 
\Line(52.3,51.3)(50.8,50.3) 
\Line(53.6,52.3)(52.3,51.3) 
\Line(55.1,53.6)(53.6,52.3) 
\Line(56.3,54.6)(55.1,53.6) 
\Line(57.6,55.6)(56.3,54.6) 
\Line(59.1,57.0)(57.6,55.6) 
\Line(60.3,58.3)(59.1,57.0) 
\Line(61.8,60.0)(60.3,58.3) 
\Line(63.1,61.7)(61.8,60.0) 
\Line(64.3,64.0)(63.1,61.7) 
\Line(65.8,67.0)(64.3,64.0) 
\Line(67.1,71.7)(65.8,67.0) 
\Line(68.3,79.1)(67.1,71.7) 
\Line(69.8,100.9)(68.3,79.1) 
\Line(69.8,100.9)(71.1,85.5) 
\Line(71.1,85.5)(72.6,70.7) 
\Line(72.6,70.7)(73.9,62.3) 
\Line(73.9,62.3)(75.1,56.3) 
\Line(75.1,56.3)(76.6,51.6) 
\Line(76.6,51.6)(77.9,47.9) 
\Line(77.9,47.9)(79.4,45.3) 
\Line(79.4,45.3)(80.6,43.6) 
\Line(80.6,43.6)(81.9,42.9) 
\Line(81.9,42.9)(83.4,42.6) 
\Line(84.6,42.9)(83.4,42.6) 
\Line(86.1,43.2)(84.6,42.9) 
\Line(87.4,43.9)(86.1,43.2) 
\Line(88.6,44.2)(87.4,43.9) 
\Line(90.1,44.9)(88.6,44.2) 
\Line(91.4,45.3)(90.1,44.9) 
\Line(92.6,45.9)(91.4,45.3) 
\Line(94.1,46.3)(92.6,45.9) 
\Line(95.4,46.6)(94.1,46.3) 
\Line(96.9,46.9)(95.4,46.6) 
\Line(98.1,47.3)(96.9,46.9) 
\Line(99.4,47.6)(98.1,47.3) 
\Line(100.9,47.9)(99.4,47.6) 
\Line(102.1,48.3)(100.9,47.9) 
\Line(103.6,48.6)(102.1,48.3) 
\Line(104.9,48.9)(103.6,48.6) 
\Line(106.1,48.9)(104.9,48.9) 
\Line(107.6,49.3)(106.1,48.9) 
\Line(108.9,49.3)(107.6,49.3) 
\Line(110.4,49.6)(108.9,49.3) 
\Line(111.7,49.9)(110.4,49.6) 
\Line(112.9,49.9)(111.7,49.9) 
\Line(114.4,50.3)(112.9,49.9) 
\Line(115.7,50.3)(114.4,50.3) 
\Line(117.2,50.3)(115.7,50.3) 
\Line(118.4,50.6)(117.2,50.3) 
\Line(119.7,50.6)(118.4,50.6) 
\Line(121.2,50.9)(119.7,50.6) 
\Line(122.4,50.9)(121.2,50.9) 
\Line(123.7,50.9)(122.4,50.9) 
\Line(125.2,51.3)(123.7,50.9) 
\Line(126.4,51.3)(125.2,51.3) 
\Line(127.9,51.3)(126.4,51.3) 
\Line(129.2,51.6)(127.9,51.3) 
\Line(130.4,51.6)(129.2,51.6) 
\Line(131.9,51.6)(130.4,51.6) 
\Line(133.2,51.6)(131.9,51.6) 
\Line(134.7,52.0)(133.2,51.6) 
\Line(135.9,52.0)(134.7,52.0) 
\Line(137.2,52.0)(135.9,52.0) 
\Line(138.7,52.0)(137.2,52.0) 
\Line(139.9,52.0)(138.7,52.0) 
\Line(141.4,52.3)(139.9,52.0) 
\Line(142.7,52.3)(141.4,52.3) 
\Line(143.9,52.3)(142.7,52.3) 
\Line(145.5,52.3)(143.9,52.3) 
\Line(146.7,52.3)(145.5,52.3) 
\Line(148.0,52.6)(146.7,52.3) 
\Line(149.5,52.6)(148.0,52.6) 
\Line(150.7,52.6)(149.5,52.6) 
\Line(152.2,52.6)(150.7,52.6) 
\Line(153.5,52.6)(152.2,52.6) 
\Line(154.7,52.6)(153.5,52.6) 
\Line(156.2,53.0)(154.7,52.6) 
\Line(157.5,53.0)(156.2,53.0) 
\Line(159.0,53.0)(157.5,53.0) 
\Line(160.2,53.0)(159.0,53.0) 
\Line(161.5,53.0)(160.2,53.0) 
\Line(163.0,53.0)(161.5,53.0) 
\Line(164.2,53.0)(163.0,53.0) 
\Line(165.7,53.0)(164.2,53.0) 
\Line(167.0,53.3)(165.7,53.0) 
\Line(168.2,53.3)(167.0,53.3) 
\Line(169.7,53.3)(168.2,53.3) 
\Line(171.0,53.3)(169.7,53.3) 
\Line(172.5,53.3)(171.0,53.3) 
\end{picture}}
\caption{Interference of the s-channel diagram $\chi\chi\rightarrow (H^0) \rightarrow Z^0Z^0$ and the t-channel diagram $\chi\chi\rightarrow (\chi^0_2) \rightarrow Z^0Z^0$.}\label{H2}
\end{minipage}
\end{figure}
\begin{figure}[h!]
\centering
\begin{minipage}{7.5cm}
{
\unitlength=1.0 pt
\SetScale{1.0}
\SetWidth{0.7}      % line    size control
\scriptsize    %  letter  size control
\begin{picture}(180,120)(0,0)
%\Text(104.9,119.7)[t]{$~o1,~o1 ->Z,Z$}
% ====================   X-axis =============
\LinAxis(37.55,33.18)(172.49,33.18)(4.000,2,-4,0.000,1.5)
\Text(37.6,27.5)[t]{$200$}
\Text(71.3,27.5)[t]{$400$}
\Text(105.1,27.5)[t]{$600$}
\Text(138.7,27.5)[t]{$800$}
\Text(172.5,27.5)[t]{$10^3$}
\Text(167.5,19.8)[rt]{$\sqrt(s)$}
% ====================   Y-axis =============
\LogAxis(37.55,33.18)(37.55,100.89)(2.277,4,1.806,1.5)
\Text(31.8,55.3)[r]{$10^-3$}
\Text(31.8,85.1)[r]{$0.01$}
\rText(6.8,100.9)[tr][l]{Cross Section [pb]}
% ============== end of axis ============
\Line(38.8,40.2)(37.6,33.2) 
\Line(40.1,44.9)(38.8,40.2) 
\Line(41.6,48.6)(40.1,44.9) 
\Line(42.8,51.6)(41.6,48.6) 
\Line(44.1,54.3)(42.8,51.6) 
\Line(45.6,56.3)(44.1,54.3) 
\Line(46.8,58.3)(45.6,56.3) 
\Line(48.3,60.3)(46.8,58.3) 
\Line(49.6,62.0)(48.3,60.3) 
\Line(50.8,63.7)(49.6,62.0) 
\Line(52.3,65.0)(50.8,63.7) 
\Line(53.6,66.4)(52.3,65.0) 
\Line(55.1,67.4)(53.6,66.4) 
\Line(56.3,68.7)(55.1,67.4) 
\Line(57.6,69.7)(56.3,68.7) 
\Line(59.1,70.4)(57.6,69.7) 
\Line(60.3,71.4)(59.1,70.4) 
\Line(61.8,71.7)(60.3,71.4) 
\Line(63.1,72.4)(61.8,71.7) 
\Line(64.3,72.4)(63.1,72.4) 
\Line(64.3,72.4)(65.8,71.7) 
\Line(65.8,71.7)(67.1,69.7) 
\Line(67.1,69.7)(68.3,61.0) 
\Line(69.8,91.8)(68.3,61.0) 
\Line(71.1,98.2)(69.8,91.8) 
\Line(71.1,98.2)(72.6,90.2) 
\Line(72.6,90.2)(73.9,87.8) 
\Line(73.9,87.8)(75.1,87.2) 
\Line(75.1,87.2)(76.6,86.8) 
\Line(77.9,86.8)(76.6,86.8) 
\Line(79.4,87.2)(77.9,86.8) 
\Line(80.6,87.5)(79.4,87.2) 
\Line(81.9,87.5)(80.6,87.5) 
\Line(83.4,87.8)(81.9,87.5) 
\Line(84.6,88.2)(83.4,87.8) 
\Line(86.1,88.5)(84.6,88.2) 
\Line(87.4,89.2)(86.1,88.5) 
\Line(88.6,89.5)(87.4,89.2) 
\Line(90.1,89.8)(88.6,89.5) 
\Line(91.4,90.2)(90.1,89.8) 
\Line(92.6,90.5)(91.4,90.2) 
\Line(94.1,90.8)(92.6,90.5) 
\Line(95.4,91.2)(94.1,90.8) 
\Line(96.9,91.5)(95.4,91.2) 
\Line(98.1,91.8)(96.9,91.5) 
\Line(99.4,91.8)(98.1,91.8) 
\Line(100.9,92.2)(99.4,91.8) 
\Line(102.1,92.5)(100.9,92.2) 
\Line(103.6,92.8)(102.1,92.5) 
\Line(104.9,93.2)(103.6,92.8) 
\Line(106.1,93.5)(104.9,93.2) 
\Line(107.6,93.5)(106.1,93.5) 
\Line(108.9,93.9)(107.6,93.5) 
\Line(110.4,94.2)(108.9,93.9) 
\Line(111.7,94.5)(110.4,94.2) 
\Line(112.9,94.5)(111.7,94.5) 
\Line(114.4,94.9)(112.9,94.5) 
\Line(115.7,95.2)(114.4,94.9) 
\Line(117.2,95.2)(115.7,95.2) 
\Line(118.4,95.5)(117.2,95.2) 
\Line(119.7,95.9)(118.4,95.5) 
\Line(121.2,95.9)(119.7,95.9) 
\Line(122.4,96.2)(121.2,95.9) 
\Line(123.7,96.2)(122.4,96.2) 
\Line(125.2,96.5)(123.7,96.2) 
\Line(126.4,96.9)(125.2,96.5) 
\Line(127.9,96.9)(126.4,96.9) 
\Line(129.2,97.2)(127.9,96.9) 
\Line(130.4,97.2)(129.2,97.2) 
\Line(131.9,97.5)(130.4,97.2) 
\Line(133.2,97.5)(131.9,97.5) 
\Line(134.7,97.9)(133.2,97.5) 
\Line(135.9,97.9)(134.7,97.9) 
\Line(137.2,98.2)(135.9,97.9) 
\Line(138.7,98.2)(137.2,98.2) 
\Line(139.9,98.2)(138.7,98.2) 
\Line(141.4,98.5)(139.9,98.2) 
\Line(142.7,98.5)(141.4,98.5) 
\Line(143.9,98.9)(142.7,98.5) 
\Line(145.5,98.9)(143.9,98.9) 
\Line(146.7,98.9)(145.5,98.9) 
\Line(148.0,99.2)(146.7,98.9) 
\Line(149.5,99.2)(148.0,99.2) 
\Line(150.7,99.6)(149.5,99.2) 
\Line(152.2,99.6)(150.7,99.6) 
\Line(153.5,99.6)(152.2,99.6) 
\Line(154.7,99.9)(153.5,99.6) 
\Line(156.2,99.9)(154.7,99.9) 
\Line(157.5,99.9)(156.2,99.9) 
\Line(159.0,100.2)(157.5,99.9) 
\Line(160.2,100.2)(159.0,100.2) 
\Line(161.5,100.2)(160.2,100.2) 
\Line(163.0,100.6)(161.5,100.2) 
\Line(164.2,100.6)(163.0,100.6) 
\Line(165.7,100.6)(164.2,100.6) 
\Line(167.0,100.6)(165.7,100.6) 
\Line(168.2,100.9)(167.0,100.6) 
\Line(169.7,100.9)(168.2,100.9) 
\Line(171.0,100.9)(169.7,100.9) 
\Line(172.5,100.9)(171.0,100.9) 
\end{picture}}
\caption{Interference of the s-channel diagram $\chi\chi\rightarrow (H^0) \rightarrow Z^0Z^0$ and the t-channel diagram $\chi\chi\rightarrow (\chi^0_3) \rightarrow Z^0Z^0$.}\label{H3}
\end{minipage}
\hspace{1cm}
\begin{minipage}{7.5cm}
{
\unitlength=1.0 pt
\SetScale{1.0}
\SetWidth{0.7}      % line    size control
\scriptsize    %  letter  size control
\begin{picture}(180,120)(0,0)
%\Text(104.9,119.7)[t]{$~o1,~o1 ->Z,Z$}
% ====================   X-axis =============
\LinAxis(37.55,33.18)(172.49,33.18)(4.000,2,-4,0.000,1.5)
\Text(37.6,27.5)[t]{$200$}
\Text(71.3,27.5)[t]{$400$}
\Text(105.1,27.5)[t]{$600$}
\Text(138.7,27.5)[t]{$800$}
\Text(172.5,27.5)[t]{$10^3$}
\Text(167.5,19.8)[rt]{$\sqrt(s)$}
% ====================   Y-axis =============
\LogAxis(37.55,33.18)(37.55,100.89)(3.762,4,8.801,1.5)
\Text(31.8,34.2)[r]{$10^-5$}
\Text(31.8,52.3)[r]{$10^-4$}
\Text(31.8,70.1)[r]{$10^-3$}
\Text(31.8,88.2)[r]{$0.01$}
\rText(6.8,100.9)[tr][l]{Cross Section [pb]}
% ============== end of axis ============
\Line(38.8,38.2)(37.6,33.2) 
\Line(40.1,41.2)(38.8,38.2) 
\Line(41.6,43.9)(40.1,41.2) 
\Line(42.8,45.9)(41.6,43.9) 
\Line(44.1,47.9)(42.8,45.9) 
\Line(45.6,49.6)(44.1,47.9) 
\Line(46.8,51.3)(45.6,49.6) 
\Line(48.3,52.6)(46.8,51.3) 
\Line(49.6,54.0)(48.3,52.6) 
\Line(50.8,55.3)(49.6,54.0) 
\Line(52.3,56.6)(50.8,55.3) 
\Line(53.6,58.0)(52.3,56.6) 
\Line(55.1,59.0)(53.6,58.0) 
\Line(56.3,60.3)(55.1,59.0) 
\Line(57.6,61.7)(56.3,60.3) 
\Line(59.1,62.7)(57.6,61.7) 
\Line(60.3,64.0)(59.1,62.7) 
\Line(61.8,65.4)(60.3,64.0) 
\Line(63.1,67.0)(61.8,65.4) 
\Line(64.3,68.7)(63.1,67.0) 
\Line(65.8,71.4)(64.3,68.7) 
\Line(67.1,74.7)(65.8,71.4) 
\Line(68.3,80.8)(67.1,74.7) 
\Line(69.8,101.2)(68.3,80.8) 
\Line(69.8,101.2)(71.1,81.1) 
\Line(71.1,81.1)(72.6,60.3) 
\Line(72.6,60.3)(73.9,48.6) 
\Line(75.1,50.6)(73.9,48.6) 
\Line(76.6,54.6)(75.1,50.6) 
\Line(77.9,57.3)(76.6,54.6) 
\Line(79.4,59.3)(77.9,57.3) 
\Line(80.6,61.0)(79.4,59.3) 
\Line(81.9,62.0)(80.6,61.0) 
\Line(83.4,63.0)(81.9,62.0) 
\Line(84.6,64.0)(83.4,63.0) 
\Line(86.1,64.7)(84.6,64.0) 
\Line(87.4,65.4)(86.1,64.7) 
\Line(88.6,66.0)(87.4,65.4) 
\Line(90.1,66.4)(88.6,66.0) 
\Line(91.4,67.0)(90.1,66.4) 
\Line(92.6,67.4)(91.4,67.0) 
\Line(94.1,67.7)(92.6,67.4) 
\Line(95.4,68.4)(94.1,67.7) 
\Line(96.9,68.7)(95.4,68.4) 
\Line(98.1,69.1)(96.9,68.7) 
\Line(99.4,69.4)(98.1,69.1) 
\Line(100.9,69.7)(99.4,69.4) 
\Line(102.1,70.1)(100.9,69.7) 
\Line(103.6,70.4)(102.1,70.1) 
\Line(104.9,70.7)(103.6,70.4) 
\Line(106.1,70.7)(104.9,70.7) 
\Line(107.6,71.1)(106.1,70.7) 
\Line(108.9,71.4)(107.6,71.1) 
\Line(110.4,71.7)(108.9,71.4) 
\Line(111.7,71.7)(110.4,71.7) 
\Line(112.9,72.1)(111.7,71.7) 
\Line(114.4,72.4)(112.9,72.1) 
\Line(115.7,72.4)(114.4,72.4) 
\Line(117.2,72.7)(115.7,72.4) 
\Line(118.4,72.7)(117.2,72.7) 
\Line(119.7,73.1)(118.4,72.7) 
\Line(121.2,73.4)(119.7,73.1) 
\Line(122.4,73.4)(121.2,73.4) 
\Line(123.7,73.7)(122.4,73.4) 
\Line(125.2,73.7)(123.7,73.7) 
\Line(126.4,74.1)(125.2,73.7) 
\Line(127.9,74.1)(126.4,74.1) 
\Line(129.2,74.4)(127.9,74.1) 
\Line(130.4,74.4)(129.2,74.4) 
\Line(131.9,74.4)(130.4,74.4) 
\Line(133.2,74.7)(131.9,74.4) 
\Line(134.7,74.7)(133.2,74.7) 
\Line(135.9,75.1)(134.7,74.7) 
\Line(137.2,75.1)(135.9,75.1) 
\Line(138.7,75.1)(137.2,75.1) 
\Line(139.9,75.4)(138.7,75.1) 
\Line(141.4,75.4)(139.9,75.4) 
\Line(142.7,75.8)(141.4,75.4) 
\Line(143.9,75.8)(142.7,75.8) 
\Line(145.5,75.8)(143.9,75.8) 
\Line(146.7,76.1)(145.5,75.8) 
\Line(148.0,76.1)(146.7,76.1) 
\Line(149.5,76.1)(148.0,76.1) 
\Line(150.7,76.4)(149.5,76.1) 
\Line(152.2,76.4)(150.7,76.4) 
\Line(153.5,76.4)(152.2,76.4) 
\Line(154.7,76.4)(153.5,76.4) 
\Line(156.2,76.8)(154.7,76.4) 
\Line(157.5,76.8)(156.2,76.8) 
\Line(159.0,76.8)(157.5,76.8) 
\Line(160.2,76.8)(159.0,76.8) 
\Line(161.5,77.1)(160.2,76.8) 
\Line(163.0,77.1)(161.5,77.1) 
\Line(164.2,77.1)(163.0,77.1) 
\Line(165.7,77.1)(164.2,77.1) 
\Line(167.0,77.4)(165.7,77.1) 
\Line(168.2,77.4)(167.0,77.4) 
\Line(169.7,77.4)(168.2,77.4) 
\Line(171.0,77.4)(169.7,77.4) 
\Line(172.5,77.8)(171.0,77.4) 
\end{picture}}
\caption{Interference of the s-channel diagram $\chi\chi\rightarrow (H^0) \rightarrow Z^0Z^0$ and the t-channel diagram $\chi\chi\rightarrow (\chi^0_4) \rightarrow Z^0Z^0$.}\label{H4}
\end{minipage}
\end{figure}\\
\\
Concluding this section, we may say that the main contribution to the large interference in the cross section of the process $\chi\chi \rightarrow Z^0Z^0$ is due to the interference between the s-channel diagrams exchanging $h^0$ and $H^0$. The interference between s-channel diagram containing $H^0$ and the t-channel diagrams with neutralino exchange may also have some impact on this behavior in the resonance region. Moreover, the t-channel diagram containing $\chi_3^0$ differs in its behavior from the other t-channel diagrams. However, it is difficult to understand all these features only by studying the cross sections of a selected subset of diagrams. Thus, finding out the main contributing diagrams in a process containing overall six diagrams is not at all a trivial task. In principle, in quantum mechanics, all diagrams interfere. Any subdivision (as here) is artificial. Still it gives good insights to understand this complex phenomenon.
\section*{Acknowledgements}
I would like to thank Nazila Mahmoudi and Gunnar Ingelman for great support, advice and interesting discussions. I am grateful to Gunnar Ingelman and Arnulf Quadt for giving me the opportunity to work in this exciting field of research in the framework of a project during my exchange semester in Uppsala, Sweden.

\end{document}